\documentclass{svjour3}                    % onecolumn
\smartqed  % flush right qed marks, e.g. at end of proof
\usepackage{amsmath}
\usepackage{amssymb}
\usepackage{graphicx}
\usepackage{bm,mathtools}
\usepackage{color}
\begin{document}

\title{Deriving hydrodynamic equations from dry active matter models in three dimensions
%\thanks{Grants or other notes
%about the article that should go on the front page should be
%placed here. General acknowledgments should be placed at the end of the article.}
}
%\subtitle{Do you have a subtitle?\\ If so, write it here}

\titlerunning{Deriving hydrodynamic equations from dry active matter models in 3D}        % if too long for running head

\author{Beno\^it Mahault  \and Aurelio Patelli \and Hugues Chat\'e
}

%\authorrunning{Short form of author list} % if too long for running head

\institute{F. Author \at
              first address \\
              Tel.: +123-45-678910\\
              Fax: +123-45-678910\\
              \email{fauthor@example.com}           %  \\
%             \emph{Present address:} of F. Author  %  if needed
           \and
           S. Author \at
              second address
              \and
}

\date{Received: date / Accepted: date}
% The correct dates will be entered by the editor

\maketitle

\begin{abstract}
We derive hydrodynamic equations from Vicsek-style dry active matter models in three dimensions (3D), building on our experience
on the 2D case using the Boltzmann-Ginzburg-Landau approach. 
The hydrodynamic equations are obtained from a Boltzmann equation expressed in terms of an expansion in spherical harmonics.
All their transport coefficients are given with explicit dependences on particle-level parameters.
The linear stability analysis of their spatially-homogeneous solutions is presented.
While the equations derived for the polar case (original Vicsek model with ferromagnetic alignment)  and their solutions
do not differ much from their 2D counterparts, the active nematics case exhibits remarkable differences:
%%%XXX this may be not so remarkable to some given that the situation is the same at equilibrium... CORRECT?
%%% XXX ---> this is true only at the non-fluctuating level, thus the real transitions can be different but we don't know yet. Therefore the point would be about the comparison between 2d and 3d instead of equilibrium vs active matter.
we find a true discontinuous transition to order with a bistability region, 
and cholesteric solutions whose stability we discuss.

 \PACS{PACS code1 \and PACS code2 \and more}
 \subclass{MSC code1 \and MSC code2 \and more}
\end{abstract}

\section{Introduction}
\label{intro}

For about twenty years now, active matter has attracted much attention from the soft matter and statistical physics communities \cite{marchetti2013hydrodynamics,ramaswamy2010mechanics}.
This field refers to assemblies of agents capable of producing systematic motion by consuming the energy present in their environment.
Many examples of active units can be found within living systems, ranging from animal groups \cite{cavagna2010scale,vicsek2012collective,attanasi2014collective} to bacteria, cells \cite{zhang2010collective,wensink2012meso,kawaguchi2017topological}, and {\it in vitro} mixtures of biofilaments and motor proteins \cite{schaller2010polar,sumino2012large,decamp2015orientational}.
But today many man-made active matter systems have been designed and studied
such as motile colloids \cite{jiang2010active,bricard2013emergence}, artificial microswimmers \cite{dreyfus2005microscopic}, or vertically-shaken granular particles \cite{deseigne2010collective,kumar2013flocking}. These two decades of work have revealed, beyond the intrinsic interest to biology, 
the emergence of new physics such as generic long-range correlations and anomalous fluctuations arising from short-range interactions, 
the possibility of long-range orientational order in 2D, and that of phase separation in the absence of explicit attractive interactions.

\begin{figure}[b!]
	\centering
	\includegraphics[scale=0.37]{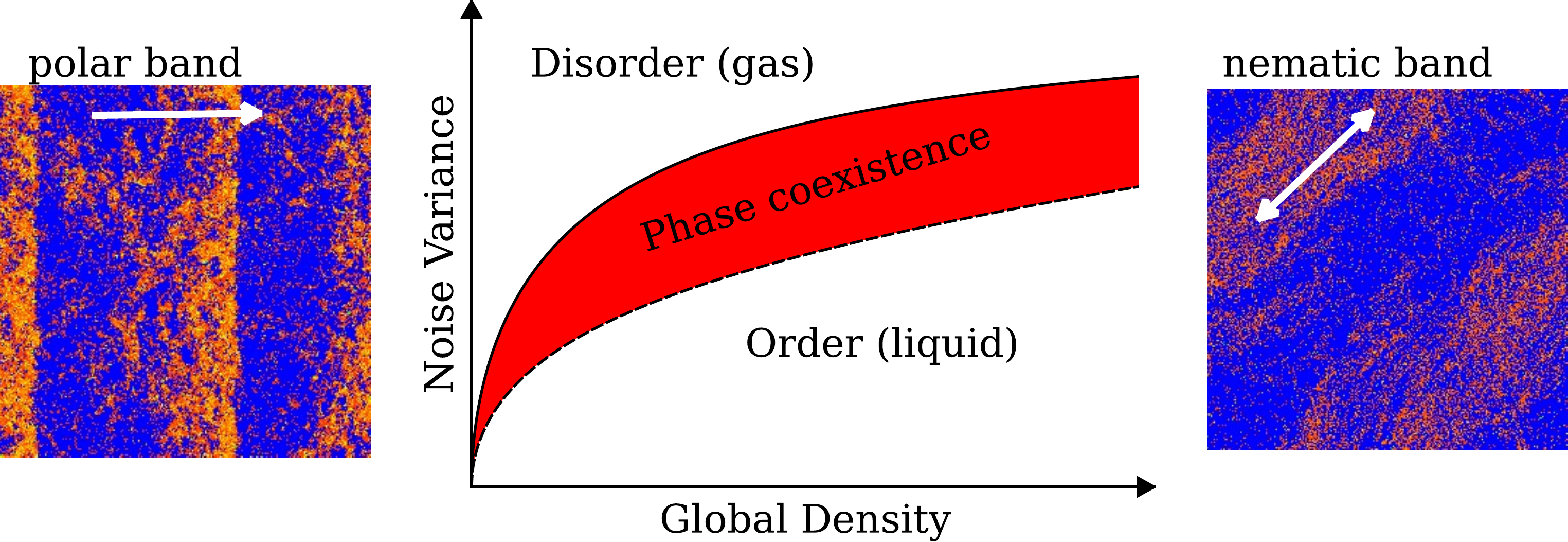}
	\caption{The central panel shows the schematic phase diagram of Vicsek-style models in two dimensions. All classes (polar, active nematics, rods) show a disordered gas phase at low densities and large noise, a (quasi)-ordered liquid phase for large densities and low noise, and a coexistence phase in between them.
	Left panel: snapshot of the coarse-grained density field in the coexistence phase of the polar class.
	Here two parallel high-density high-order bands travel from left to right above a residual disorder gas.
	Right panel: same as left panel but in the active nematics class. Here the nematic order is along the band. Such nematic bands do not travel ballistically as in the polar class, but they are known to be linearly unstable, so that the coexistence phase consists in a spatiotemporal chaos of bands.
	 }
	\label{fig:Vicsek_phase_diag}
\end{figure}

All active particles evolve surrounded by a fluid. This fluid, though, can be safely neglected in a number of situations such as granular particles. 
This ``dry limit'' is anyway worth studying because it is typically simpler. 
Arguably the simplest models of dry active matter dominated by alignment interactions are Vicsek-style models designed after the
1995 seminal work of Vicsek {\it et al.}\cite{vicsek1995novel}. 
Vicsek-style models consist in constant-speed point-like particles that, in the presence of noise, align their velocity with that of neighbors. 
The original Vicsek model consists of polar particles aligning ferromagnetically and is a member of the ``polar" class for which global polar order can emerge. 
Particles aligning nematically (i.e. roughly anti-aligning when incoming at obtuse angle), on the other hand, give rise to global nematic order, and are customarily
divided into two classes depending on whether they have a finite velocity reversal rate (active nematics) or not (``self-propelled rods''). 
Although the Vicsek framework may seem too restrictive, it has allowed to study in depth many of the fascinating properties of dry aligning active matter
\cite{chate2006simple,chate2008collective,ginelli2010large,ngo2014large}. 
In particular, work in 2D at both levels of microscopic simulations and hydrodynamic equations  has revealed that the transition of orientational order
is best described as a phase separation scenario between a disordered gas and an ordered fluid, so that there is usually no direct transition to order 
due to the presence of a coexistence phase where dense ordered structures evolve in the gas (Fig.~\ref{fig:Vicsek_phase_diag})\cite{ginelli2010large,ngo2014large,gregoire2004onset,grossmann2016mesoscale}.

%%%%%%%%%%%%%%%%%%%%%%%%%%%
%%%%%%%%%%%%%%%%%%%%%%%%%%%

Field theories describing the evolution of the slow modes of the dynamics in these models were first written on the basis of symmetry arguments \cite{toner1995long,toner1998flocks,ramaswamy2003active,toner2012reanalysis,mishra2010dynamic}.
Such a phenomenological approach can succeed in capturing the large-scale behavior of active matter systems, but is strongly impaired
by the lack of connection between the parameters governing the microscopic dynamics and the (typically) many transport coefficients present
at the hydrodynamic level.

Several direct coarse-graining approaches have been proposed to derive hydrodynamic equations from simple microscopic models,
which allows to keep a connection with microscopic-level parameters at hydrodynamic level. 
They have been successful in 2D, allowing, in the best cases, to recover at hydrodynamic level the phase diagram determined at particle level\cite{bertin2006boltzmann,baskaran2008enhanced,bertin2009hydrodynamic,ihle2011kinetic,farrell2012pattern,peshkov2012nonlinear,bertin2013mesoscopic,grossmann2013self,bertin2015comparison}. 
They are usually based on a kinetic equation for the single particle distribution, which is expressed in terms of Fourier angular modes and then truncated and closed. 
Among them, the ``Boltzmann-Ginzburg-Landau'' (BGL) framework \cite{peshkov2014boltzmann} is now known to allow for a controlled derivation of well-behaved nonlinear hydrodynamic equations.
In particular, the BGL approach is able to reproduce qualitatively the microscopic phase diagrams of Vicsek-style models, and provides a simple theoretical understanding of the phase-separation scenario  \cite{ngo2014large,peshkov2012nonlinear,caussin2014emergent,solon2015pattern,solon2015phase}.

Again, the above successes were all obtained in 2D. Not much is known in 3D. In particular, the connection between the microscopic and hydrodynamic levels
remains essentially unexplored. 
%(One exception is the work of Reinken {\it et al.} which, however, considers a suspensions of swimmers \cite{reinken2017derivation,Others?}, but no systematic study of the phase diagram has been performed.
%XXX----> So we don't cite this one?
In this paper, we apply the BGL formalism in three dimensions in order to derive hydrodynamic equations for the basic classes of dry aligning active matter,
starting from Vicsek-style models. We study the linear stability of the spatially-homogeneous solutions to these equations. All results are compared to the 2D case. 

In Section~\ref{sec:Micro} we define the microscopic models that we use as a starting point.
Section~\ref{sec:BGL} introduces the general framework for the BGL approach in three dimensions. In~\ref{subsec:Kinetic} we build the Boltzmann equation for the single particle distribution starting from the microscopic dynamics.
The decomposition of this distribution in terms of spherical harmonics and the connections to physical fields 
is presented in~\ref{subsec:SH}. The expression of the Boltzmann equation in terms of spherical harmonics is given in~\ref{subsec:Decomposition}.
Sections~\ref{sec:HE_polar} and \ref{sec:HE_active_nematic} are dedicated to the derivation of the hydrodynamic equations and the linear stability analysis of their homogeneous solutions for, respectively, systems with ferromagnetic and nematic alignment.
A brief discussion of our results and an outlook of future work can be found in Section~\ref{sec:conclusion}.

\section{Microscopic models}
\label{sec:Micro}

Vicsek-style models consist of $N$ pointlike particles moving at a constant speed $v_0$ in a periodic domain of volume ${\cal V}$.
The only interaction, competing with noise, is the local alignment of particle velocities.
Here velocity and position of particle $i$ are updated at discrete time steps following:
\begin{eqnarray}
	\vec{v}_i(t+1) &=& \left( {\cal R}_\eta \circ \vartheta \right)\langle \vec{v}(t) \rangle_{i} \\
	\vec{r}_i(t+1) &=& \vec{r}_i(t) +\varepsilon(t) \vec{v}_i(t +1)\;,
\end{eqnarray}
where
$\vartheta$ is an operator returning unit vectors ($\vartheta(\vec{u})=\vec{u}/\|\vec{u}\|$),
and ${\cal R}_\eta\vec{v}$ rotates the vector $\vec{v}$ by a random angle drawn from a uniform distribution inside the cap of surface $2\pi(1-\cos(\eta))$ (an arc of length $2\pi\eta$ in 2D) centered on $\vec{v}$.
In the polar and rods cases, $\varepsilon=1$, while in the active nematics case where velocity reversals occur, $\varepsilon=\pm1$ and 
changes sign with probability $\alpha$. 
Finally, $\langle \cdot \rangle_i$ stands for the (equal-weight) average of the polarities of all particles present in
 the local neighborhood of particle $i$ (including $i$ itself), $\partial_i$, defined in 3D (2D) by the sphere (disk) of radius $r_0$ centered on ${\vec r}_i$.
Particles interacting ferromagnetically align their polarities while nematic symmetry involves anti-alignment of polarities that initially point in opposite directions:
\begin{equation}
\langle \vec{v}(t) \rangle_i^{\rm f} = \sum_{j\in \partial_i} \vec{v}_j(t) \quad ; \quad \langle \vec{v}(t) \rangle_i^{\rm n} = \sum_{j\in \partial_i} {\rm sign}[\vec{v}_i(t)\cdot \vec{v}_j(t)  ] \vec{v}_j(t) \; .
\end{equation}
 
It is now well-known that the main parameters of these models are
the global number density $\rho_0 = N/{\cal V}$ and the noise strength $\eta$, while the speed of particles $v_0$ and the 
reversal rate $\alpha$ play only minor roles. As already shown previously \cite{bertin2013mesoscopic}, $r_0$ and $v_0$ can be set to unity
at the kinetic level, and thus also at hydrodynamic level. Anticipating our results, we will confirm that in 3D also, the reversal rate $\alpha$ 
has no qualitative influence (at least at the deterministic, linear level considered below), leaving us with the usual two main parameters $\rho_0$ and $\eta$.
In this parameter plane, the phase diagrams of these models for the 2D case all take the form depicted in Fig.~\ref{fig:Vicsek_phase_diag}.
In the polar case (ferromagnetic alignment and no velocity reversals), the liquid phase has true long-range polar order, and the coexistence phase is a smectic arrangement of dense, ordered, traveling bands moving in a disordered gas (left panel of Fig.~\ref{fig:Vicsek_phase_diag}). With nematic alignment, the liquid shows global nematic order 
(quasi-long-range for active nematics and possibly truly long-range for rods), and a spatiotemporal chaos of dense, ordered, nematic bands is observed in 
the coexistence phase (right panel of Fig.~\ref{fig:Vicsek_phase_diag}) \cite{ginelli2010large,ngo2014large,gregoire2004onset,peshkov2012nonlinear,solon2015phase}. 

%At large noise amplitude and/or low density the dynamics is dominated by fluctuations, thus the system settles in a disordered phase with short range correlations in space and time.
%When $\eta$ decreases and/or $\rho_0$ increases, particles start to move collectively along a given direction giving rise in 2D to long range order in the polar and rods cases \cite{ginelli2010large,toner1995long,toner1998flocks,toner2012reanalysis} and quasi long range order for active nematics \cite{chate2006simple,mishra2010dynamic,shankar2017low}.
%However this (quasi-)order is known to be unstable at the onset of collective motion, which leads to phase coexistence between a disordered gas and a smectic arrangement of traveling bands (polar) or nematic structures unstable to long wavelength perturbations inducing chaos (active nematics, rods) \cite{ginelli2010large,ngo2014large,gregoire2004onset,peshkov2012nonlinear}.
%We know that this phase separation scenario holds in the three dimensional version of the polar case \cite{chate2008collective}, with traveling structures extending the bands in the third dimension.
%However whether these objects have a finite length as in 2D or grow with system size remains an open question, so is the nature of the transition in the other classes.

\section{Boltzmann-Ginzburg-Landau approach in $3$ dimensions}
\label{sec:BGL}

In this section we describe the implementation of the Boltzmann-Ginburg-Landau (BGL) approach in three dimensions.
The path followed is the same as in two dimensions, and was described in detail in \cite{peshkov2014boltzmann}.
The aim is to derive hydrodynamic equations from microscopic models of dry aligning active matter, 
keeping track of the particle-level parameters in the transport coefficients. 
The starting point is a Boltzmann equation, considered a good approximation in the dilute regime, although
the results obtained in two dimensions have proven to remain qualitatively good even at high densities
as long as steric interactions do not become dominant. 
The Boltzmann equation governs a one-body density.
Expanding it in term of spherical harmonics modes of the orientations, a hierarchy of field equations is obtained.
A scaling ansatz valid near the onset of orientational order is then used to truncate and close this hierarchy, 
keeping only the slow modes.

\subsection{Building blocks of the Boltzmann equation}
\label{subsec:Kinetic}

The easiest way to transform the microscopic model in a time-continuous model is to consider that the tumbling events, given by the angular noise, become probabilistic with a time rate $\lambda \sim \frac{1}{\Delta t} = 1$.
This preserves the statistical properties of the angular noise of the microscopic model.
Therefore, a particle experiences a random variation of its direction of motion, drawn from a distribution of width $\sigma$
that plays the role of the microscopic angular noise strength $\eta$.

In the dilute limit and assuming the molecular chaos hypothesis~\cite{cercignani1969mathematical}, 
the evolution of the system can be reduced to the study of the evolution of the 
coarse-grained single particle distribution function $f(\vec{r},\vec{v},t)$, 
which counts the density of particles in a phase space domain of mesoscopic dimensions centered on $(\vec{r},\vec{v})$
where $\vec{r}$ is the spatial location of the particles and $\vec{v}= v_0\vec{e}$ is their velocity, with $\vec{e}$ a unit vector.
The dynamics of $f(\vec{r},\vec{v},t)$ is governed by the generic Boltzmann equation~\cite{peshkov2014boltzmann}
\begin{eqnarray}
	\partial_t f(\vec{r},\vec{v},t) +  \bar{v}\, \vec{e}\cdot\vec{\nabla} f(\vec{r},\vec{v},t) & = & D_0\Delta f(\vec{r},\vec{v},t) + D_1q_{\alpha\beta}\partial_\alpha\partial_\beta f(\vec{r},\vec{v},t) \nonumber \\
	& & - a\left[f(\vec{r},\vec{v},t)- f(\vec{r},-\vec{v},t)\right] + I_{{\rm sd}}[f]+I_{{\rm col}}[f]\;,
	\label{Boltzmann}
\end{eqnarray} 
where  $\vec{\nabla}= (\partial_x,\partial_y,\partial_z)$. 

The first line of equation \eqref{Boltzmann} is the general form of the free motion contribution, which we derive in detail now~\cite{peshkov2014boltzmann}.
In the case of polar particles, or when velocity reversals occur on timescales larger than the ones resolved by the Boltzmann equation, we trivially have $\bar{v} = v_0$ and $D_0 = D_1 = 0$.
On the other hand if the reversal rate of velocities is sent to infinity the free motion of particles is apolar at the kinetic level.
In that case particles update their positions with a random displacement $\tilde{\vec{v}}\Delta t$ drawn from the distribution
\begin{equation}
\Phi(\tilde{\vec{v}} - v_0\vec{e}) = \frac{1}{2}\left[\delta^{(3)}(\tilde{\vec{v}} - v_0\vec{e}) + \delta^{(3)}(\tilde{\vec{v}} + v_0\vec{e}) \right] \;.
\end{equation}
The corresponding evolution of $f$ can then be computed from It\^o calculus to second order and reads
\begin{equation}
\partial_t f(\vec{r},\vec{v},t) = \frac{v_0^2\Delta t}{6} \left(\Delta f(\vec{r},\vec{v},t) + 3q_{\alpha\beta}\partial_\alpha\partial_\beta f(\vec{r},\vec{v},t)\right)\;,
\end{equation}
where $q_{\alpha\beta} = e_\alpha e_\beta - \delta_{\alpha\beta}/3$ and summation over repeated indices is assumed.
We thus obtain the free transport terms in \eqref{Boltzmann} with $\bar{v} = 0$, $D_0 = v_0^2\Delta t/6$ and $D_1 = 3D_0$.
We note that these relations only hold for simple Vicsek-style dynamics while in more complicated cases $\bar{v}$, $D_0$ and $D_1$ can be different.
However as we will see in the following $\bar{v}$ can be eliminated by nondimensionalizing the Boltzmann equation and the diffusion coefficients do not affect qualitatively the results, therefore and for simplicity we focus this work on simple Vicsek-like models. 

The second line of equation \eqref{Boltzmann} regroups terms that account for the dynamics of velocities. 
From left to right we find an exchange term that models reversal of $\vec{v}$ at a finite (small) rate $a$ 
and the integrals describing angular self diffusion of velocities and collision events.

The integrals $I_{\rm sd}$ and $I_{\rm col}$ depend on the microscopic model as well. 
The microscopic dynamics can be described in terms of rotations of the direction of motion of the particles.
Rotation transformations in three dimensions belong to the $SO(3)$ group, which can be parameterized by the three Euler's angles.
Rotations of a generic vector $\vec{v}$ are thus obtained by
\begin{equation}
	\vec{v}' = R(\alpha,\theta,\psi)\vec{v} = R_z(\psi)R_y(\theta)R_z(\alpha)\vec{v}\;,
	\label{eq:Vel_Euler_angle}
\end{equation}
where $R_{i}(\varphi)$ represents a rotation around the $i$ axis of an angle $\varphi$ and $\alpha,\theta,\psi$ are the Euler's angles.
Any velocity vector can be obtained from rotations of the north pole $\vec{e}_z$, 
which is the unit vector pointing along the $z$ axis.
Such a rotation is only given by the two zenith and azimuthal angles $(\theta,\psi)$ 
(position on the unit sphere, $0\le\theta\le\pi$, $0\le\psi<2\pi$) since the north pole is invariant under rotations around the $z$ axis.
In order to lighten the notations we define the two angles as $\Omega = (\theta,\psi)$, and a velocity vector is represented by
\begin{equation}
	\vec{v}(\Omega) = v_0 \vec{e}(\Omega) = v_0 R(\Omega)\vec{e}_z \;.
\end{equation}

In order to map the action of the microscopic angular noise in the self-diffusion integral $I_{\rm sd}[f]$ 
we consider tumbling events that are rotations of the velocity of a particle.
We define the angular noise operator as 
\begin{equation}
	{\cal{P}}_{\Lambda}\vec{v}(\Omega)=R(\Omega)R(\Lambda)R^{-1}(\Omega)\vec{v}(\Omega)=v_0R(\Omega)R(\Lambda)\vec{e}_z \;,
\end{equation}
where the couple of angles $\Lambda$ is drawn from a probability distribution $P_{\rm sd}(\Lambda)$.
The noise operator first rotates the velocity direction to a reference direction $\vec{e}_z$, then it adds the noise in the form of a rotation and finally it rotates back the vector.
This way of applying the noise was chosen so that $P_{\rm sd}$ does not depend on the current orientation $\Omega$.
The self-diffusion operator is then given by the sum of a loss and a gain terms
\begin{eqnarray}
	I_{{\rm sd}}[f] &=& -\lambda f(\Omega)+\lambda\int d\Omega'\int d\Lambda P_{\rm sd}(\Lambda)f(\Omega')\delta^{(3)}\left(\vec{v}(\Omega) - {\cal{P}}_{\Lambda}\vec{v}(\Omega')\right) \nonumber \\
	&=& -\lambda f(\Omega)+\lambda\int d\Omega' f(\Omega')P_{\rm sd}(\text{Arg}[R^{-1}(\Omega')R(\Omega)\vec{e}_z]) \nonumber \\
	&=& -\lambda f(\Omega) + \lambda(f * P_{\rm sd})(\Omega) \;,
	\label{self_diff}
\end{eqnarray}
where $\int d\Omega$ stands for $\int_0^\pi \sin(\theta)d\theta \int_0^{2\pi} d\phi$.
The short notation $f(\Omega)$ stands for $f(\vec{r},\vec{v}(\Omega))$ where we have hidden the spatial dependency in order to lighten the notations.
The $\text{Arg}$ function returns the couple of angles defining the direction of a vector on the unit sphere, thus $\text{Arg}[\vec{e}(\Omega)] = \Omega$ .
The gain term of the self-diffusion corresponds to a kind of angular convolution\footnote{In our notation using both rotations and angles, the convolution operation between two functions $A(\Omega),B(\Omega)$ corresponds to $(A*B)(\Omega) = \int d\Omega_1 A(\Omega_1)B(\text{Arg}[R^{-1}(\Omega_1)R(\Omega)\vec{e}_z])$ where $\vec{e}_z$ is the north pole. In $2$ dimensions it correspond to the usual convolution operator because $\text{Arg}[R^{-1}(\Omega_1)R(\Omega)\vec{e}_z] = \theta - \theta_1$.}~\cite{AngularConvolution,baddour2010operational} between the distribution function $f$ and the noise probability $P_{\rm sd}$.

The collision integral is also the sum of a loss and a gain terms, considering the two processes by which the direction of motion of a reference particle can move away from or reach  $\Omega$ during a two-body collision
\begin{equation}
	I_{{\rm col}}[f]=-I_{{\rm col,loss}}[f]+I_{{\rm col,gain}}[f] \;.
	\label{coll}
\end{equation}
The loss part of the integral counts all the collisions of a reference particle moving initially along the direction $\Omega$ with the other particles at distance $r_0$ and moving in a direction parameterized by $\Omega'$
\begin{equation}
	I_{{\rm col,loss}}[f]=f(\Omega)\int d\Omega'f(\Omega')K(\Omega,\Omega') \;,
	\label{coll_loss}
\end{equation}
where the function $K(\Omega,\Omega')$ is the \textit{collision kernel} that measures the frequency of collisions.
The gain part of the integral, on the contrary, counts all the collisions after which the reference particle aligns its motion along the direction $\Omega$
\begin{eqnarray}
	I_{{\rm col,gain}}[f] &=&\int d\Omega_1\int d\Omega_2\int d\Lambda P_{\rm col}(\Lambda)f(\Omega_1)f(\Omega_2)K(\Omega_1,\Omega_2)\times \nonumber \\
	&\,& \delta^{(3)}\left(\left[R(\Omega)-{\cal{P}}_\Lambda R(\Psi(\Omega_1,\Omega_2))\right]\vec{e}_z\right) \; .
	\label{coll_gain}
\end{eqnarray}
The function $\Psi(\Omega_1,\Omega_2)$ returns the direction of the post collision state of the two aligned particles and $P_{\rm col}$ is the probability distribution of the collisional noise.

The collision integral varies with the symmetries of the microscopic dynamics that define the different classes.
First of all, the kernel is proportional to the relative velocity between the two colliding particles, and depends 
on how the two particles approach each other in the microscopic free motion.
In the propagative cases with polar particles not reversing their velocity, such as in the standard Vicsek and rods models, the kernel reads
$K_{\rm p}(\Omega_1,\Omega_2)=\pi r_0^2v_0\vert (R(\Omega_1)-R(\Omega_2))\vec{e}_z \vert$.
Thanks to its invariance under global rotations it depends only on the relative zenith angle between the particles
\begin{equation}
K_{\rm p}(\Omega_1,\Omega_2)= \pi r_0^2v_0\left|({Id}-R(\bar{\Omega}))\vec{e}_z\right|=\widetilde{K}_{\rm p}(\bar{\Omega}) = 2\pi r_0^2v_0\left|\sin\left(\frac{\bar{\theta}}{2}\right)\right| \,,
	\label{Kernel}
\end{equation}
where $R(\bar{\Omega})=R^{-1}(\Omega_1)R(\Omega_2)$.
Conversely, when particles reverse their velocity at some finite rate (e.g. in active nematics),
they locally diffuse and the same collision cannot discriminate if the colliding particle comes from the front or from the back.
In this case the collisional kernel reads
\begin{equation}
	K_{\rm a}(\Omega_1,\Omega_2)\sim |\vec{e}_1-\vec{e}_2|+|\vec{e}_1+\vec{e}_2| \;,
\end{equation}
since the particles move forward or backward to their velocity director with equal probability. 
In the reference frame of the particle $1$, the kernel for apolar particles reads
\begin{equation}
	\widetilde{K}_{\rm a}(\bar{\Omega})=\pi r_0^2v_0\left(\left|\sin\left(\frac{\bar{\theta}}{2}\right)\right|+\left|\cos\left(\frac{\bar{\theta}}{2}\right)\right| \right) \;,
	\label{apolar_kernel}
\end{equation}
where $\bar{\theta}$ is the relative zenith angle defining the orientation of the particle $2$ with respect to $1$.

The post collisional state of the particles is encoded in the function $\Psi$ that depends on the alignment rules.
In the case of polar (ferromagnetic) alignment the function $\Psi(\Omega_1,\Omega_2)$ returns the mean direction between 
$R(\Omega_1)\vec{e}_z$ and $R(\Omega_2)\vec{e}_z$.
Using the rotation properties of $\Psi$, the alignment rule is
\begin{equation}
	R\left(\Psi(\Omega_1,\Omega_2)\right)=R(\Omega_1)R\left(\widetilde{\Psi}(\bar{\Omega})\right) \;,
	\label{Alignment}
\end{equation}
where the aligned angle for ferromagnetic alignment is
\begin{equation}
	\widetilde{\Psi}_{\rm f}(\Omega)=\left(\frac{\theta}{2},\psi\right) \;.
	\label{eq:alignment_term}
\end{equation}
It corresponds to the mean direction of the colliding particles in the reference frame of the vector $\vec{v}(\Omega_1)$.
In the case of nematic alignment the post collisional direction is
\begin{equation}
\widetilde{\Psi}_{\rm n}(\Omega)=\left(\frac{\theta}{2}+h(\theta),\psi\right)\;\text{with}\left\{\begin{matrix}h(\theta)=0\;\text{if}\;0\le\theta\le\frac{\pi}{2}\\
h(\theta)=\frac{\pi}{2}\;\text{if}\;\frac{\pi}{2}<\theta\le\pi\end{matrix}\right.
\label{nematic_alignment}
\end{equation}
which is polar alignment at small relative angles and anti-alignment when the relative angle between the particles is larger than $\frac{\pi}{2}$.

In order to simplify the derivation of the hydrodynamic equations we consider that the distributions of the angular noise in the self-diffusion and collision processes are identical and that they are isotropic with respect to the azimuthal angle
\begin{equation}
	P_{\rm sd}(\Omega) = P_{\rm col}(\Omega) = P(\Omega) = P(\theta) \; .
	\label{eq:noise_distribution}
\end{equation}
The same symmetry is also present in the collision kernel $\widetilde{K}$.
Finally, we nondimensionalize the Boltzmann equation by rescaling space, time and the homogeneous density.
This is equivalent to set the speed $\bar{v} = v_0$ and the rate of angular self-diffusion $\lambda$ to unity without loss of generality (equivalently for apolar particles $\bar{v} = 0$ and we set $D_0$ to $\frac{1}{3}$). 
The interaction radius $r_0$ is eliminated defining the nondimensional homogeneous density $\tilde{\rho}_0 = \frac{2\pi r_0^2v_0}{\lambda}\rho_0$ (we remove the tilde in the following).
We are left with only three free parameters: the nondimensional homogeneous density $\rho_0$, the noise strength $\sigma$ and the nondimensional velocity reversal rate $a$.

\subsection{Generalities on spherical harmonics}
\label{subsec:SH}

In the following we manipulate functions (distributions) that depend on the two angles parameterizing the velocity (while the speed is kept constant).
This motivates us to decompose the distribution using Laplace's spherical harmonics (SH)~\cite{SphericalHarmonics}
\begin{equation}
	\begin{split}
	&f(\theta,\psi)=\sum_{l=0}^\infty\sum_{m=-l}^{l}\hat{f}_m^lY_l^m(\theta,\psi)=\sum_{l=0}^\infty\sum_{m=-l}^{l}\hat{f}_m^lY_l^m(\Omega) \;,\\
	&\hat{f}_m^l=\int_{0}^{\pi}\sin(\theta)d\theta\int_{0}^{2\pi}d\psi Y_l^{m^*}(\theta,\psi)f(\theta,\psi)=\int d\Omega Y_l^{m^*}(\Omega)f(\Omega)\;,
	\label{SH_decomposition}
	\end{split}
\end{equation}
where $\hat{f}^l_m$ are called hereafter the modes of the SH decomposition of the function $f(\Omega)$, or shortly the \textit{modes}.
The functions $Y_l^m$ are the spherical harmonics. They are defined by
\begin{equation}
	Y_l^m(\theta,\psi)=A_l^m{\cal L}_l^m(\cos(\theta))e^{\imath m\psi} \;,
	\label{SH}
\end{equation}
where $A_l^m=\sqrt{\frac{(2l+1)(l-m)!}{4\pi(l+m)!}}$ is a normalization constant and ${\cal L}_l^m$ is the associated Legendre polynomial of degree $l$ and order $m$ defined by:
\begin{equation}
	{\cal L}_l^m(x) = \frac{(-1)^m}{2^l l!} (1-x^2)^{\frac{m}{2}} \frac{d^{l+m}}{dx^{l+m}}(x^2-1)^l \;.
\end{equation}
From this definition and after lengthy but straightforward algebra we obtain the following useful recurrence relations
\begin{subequations}
\begin{eqnarray}
x{\cal L}_l^m(x) & = & \frac{1}{2l+1}\left((l-m+1){\cal L}_{l+1}^{m}(x)+(l+m){\cal L}_{l-1}^{m}(x)\right)\\
\sqrt{1-x^2}{\cal L}_l^m(x) & = & \frac{1}{2l+1}\left((l-m+1)(l-m+2){\cal L}_{l+1}^{m-1}(x) \nonumber \right. \\
& & \quad\quad\quad\quad\quad\quad\quad\quad\quad \left.-(l+m-1)(l+m){\cal L}_{l-1}^{m-1}(x)\right)\\
\sqrt{1-x^2}{\cal L}_l^m(x) & = & \frac{1}{2l+1}\left({\cal L}_{l-1}^{m+1}(x)-{\cal L}_{l+1}^{m+1}(x)\right) \;,
\end{eqnarray}
\label{Legendre}
\end{subequations}
for all $l\ge0, -l\le m \le l$ and $x\in [-1;1]$.

As a natural basis of $L^2(\mathcal{S}^2)$, the spherical harmonics are an orthogonal and normalized set:
\begin{equation}
\int d\Omega \, Y_{l_1}^{m_1}(\Omega) Y_{l_2}^{m_2^*}(\Omega) = \delta_{l_1l_2}\delta_{m_1m_2} \;.
\label{SH_norm}
\end{equation}
The SH decomposition of distributions (real, positive and normalizable functions) induces relations between the modes.
The \textit{reality} of the distribution implies that
\begin{equation}
	\hat{f}_{-m}^l = (-1)^m \hat{f}_m^{l*} \;.
\end{equation}
The \textit{positivity} of the distribution implies a bound on the modes
\begin{equation}
	\vert \hat{f}_m^l \vert  \le \int d\Omega \vert Y^{m*}_l(\Omega) \vert \vert f(\Omega) \vert = \int d\Omega \vert Y^{m*}_l(\Omega) \vert f(\Omega) \le \frac{A^m_l}{A_0^0} \hat{f}_0^0  \;,
\end{equation}
since the associated Legendre polynomials are functions of $\cos(\theta)$ and bounded to $1$ in the window $[-1:1]$.
This relation allows us to separate physical and unphysical solutions, since all the modes must be smaller or equal to the zero mode times a constant.
\footnote{The \textit{normalization} condition means that the distribution $f(\vec{r},\theta,\phi,t)$ is $L^1(\mathbb{R}^3\times\mathcal{S}^2)$ and the SH decomposition requires the modes to be $L^2(\mathbb{R}^3)$ integrable.}

The rotation of a spherical harmonic of degree $l$ is simply given by a linear combination in terms of spherical harmonics of same degree.
Denoting $R(\Omega')\vec{e}_z = R^{-1}(\alpha,\theta,\psi)R(\Omega)\vec{e}_z$, with $R(\alpha,\theta,\psi)$ defined in Eq.~\eqref{eq:Vel_Euler_angle}, we have
\begin{equation}
Y_l^m(\Omega') = \sum_{m' = -l}^{l} D^l_{m',m}(\alpha,\theta,\psi)Y_l^{m'}(\Omega) \;,
\label{SH_rot}
\end{equation}
where $D^l_{m',m}(\alpha,\theta,\psi) = d^l_{m',m}(\theta)\exp(-\imath m\alpha - \imath m' \psi)$ are the Wigner D-matrices, with
\begin{eqnarray}
d^l_{m',m}(\theta) & = & \left[(l+m')!(l-m')!(l+m)!(l-m)!\right]^{\frac{1}{2}} \times  \nonumber \\
& &  \sum_k \frac{(-1)^k \left[\cos\left(\frac{\theta}{2}\right)\right]^{2l + m - m' - 2k} \left[\sin\left(\frac{\theta}{2}\right)\right]^{m'-m + 2k}}{(l+m-k)!k!(m'-m+k)!(l-m'-k)!}
\end{eqnarray}
and the sum over $k$ is taken such that the factorials are non negative.
Note that in the following we will consider only the case corresponding to $\alpha = 0$ since we deal with rotations of vectors on the sphere. 
We therefore introduce the notation $D^l_{m',m}(\Omega)$ for $D^l_{m',m}(0,\theta,\psi)$, which are related to the spherical harmonics from
\begin{equation}
D^l_{m,0}(\Omega) = \sqrt{\frac{4\pi}{2l+1}}Y_l^{m^*}(\Omega) \;,
\label{SH_WD}
\end{equation}
and follow the condition
\begin{eqnarray}
\int d\Omega\, D^{l_1}_{m'_1,m_1}(\Omega) D^{l_2}_{m'_2,m_2}(\Omega) D^{l_3^*}_{m'_3,m_1 + m_2}(\Omega) & = & \frac{4\pi}{2l_3 + 1} \langle l_1\,l_2\,m'_1\,m'_2|l_3\,m'_3\rangle \times \nonumber \\ 
& &  \langle l_1\,l_2\,m_1\,m_2|l_3\,m_1+m_2\rangle \;,
\label{WD_norm}
\end{eqnarray}
with $\langle l_1\,l_2\,m_1\,m_2|l_3\,m_3\rangle$ the Clebsch-Gordan coefficient~\cite{Clebsch-Gordan} which is non zero only if $|l_1-l_2|\le l_3 \le l_1 + l_2$ and $m_1 + m_2 = m_3$.

\subsection{Relations between the spherical harmonics modes and the physical fields} 
\label{subsec:relation_fields_SH}

We first define the decomposition of the useful observable fields we are interested in.
In order to accomplish this we simplify the notations considering the functions
\begin{equation}
	\hat{g}^l_m(\vec{r},t)\equiv\frac{\hat{f}^l_m(\vec{r},t)}{A_l^m} \;.
\end{equation}
The density field is the zero mode
\begin{equation}
	\rho(\vec{r},t) = \frac{\hat{f}_0^0(\vec{r},t)}{\sqrt{4\pi}} = \hat{g}_0^0(\vec{r},t) \;.
	\label{eq:density_decomposition}
\end{equation}
The polar field as function of $\{\hat{g}^1_m\}$ reads
\begin{equation}
	\vec{w}(\vec{r},t)=\int d\vec{v}f(\vec{r},\vec{v},t)\vec{v}=\begin{pmatrix} 2\Re(\hat{g}_{-1}^{1}) \\ 2\Im(\hat{g}_{-1}^{1})\\ \hat{g}_0^{1}\end{pmatrix}(\vec{r},t) \;.
	\label{eq:polar_field_decomposition}
\end{equation}
Therefore when the global polar order points in the $z$ direction, the scalar order parameter is simply given by $\vert\hat{g}_0^1\vert$. 

The nematic tensor $\textbf{q}$ in terms of the director $\vec{n}$ is defined by
\begin{equation}
	\textbf{q}(\theta,\phi)=\vec{e}(\theta,\psi)^t\vec{e}(\theta,\psi)-\frac{1}{3}\textbf{I} \;,
	\label{eq:Nematic_definition_real}
\end{equation}
with $\textbf{I}$ the identity matrix.
Note that this tensor is traceless and symmetric.
Like the polar field, $\textbf{q}$ can be expressed in term of the spherical harmonics of order $l=2$.
Thus the nematic order parameter is related to the $\{\hat{g}_m^2\}$ modes by
\begin{equation}
	\textbf{Q}=\int d\Omega\, \textbf{q}(\Omega) f(\Omega)=2\begin{pmatrix} 2\Re(\hat{g}^{2}_{-2})-\frac{1}{6}\hat{g}^2_0 & \; 2\Im(\hat{g}^{2}_{-2}) & \;\; \Re(\hat{g}^{2}_{-1})\\
	2\Im(\hat{g}^{2}_{-2}) & \; -2\Re(\hat{g}^{2}_{-2})-\frac{1}{6}\hat{g}^2_0 &  \;\; \Im(\hat{g}^{2}_{-1})\\
	\Re(\hat{g}^{2}_{-1}) & \; \Im(\hat{g}^{2}_{-1}) & \;\; \frac{1}{3}\hat{g}^2_0
	\end{pmatrix} \;.
	\label{eq:Nem_order_parameter}
\end{equation}
The scalar nematic order parameter is usually defined as the largest eigenvalue (in absolute value) of the tensor $\textbf{Q}$ \cite{degennes1995physics}. 
In the reference frame where the nematic order lies along the $z$ direction, $\textbf{Q}$ is diagonal and the scalar nematic order parameter is given by $\hat{g}_0^2$. 

\subsection{Spherical harmonics decomposition of the Boltzmann equation}
\label{subsec:Decomposition}

Using the SH decomposition \eqref{SH_decomposition}, the advection part of the free transport operator, $-\vec{e}(\Omega)\cdot\vec{\nabla}$, can be recast into the following form
\begin{equation}
-\vec{e}(\Omega)\cdot\vec{\nabla}=\sqrt{\frac{2\pi}{3}}\left[Y_1^{1}(\Omega)\nabla^*-Y_1^{-1}(\Omega)\nabla-\sqrt{2}Y_1^{0}(\Omega)\partial_z\right] \;,
\end{equation}
where $\nabla = \partial_x + \imath \partial_y$ and $\nabla^* = \partial_x - \imath \partial_y$.
The decomposition of the corresponding term in the Boltzmann equation reads 
\begin{eqnarray}
 {\cal T}_m^l\left[\left\{\hat{g}_m^l\right\}\right] &  & \equiv -\frac{1}{A_l^m}\int d\Omega \, Y_l^{m^*}(\Omega)\vec{e}(\Omega)\cdot\vec{\nabla} f(\Omega)=\frac{1}{2(2l+1)}\nabla\left[\hat{g}^{l+1}_{m+1}-\hat{g}^{l-1}_{m+1}\right] \nonumber \\
	& &+\frac{1}{2(2l+1)}\nabla^*\left[(l+m-1)(l+m)\hat{g}^{l-1}_{m-1}-(l-m+1)(l-m+2)\hat{g}^{l+1}_{m-1}\right] \nonumber \\
	& &-\frac{1}{2l+1}\partial_z\left[(l-m+1)\hat{g}^{l+1}_{m}+(l+m)\hat{g}^{l-1}_{m}\right] \;,
	\label{kin_drift}
\end{eqnarray}
which follows from the recurrence relations \eqref{Legendre} between associated Legendre polynomials.

The Laplacian operator is isotropic and thus commutes with the SH decomposition.
On the contrary, the term associated to the anisotropic spatial diffusion operator,
$q_{\alpha\beta}\partial_\alpha\partial_\beta$, is transformed into
\begin{eqnarray}
	&\,&  {\cal D}_m^l\left[\left\{\hat{g}_m^l\right\}\right] \equiv \frac{1}{A_l^m}\int d\Omega\, Y_l^{m^*}(\Omega)q_{\alpha\beta}(\Omega)\partial_\alpha\partial_\beta f(\Omega)= \left[\frac{1}{(2l+1)(2l+3)}\right.\times \nonumber\\
	&\,&\left(\frac{(l-m+4)!}{4(l-m)!}\nabla^{*^2}\hat{g}^{l+2}_{m-2}+\frac{(l-m+3)!}{(l-m)!}\partial_z\nabla^*\hat{g}^{l+2}_{m-1}+\frac{(l-m+2)!}{2(l-m)!}\Box \hat{g}^{l+2}_m\right. \nonumber\\
&\,&\left.-(l-m+1)\partial_z\nabla \hat{g}^{l+2}_{m+1}+\frac{1}{4}\nabla^{2}\hat{g}^{l+2}_{m+2}\right) \nonumber \\
&\,&-\frac{1}{(2l-1)(2l+3)}\left(\frac{(l-m+2)!(l+m)!}{2(l-m)!(l+m-2)!}\nabla^{*^2}\hat{g}^{l}_{m-2}-\frac{l^2+l-3m^2}{3}\Box \hat{g}^{l}_m\right. \nonumber\\
&\,&\left.\left.+(2m-1)(l-m+1)(l+m)\partial_z\nabla^* \hat{g}^{l}_{m-1}+(2m+1)\partial_z\nabla \hat{g}^{l}_{m+1}+\frac{1}{2}\nabla^{2}\hat{g}^{l}_{m+2}\right)\right. \nonumber\\
&\,&+\frac{1}{(2l+1)(2l-1)}\left(\frac{(l+m)!}{4(l+m-4)!}\nabla^{*^2}\hat{g}^{l-2}_{m-2}-\frac{(l+m)!}{(l+m-3)!}\partial_z\nabla^* \hat{g}^{l-2}_{m-1}\right. \nonumber\\
&\,& \left.\left.+\frac{(l+m)!}{2(l+m-2)!}\Box \hat{g}^{l-2}_m +(l+m)\partial_z\nabla \hat{g}^{l-2}_{m+1}+\frac{1}{4}\nabla^{2}\hat{g}^{l-2}_{m+2}\right)\right] \;,
\label{kin_dif}
\end{eqnarray}
where $\Box = 2\partial^2_{zz}-\partial^2_{xx}-\partial^2_{yy}$.

Considering $l=0$ we get the continuity equation
\begin{eqnarray}
	\partial_t \rho = & &- \left[2\Re(\nabla^*\hat{g}_{-1}^1)+\partial_z\hat{g}_0^1\right] \nonumber\\
	& & + D_0\Delta \rho + D_1\left[4\Re\left(\nabla^{*^2}\hat{g}_{-2}^2\right) + 4\Re\left(\partial_z\nabla^*\hat{g}_{-1}^2\right) + \frac{1}{3}\Box \hat{g}_0^2\right] \;,
\end{eqnarray}
which exhibits similar terms as in 2D: if the dynamics of particles is propagative the density will be advected by the polar field as pointed out on the first line.
On the other hand the terms in the second line reflect a diffusive dynamics with isotropic spatial diffusion of the density and advection by the curvature induced current generated by the nematic field.

The SH decomposition of the velocity reversal term is straightforward knowing how spherical harmonics transform under parity: $Y_l^m(\Omega) \rightarrow Y_l^m({\rm\Pi}\Omega) = (-1)^lY_l^m(\Omega)$ with ${\rm\Pi}\Omega = (\pi-\theta,\psi+\pi)$.
This relation yields
\begin{equation}
-\frac{a}{A_l^m}\int d\Omega\, Y_l^{m^*}(\Omega) \left(f(\Omega) - f({\rm\Pi}\Omega) \right) = -a\left(1 - (-1)^l\right)\hat{g}^l_m \;.
\end{equation}

The angular self-diffusion operator, defined in \eqref{self_diff}, can be seen as a convolution between the distribution and the noise probability.
Therefore as in the $2$ dimensional case, the SH decomposition of the self-diffusion term of the Boltzmann equation 
is simply given by the multiplication of the two modes of the convoluted functions.
\begin{equation}
	\frac{1}{A_l^m}\int d\Omega Y_l^{m^*}(\Omega)I_{\text{\rm sd}}[f]=-\hat{g}^l_m+\sqrt{\frac{4\pi}{2l+1}}\hat{P}^{l}_{0}\hat{g}^{l}_{m}=-\hat{g}^l_m+\widetilde{P}^{l}_{0}\hat{g}^{l}_{m} \;,
	\label{kin_self}
\end{equation}
where $\widetilde{P}^{l}_{0}=\hat{P}^{l}_{0}/A_l^0$ and $\hat{P}^l_0$ is the SH mode of degree $l$ of the noise distribution \eqref{eq:noise_distribution}.
Note that from the symmetry of $P$ its modes are different from $0$ only for $m = 0$.

We now turn to the spherical harmonics decomposition of the collision gain term \eqref{coll_gain}.
In the following and for simplicity we will use a slight abuse of notations and denote by $R(\Omega_1)\Omega_2$ the orientation of the vector given by $R(\Omega_1)R(\Omega_2)\vec{e}_z$.
Using the rotational symmetry properties of the kernel and alignment rule \eqref{Kernel} and \eqref{Alignment},
with $\bar{\Omega} = R^{-1}(\Omega_1)\Omega_2$, the collision integral reads
\begin{eqnarray}
	I_{{\rm col,gain}}[f] &=&\int d\Omega_1\int d\bar{\Omega}\, P\left(R^{-1}\left(\widetilde{\Psi}(\bar{\Omega})\right)R^{-1}(\Omega_1)\Omega\right)f(\Omega_1)f(R(\Omega_1)\bar{\Omega})\widetilde{K}(\bar{\Omega}) \;.\;\;\;\;\;\;\;
\end{eqnarray}
Expanding the distributions in spherical harmonics modes, with the rotation identity \eqref{SH_rot} and the axial symmetry of $P$ \eqref{eq:noise_distribution}, gives
\begin{eqnarray}
	\frac{1}{A_l^m} \int d\Omega\, Y_l^{m^*}(\Omega) I_{{\rm col,gain}}[f] &=& \frac{1}{A_l^m} \sum_{l_1,m_1} A_{l_1}^{m_1} \hat{g}^{l_1}_{m_1} \sum_{l_2,m_2} A_{l_2}^{m_2} \hat{g}^{l_2}_{m_2} \sum_{l_3} \hat{P}_{0}^{l_3} \times \nonumber \\
	& & \sum_{m'_2, m'_3, m''_3} \int d\Omega\, Y_l^{m^*}(\Omega) Y_{l_3}^{m''_3}(\Omega) \times \nonumber \\
	& & \int d\Omega_1 Y_{l_1}^{m_1}(\Omega_1) D^{l_2^*}_{m_2,m'_2}(\Omega_1) D^{l_3}_{m''_3,m'_3}(\Omega_1) \times \nonumber \\
	& & \int d\bar{\Omega}\, Y_{l_2}^{m'_2}(\bar{\Omega}) D^{l_3}_{m'_3,0}\left(\widetilde{\Psi}(\bar{\Omega})\right)\widetilde{K}(\bar{\Omega}) \;,
\end{eqnarray}
where $\sum_{l,m}$ is the shortened form of $\sum_{l=0}^{\infty}\sum_{m=-l}^l$ and the sum over integers $m$ with index $i$ are taken between $-l_i$ and $l_i$.
We now make use of the orthogonality relation \eqref{SH_norm} and the identity \eqref{WD_norm} together with the correspondence between spherical harmonics and Wigner D-matrices \eqref{SH_WD}, this yields
\begin{eqnarray}
	\frac{1}{A_l^m} \int d\Omega\, Y_l^{m^*}(\Omega) I_{{\rm col,gain}}[f] &=& \frac{\hat{P}_{0}^l}{A_l^0} \sum_{l_1,m_1}\sum_{l_2,m_2} \frac{A_{l_1}^{m_1}A_{l_2}^{m_2}}{A_l^m}\sqrt{\frac{2l_1+1}{2l+1}} \langle l_1\,l_2\,m_1\,m_2|l\,m\rangle \times \nonumber \\   
	& & \left[\sum_{m'_2 = -{\rm Min}(l,l_2)}^{{\rm Min}(l,l_2)}\widetilde{K}^{l_2,l}_{m'_2}\langle l_1\,l_2\,0\,m'_2|l\,m'_2\rangle\right]\hat{g}^{l_1}_{m_1}\hat{g}^{l_2}_{m_2} \;,
\end{eqnarray}
where, from the properties of the Clebsch-Gordan coefficients, the sum over $l_2$ is now taken between $|l-l_1|$ and $l + l_1$, and 
\begin{equation}
\widetilde{K}^{l_1,l_2}_{m_2} = \sqrt{\frac{4\pi}{2l_2+1}}\int d\bar{\Omega}\, Y_{l_1}^{m_2}(\bar{\Omega})Y^{m_2^*}_{l_2}\left(\widetilde{\Psi}(\bar{\Omega})\right) \;.
\end{equation}
Note that the decomposition of the angular self diffusion integral \eqref{kin_self} can be obtained using the same properties of spherical harmonics as the ones that have been employed for this calculation. 
After computation of the SH decomposition of the loss part of the collision integral, which follows straightforwardly from this derivation and is thus not detailed here,
the decomposition of the full collision term reads
\begin{equation}
	\frac{1}{A_l^m}\int d\Omega\, Y_l^{m^*}(\Omega)I_{\text{\rm col}}[f]=\sum_{l_1,m_1}\sum_{l_2=|l-l_1|,m_2}^{l+l_1}J^{\,l,l_1,l_2}_{m,m_1,m_2}\hat{g}_{m_1}^{l_1}\hat{g}_{m_2}^{l_2} \;,
	\label{kin_coll}
\end{equation}
where 
\begin{equation}
	J^{\,l,l_1,l_2}_{m,m_1,m_2}=\widetilde{P}_{0}^l\widetilde{J}^{\,l,l_1,l_2}_{m,m_1,m_2}-\widetilde{I}^{\,l,l_1,l_2}_{m,m_1,m_2} \;,
\end{equation}
and
\begin{eqnarray*}
	\widetilde{J}^{\,l,l_1,l_2}_{m,m_1,m_2} &=&\frac{A_{l_1}^{m_1}A_{l_2}^{m_2}}{A_l^m}\sqrt{\frac{2l_1+1}{2l+1}}\langle l_1\,l_2\,m_1\,m_2|l\,m\rangle \delta_{m,m_1+m_2}\times \nonumber \\
	&\,& \left[\sum_{m'_2=-{\rm Min}{(l_2,l)}}^{{\rm Min}{(l_2,l)}}\widetilde{K}^{l_2,l}_{m'_2}\langle l_1\,l_2\,0\,m'_2|l\,m'_2\rangle\right]\, , \\
	\widetilde{I}^{\,l,l_1,l_2}_{m,m_1,m_2}&=& \frac{A_{l_1}^{m_1}A_{l_2}^{m_2}}{A_l^m}\sqrt{\frac{2l_1+1}{2l+1}}\hat{K}^{l_2}_{0}\langle l_1\,l_2\,m_1\,m_2|l\,m\rangle\langle l_1\,l_2\,0\,0|l\,0\rangle\;.
\end{eqnarray*}
The function $\hat{K}^{l}_{m}$ is the SH decomposition of the collision kernel and it is zero $\forall\, m\neq0$ thanks to the global rotation invariance of the system.

Collecting all the SH transformed terms of the kinetic equation we obtain the mode decomposition of the $3$ dimensional Boltzmann equation \eqref{Boltzmann}
\begin{eqnarray}
	\partial_tg^l_m & = & {\cal T}_m^l\left[\left\{g_m^l\right\}\right] + D_0\Delta g_m^l + D_1{\cal D}_m^l\left[\left\{g_m^l\right\}\right] -a\left(1 + (-1)^l\right)g_{m}^l \nonumber \\
	& & +\left[P_{0}^l-1\right]g_{m}^l+\sum_{l_1,m_1}\sum_{l_2=|l-l_1|,\,m_2}^{l+l_1}J^{\,l,l_1,l_2}_{m,m_1,m_2}g_{m_1}^{l_1}g_{m_2}^{l_2} \;,
	\label{Kin_eq}
\end{eqnarray}
where we have removed the hats and tildes in order to simplify the notations. 

Although this computation holds for any axisymmetric noise distribution, numerical evaluations of the coefficients of the hydrodynamic equations derived in the following have been done using Gaussian weights $P_0^l = \exp(-l^2\sigma^2/2)$.
We have checked that results that are presented are not qualitatively influenced by the precise form of the distribution.

%%%%%%%%%%%%%%%%%%%%%%%%%%%
\section{Hydrodynamic equations for ferromagnetic alignment}
\label{sec:HE_polar}

In this section we derive hydrodynamic equations for particles which align their velocities in a ferromagnetic way.
In the restricted Vicsek framework, this symmetry of the interaction requires the particles to exhibit polar free motion in order to be able to generate spontaneous order.
Therefore, we only consider Vicsek-like particles moving at constant speed with no velocity reversal.

%%%%%%%%%%%%%%%%%%%%%%%%%%%
\subsection{Derivation of the hydrodynamic equations}
\label{subsec:HE_der_polar}
%%%%%%%%%%%%%%%%%%%%%%%%%%%

As argued in Sec.~\ref{subsec:Kinetic}, the Boltzmann equation does not show any spatial diffusion nor velocity reversal term:
\begin{equation}
	\partial_tg^l_m = {\cal T}_m^l\left[\left\{g_m^l\right\}\right] +\left[P_{0}^l-1\right]g_{m}^l+\sum_{l_1,m_1}\sum_{l_2=|l-l_1|,\,m_2}^{l+l_1}J^{\,l,l_1,l_2}_{m,m_1,m_2}g_{m_1}^{l_1}g_{m_2}^{l_2} \;.
	\label{Kin_eq_polar}
\end{equation}
The system of equations~\eqref{Kin_eq_polar} exhibits a trivial solution: the homogeneous disordered (HD) solution,
existing for any value of the microscopic parameters $(\rho_0,\sigma)$.
For this solution the zero mode is equal to the homogeneous density: $g^0_0=\rho_0$, and all the other modes vanish: $g^l_m = 0$ for all $l > 0$ and $m$.
Linearizing the Boltzmann hierarchy~\eqref{Kin_eq_polar} around the HD state with respect to homogeneous perturbations (space independent), the modes evolve as
\begin{subequations}
\begin{eqnarray}
	g_0^0&=&\rho=\rho_0+\delta \rho, \qquad g_m^l=\delta g_m^l \quad \forall l>0, \\
	\partial_t\delta g^l_m&=&\left[\left(P_{0}^l-1\right)+\left(J^{\,l,0,l}_{m,0,m}+J^{\,l,l,0}_{m,m,0}\right)\rho_0\right]\delta g^l_{m},\\
	&\equiv&\mu^l_{m}[\rho_0]\delta g^l_{m},
\end{eqnarray}
	\label{lin_Kin_eq}
\end{subequations}
%%%%%%%%%%%%%%%%%%%%%%%%%%%
\begin{figure}[t!]
	\centering
	\includegraphics[scale=0.5]{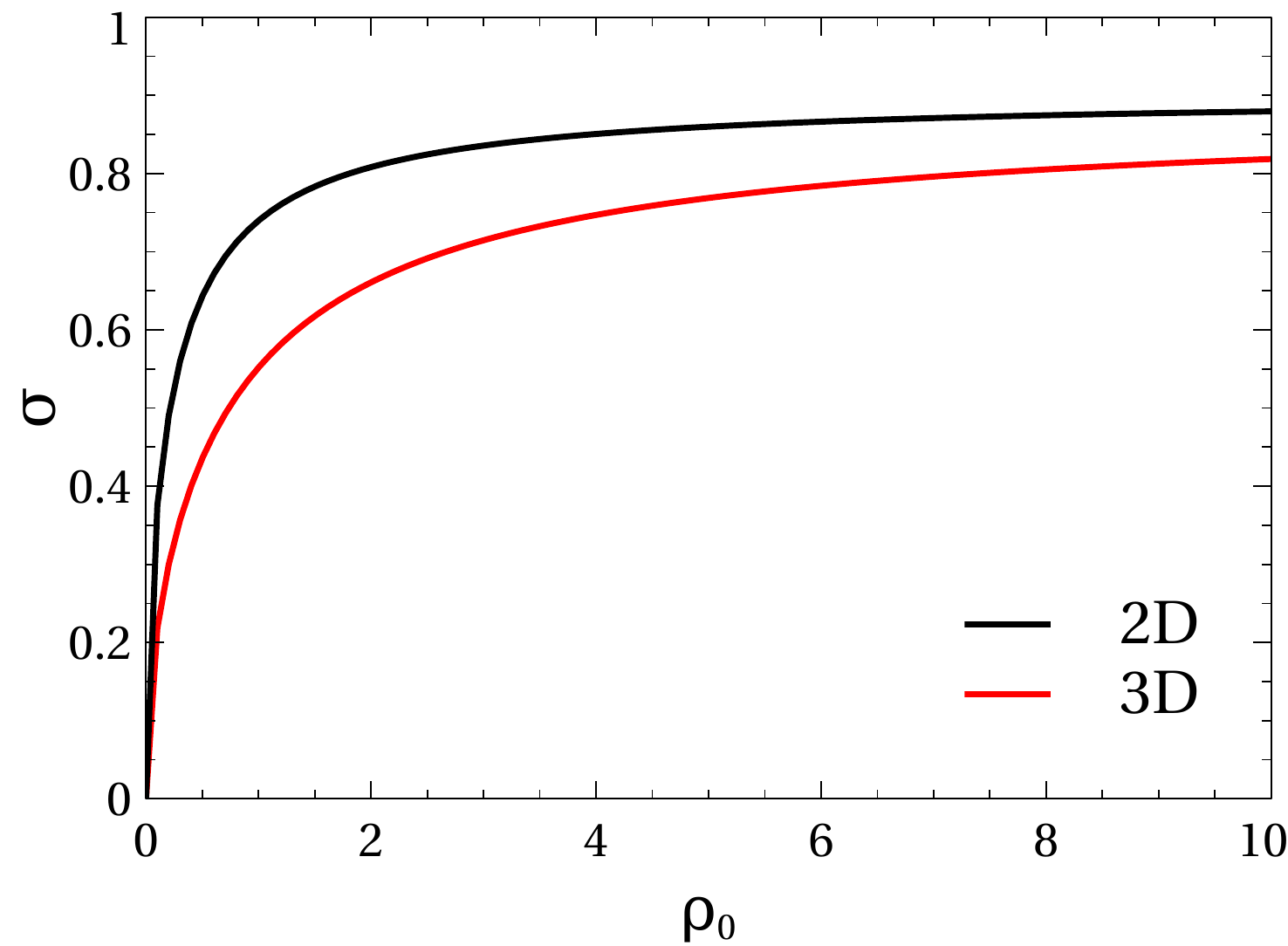}
	\caption{Linear stability limit of the homogeneous disordered solution in the density-noise plane for the polar class.
	The transition line given by $\mu^1=0$, the stable region is above. The black and red lines, below which the disordered solution is unstable to homogeneous perturbations, correspond respectively to the 2D and 3D cases.}
	\label{fig:Transitions}
\end{figure}
%%%%%%%%%%%%%%%%%%%%%%%%%%%
where the linear coefficients $\mu^l_m$ are given by
\begin{equation}
\mu^l_m[\rho_0] = P_0^l - 1 + \frac{\rho_0}{\sqrt{4\pi}}\left[P_0^l \left(\frac{1}{\sqrt{2l+1} }\sum_{m'=-l}^l \tilde{K}^{l,l}_{m'} + \tilde{K}^{0,l}_0 \right) - \frac{\hat{K}^l_0}{\sqrt{2l+1}} - \hat{K}^0_0   \right] \;.
\end{equation}
Note that, as a consequence of global rotational invariance, they do not depend on the index $m$, which we omit in the following.
The HD state is linearly stable when all the linear coefficients $\mu^l$ are negative, 
while the physical field associated to $g^l$ grows when $\mu^l$ becomes positive.
For $l=0,1,2$ the coefficients are
\begin{subequations}
\begin{eqnarray}
	\mu^0&=&0 \;,\\
	\mu^1&=&(P^1_{0}-1)+\left(\frac{\pi}{4}P^1_{0}-\frac{8}{15}\right)\rho_0 \;,\\
	\mu^2&=&(P^2_{0}-1)+\left(\frac{2}{15}P^2_{0}-\frac{68}{105}\right)\rho_0\quad ( \le 0) \;.
\end{eqnarray}
\end{subequations}
The linear coefficient $\mu^1$ associated to the polar field becomes positive at large densities $\rho_0$ and 
small angular noise strength $\sigma$.
Figure~\ref{fig:Transitions} shows the line $\sigma_t(\rho_0)$ along which the coefficient $\mu^1=0$ in the ($\rho_0,\sigma$) plane,
comparing the $2$ dimensional~\cite{bertin2009hydrodynamic} and the $3$ dimensional results.
The HD state is stable above this line. 
Below, the Boltzmann equation possesses another homogeneous solution with polar order that we call the homogeneous ordered (HO) state.
We checked until $l=10$ that the linear coefficients $\mu^l$ are more and more negative with $l$. 
We assume that for larger values of $l$ they do the same.

%%%%%%%%%%%%%%%%%%%%%%%%%%%
\begin{figure}[t!]
	\centering
	\includegraphics[scale=0.5]{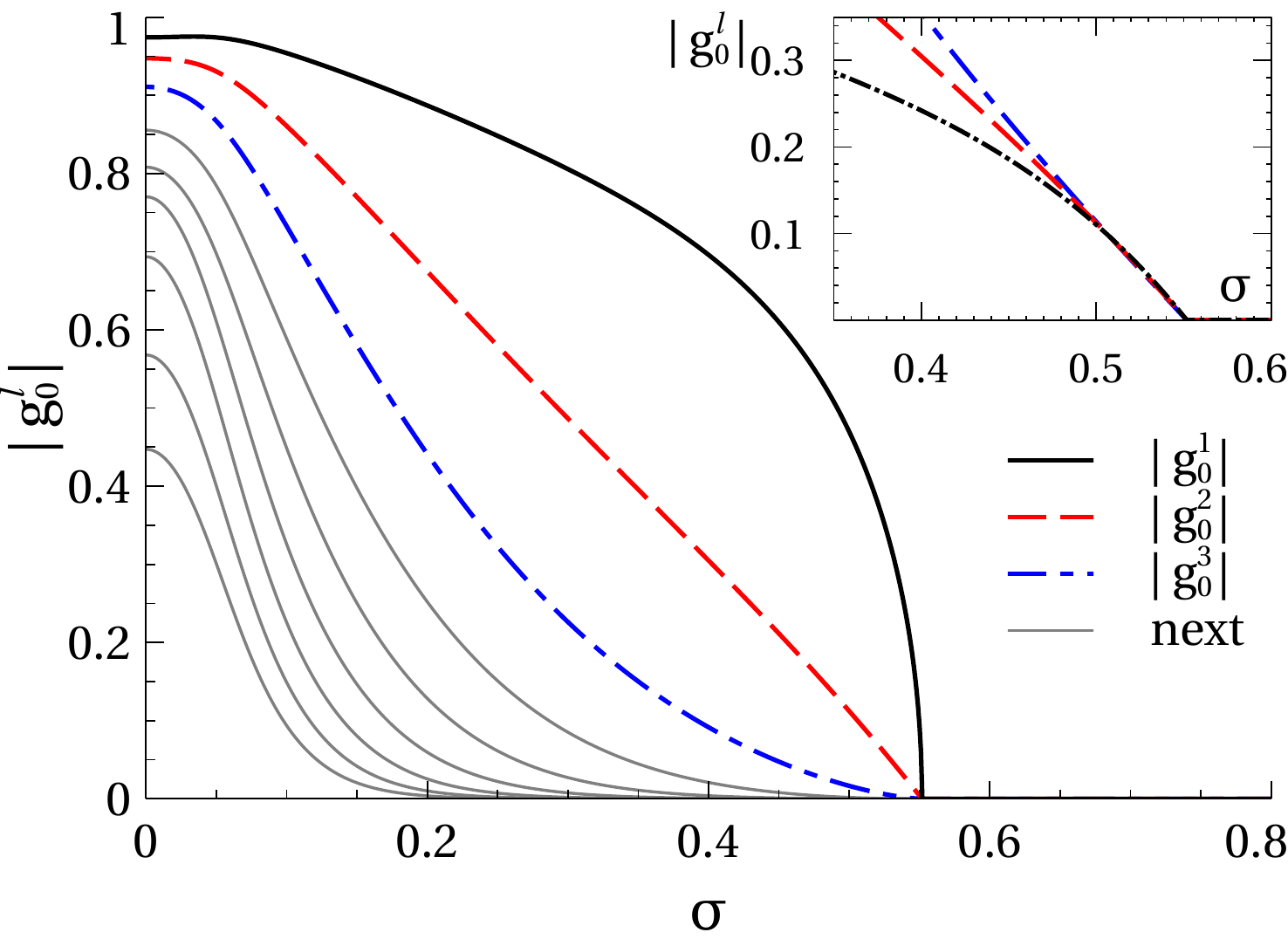}
	\caption{{Numerical evaluation of the homogeneous ordered solution of the Boltzmann equation for the polar class truncating the hierarchy at the $10^{\rm th}$ mode
			(density $\rho_0=1$).
			The solid black line indicates the first mode $\vert g^1_0 \vert$, saturating at a finite value.
			The dashed red and dashed-dotted blue lines correspond respectively to the second and third modes.
			The next modes from $l=4$ to $l=9$ are shown with grey lines.
			The inset shows the scaling between $\sim \vert g^1_0 \vert^2$, $\vert g^2_0 \vert$ and $\sim \vert g^3_0 \vert^\frac{2}{3}$ close to the transition point.}}
	\label{fig:Homogeneous_modes}
\end{figure}
%%%%%%%%%%%%%%%%%%%%%%%%%%%

From the global rotational invariance of the system we can choose the reference frame such that the global polar order is along the $z$ axis.  
In Section~\ref{subsec:relation_fields_SH} we have shown that this comes down to set the $g^1_m$ fields to $0$ for any $m \ne 0$.
Then from the property of the nonlinear coefficients in the Boltzmann equation, $J^{\,l,l_1,l_2}_{m,m_1,m_2} = J^{\,l,l_1,l_2}_{m,m_1,m_2}\delta_{m,\,m_1 + m_2}$, the numerical evaluation of its HO solution can be done considering only real $g^l_0$ fields for all $l$.

%We calculated numerically the HO state close to and below the transition line $\sigma_t(\rho_0)$.
Figure~\ref{fig:Homogeneous_modes} shows the result for the Boltzmann equation~\eqref{Kin_eq} 
truncated at the $10^{\rm th}$ mode with homogeneous density $\rho_0=1$, varying the noise strength $\sigma$.
(We checked that results are stable considering up to 20 modes.)
As expected, $|g^1_0|$ grows below the critical noise $\sigma_c\simeq0.551$ like  $\sqrt{\sigma_{\rm c}-\sigma}$,
and all the others modes grow consequently due to nonlinear couplings in the Boltzmann equation.
 The inset of Figure~\ref{fig:Homogeneous_modes} shows that the first three modes exhibit a scaling behavior in the vicinity of the transition: 
comparing $\vert g^2_0 \vert$ with $\vert g^1_0 \vert^2$ and $\vert g^3_0 \vert^\frac{2}{3}$ we show that they fall on the same curve. 
(This is also true for the next modes, but they are not shown for clarity.)
Defining the scaling small parameter $\varepsilon$ by the amplitude of the polar field ($|g^1|\approx\varepsilon$), we thus find that
\begin{equation}
	g^l_m \approx \varepsilon^{l}, \qquad \forall l>0, \quad \forall m \;,
\end{equation}
which we assume to be true in all generality.

Moreover, it is assumed that the spatial and the temporal variations of the modes are small and comparable in magnitude to the small parameter introduced above.
For systems such as the $3$ dimensional polar class considered here, one uses 
the \textit{propagative ansatz} where the temporal variations are of the same order as the spatial variations:
\begin{equation}
	\nabla \approx \nabla^* \approx \partial_x \approx \partial_z \approx \partial_t \approx \varepsilon \;,
	\label{eq:PropagativeAnsatz}
\end{equation}
implying that the density variation goes as $\delta \rho\approx\varepsilon$ in order to balance the first equation of the hierarchy.

Using this scaling ansatz, we expand the Boltzmann hierarchy in series of $\varepsilon$ and we truncate it at the first
non-trivial order, $\varepsilon^3$. 
This leads to equations for the density, polar and nematic fields.
The density and polar fields are governed by
\begin{subequations}
\begin{eqnarray}
\partial_t\rho&=&-\nabla^*g^1_{-1}+\frac{1}{2}\nabla g^1_1-\partial_zg^1_0\\
\partial_tg_{-1}^1&=&-2\nabla^*g^2_{-2}+\frac{1}{6}(\nabla g^2_0-\nabla\rho)-\partial_zg^2_{-1}+\mu^1[\rho]g^1_{-1}+X^{1\,1\,1}_{-1\,-1\,0}\;g^1_0g^1_{-1} \nonumber \\
&\,&+X^{1\,1\,2}_{-1\,-1\,0}\;g^2_0g^1_{-1}+X^{1\,1\,2}_{-1\,0\,-1}\;g^1_0g^2_{-1}+X^{1\,1\,2}_{-1\,1\,-2}\;g^1_1g^2_{-2}\\
\partial_tg^1_0&=&-\nabla^*g^2_{-1}+\frac{1}{6}\nabla g^2_1-\frac{2}{3}\partial_zg^2_0-\frac{1}{3}\partial_z\rho+\mu^1[\rho]g^1_0+X^{1\,1\,1}_{0\,-1\,1}\;g^1_{-1}g^1_{1} \nonumber \\
&\,&+X^{1\,1\,2}_{0\,-1\,1}\;g^1_{-1}g^2_{1}+J^{1\,1\,1}_{0\,0\,0}\;g^1_{0}g^1_{0}+X^{1\,1\,2}_{0\,0\,0}\;g^1_{0}g^2_{0}+X^{1\,1\,2}_{0\,1\,-1}\;g^1_{1}g^2_{-1}
\end{eqnarray}
	\label{eq:Slow_Variables}
\end{subequations}
with $X^{l\,l_1\,l_2}_{m\,m_1\,m_2}=J^{l\,l_1\,l_2}_{m\,m_1\,m_2}+J^{l\,l_2\,l_1}_{m\,m_2\,m_1}$.
The nematic field $g^2$, at this order, is slaved to the density and the polar fields, as in the $2$ dimensional case~\cite{bertin2009hydrodynamic}:
\begin{equation*}
	\begin{split}
	g^2_{-2}&=\frac{1}{10\mu^2}\nabla g^1_{-1}-\frac{J^{2\,1\,1}_{-2\,-1\,-1}}{\mu^2}\;g^1_{-1}g^1_{-1} \;,\\
	g^2_{-1}&=\frac{1}{10\mu^2}\nabla g^1_{0}+\frac{1}{5\mu^2}\partial_zg^1_{-1}-\frac{X^{2\,1\,1}_{-1\,-1\,0}}{\mu^2}\;g^1_{-1}g^1_{0} \;,\\
	g^2_{0}&=-\frac{2}{5\mu^2}\Re(\nabla^*g^1_{-1})+\frac{2}{5\mu^2}\partial_zg^1_0+2\frac{X^{2\,1\,1}_{0\,-1\,1}}{\mu^2}\;|g^1_{-1}|^2-\frac{J^{2\,1\,1}_{0\,0\,0}}{\mu^2}\;g^1_{0}g^1_{0} \;.
	\end{split}
\end{equation*}
Injecting these relations into Eq.~\eqref{eq:Slow_Variables} we obtain a closed set of equations for the density and the polar fields.
After some algebra and going back to the representation in the real space~\eqref{eq:polar_field_decomposition}, the hydrodynamic equations for the $3$ dimensional polar class are
\begin{subequations}
\begin{eqnarray}
	\label{eq:hydro_continuity}
	\partial_t\rho&=&-\vec{\nabla}\cdot\vec{w} \;,\\
	\partial_t\vec{w}&=& \left(\mu^1[\rho]-\frac{\xi}{4}\vert\vec{w}\vert^2 \right)\vec{w}-\frac{1}{3}\vec{\nabla}\rho+D_{\text{B}}\vec{\nabla}\left(\vec{\nabla}\cdot\vec{w}\right)+D_{\text{T}}\Delta\vec{w} \nonumber\\ 
	&\,&-\lambda_1\left(\vec{w}\cdot\vec{\nabla}\right)\vec{w}-\lambda_2\left(\vec{\nabla}\cdot\vec{w}\right)\vec{w}-\lambda_3\vec{\nabla}\left(\vert\vec{w}\vert^2 \right) \;.
	\label{eq:hydro_polar}
\end{eqnarray}
\label{eq:HYDRO_REAL}
\end{subequations}
These equations are nothing but the Toner-Tu equations. They are formally the same as those derived in 2D using the same method.
The definition of all the hydrodynamic parameters are given in Table~\ref{tab:Polar_parameters} and their dependency on the local fields is made explicit with the use of the square brackets.
%%%%%%%%%%%%%%%%%%%%%%%%%%%
\begin{table}
	\centering
\begin{tabular}{ l | c | c }
   & $3$D & $2$D \\
  \hline	
   & & \\
   $\mu^1[\rho]$ & $P_0^1-1+\left(\frac{\pi}{4}P_0^1-\frac{8}{15}\right)\rho$  &  $P_1-1+\frac{8}{\pi}\left(P_1-\frac{2}{3}\right)\rho$ \\
   & & \\
   $\xi{ \, (>0)}$ & $\frac{2(128-105\pi P^1_0)(2+7P^2_0)}{25(-105-68\rho_0+7(15+2\rho_0)P^2_0)}$ & $\frac{240}{\pi}\frac{15(5P_1-2)(3P_2+1)}{-15\pi + 112\rho_0 +(15\pi+80\rho_0)P_2} $\\
   & & \\
   $D_T{ \, (>0)}$ & $\frac{42}{2(105+68\rho_0-7(15+2\rho_0)P^2_0)}$ & $\frac{15\pi}{4\left( -15\pi + 112\rho_0 +(15\pi+80\rho_0)P_2 \right)}$ \\
   & & \\
   $D_B$ & $\frac{D_T}{3}$ & $0$ \\
   & & \\
   $\lambda_1$ & $-\frac{3(352-105\pi P^1_0+784P^2_0)}{40(-105-68\rho_0+7(15+2\rho_0)P^2_0)}$ & $\frac{4(16+30P_2-15P_1)}{ -15\pi + 112\rho_0 +(15\pi+80\rho_0)P_2 }$ \\
   & & \\
   $\lambda_2$ & $-\frac{208+105\pi P^1_0+1176P^2_0}{20(-105-68\rho_0+7(15+2\rho_0)P^2_0)}$ & $-\frac{4(4+30P_2+15P_1)}{ -15\pi + 112\rho_0 +(15\pi+80\rho_0)P_2 }$ \\
   & & \\
   $\lambda_3$ & $\frac{224(2+7P^2_0) + (384-315\pi P^1_0)}{80(105+68\rho_0-7(15+2\rho_0)P^2_0)}$ & $-\frac{\lambda_2}{2}$ \\
\end{tabular}
	\caption{ Comparison of the hydrodynamic parameters of the polar class between the $3$ dimensional case and the $2$ dimensional case \cite{bertin2009hydrodynamic}. 
	The main difference between the two cases lies in the fact that there is an anisotropic diffusion term 
	and $\lambda_2 \ne -2\lambda_3$ in three dimensions.
	In the right column the $P_i$ parameters are the moments of the angular noise distribution in 2D, analogous to the $P^i_0$ in 3D.}
	\label{tab:Polar_parameters}
\end{table}
%%%%%%%%%%%%%%%%%%%%%%%%%%%

The non conservation of the momentum in the microscopic model forces the hydrodynamic equations to reflect the absence of
Galilean invariance which would have required $\lambda_1 = \frac{1}{\rho_0}$ and $\lambda_2 = \lambda_3 = 0$.
Moreover, although breakdown of Galilean invariance is also allowed at equilibrium, derivation of Equations \eqref{eq:HYDRO_REAL} from a free energy implies
that $\lambda_3 = -2 \lambda_2$ \cite{marchetti2013hydrodynamics}.
This relation holds in the $2$ dimensional equations for the Vicsek model without steric interactions, derived from the BGL approach, but 
remarkably it is no more the case in $3$ dimensions as shown in Table~\ref{tab:Polar_parameters}.
Another difference is the presence of the anisotropic diffusion term 
$\vec{\nabla}\left(\vec{\nabla}\cdot\vec{w}\right)$ which is allowed in 2D although it has a zero coefficient \cite{bertin2009hydrodynamic}.
Finally, similarly to the $2$ dimensional result, the isotropic pressure term is not modified by activity and remains linear in $\rho$.

%%%%%%%%%%%%%%%%%%%%%%%%%%%
\subsection{Homogeneous solutions}
\label{subsec:Homogeneous_polar}
%%%%%%%%%%%%%%%%%%%%%%%%%%%

The homogeneous solutions of \eqref{eq:HYDRO_REAL} satisfy $\rho=\rho_0$ and evolve according to the Ginzburg-Landau equation
\begin{equation}
	\partial_t w_0 = \left(\mu^1[\rho_0]-\frac{\xi}{4}w_0^2 \right)w_0 \;,
\end{equation}
where $w_0=\vert\vec{w}\vert$.

%%%%%%%%%%%%%%%%%%%%%%%%%%%
\begin{figure}[t!]
	\centering
	\includegraphics[scale=0.4]{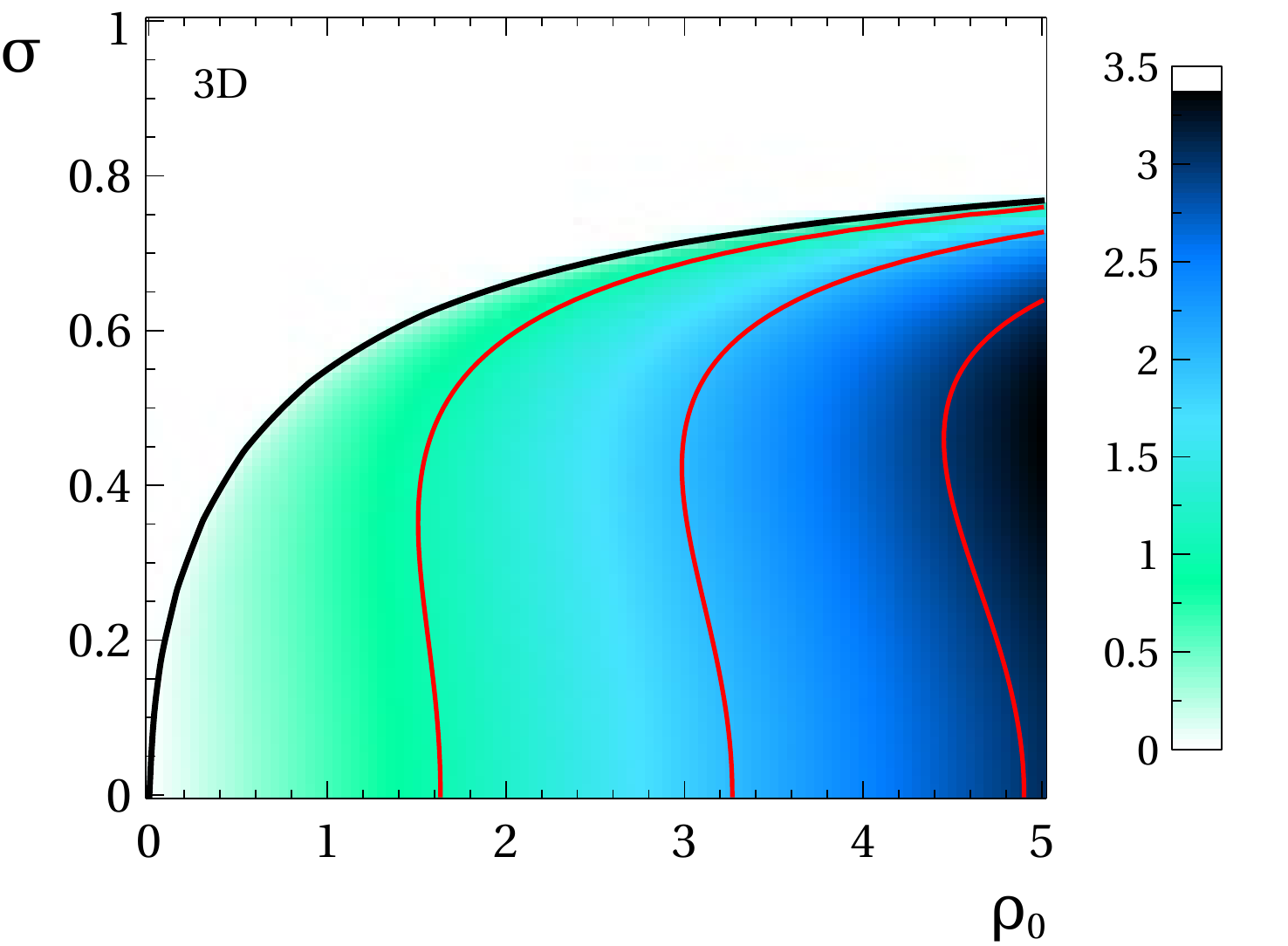}
	\includegraphics[scale=0.4]{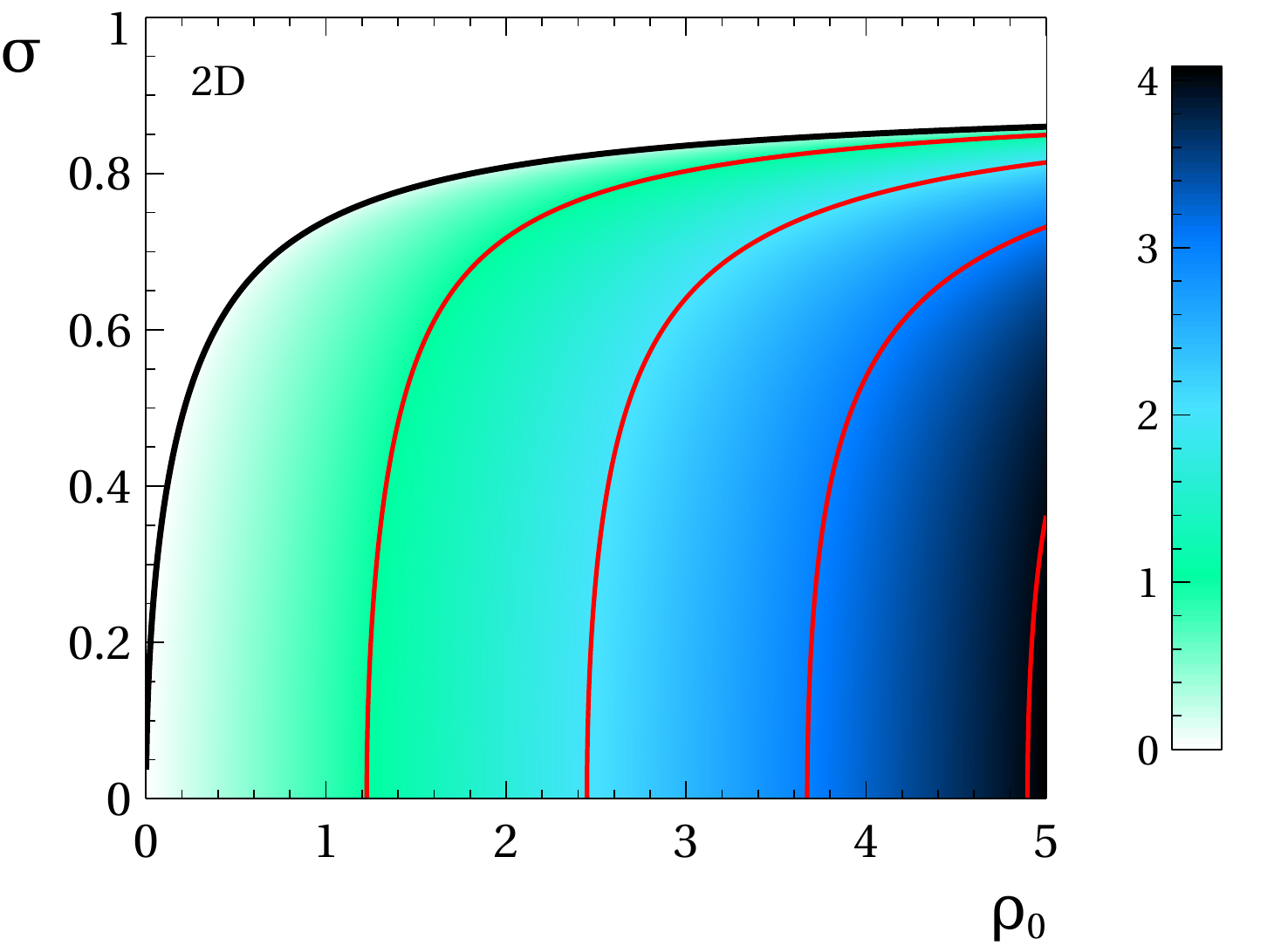}
	\caption{Phase diagram in the density-noise plane of the hydrodynamic equations for the polar class. 
	The color codes for $w_0$, the strength of the polar order of the homogeneous ordered solution. 
 Solid black line: $\mu^1 [ \rho ] =0$, limit of linear stability of the homogeneous disordered solution. 
				Red lines:  contours  $w_0=1,2,3$.
				Left panel: 3D case for which $w_0$ is not monotonously varying with $\sigma$.
				Right panel: 2D case, with monotonous variation of $w_0$.}
	\label{fig:HO}
\end{figure}
%%%%%%%%%%%%%%%%%%%%%%%%%%%

The homogeneous disordered solution $w_0=0$ becomes unstable whenever $\mu^1$ becomes positive.
Below the transition line in the $(\rho_0, \sigma)$ parameter plane, 
the polar field grows until the cubic nonlinearity is saturated, 
and the homogeneous ordered solution is found:
\begin{equation}
	\rho=\rho_0\quad w_0=2\sqrt{\frac{\mu^1[\rho_0]}{\xi}} \;.
\end{equation}
When it exists, the HO solution is always physical, being smaller than the homogeneous density $\rho_0$ 
in all the parameter space. This is shown in Figure~\ref{fig:HO}
where the red isolines correspond to the values $w_0=1,2,3$.
Remarkably, the global order $w_0$ is not monotonous and shows a maximum at finite noise at any fixed density,
in contrast with the $2$ dimensional case.
This effect depends on the distribution of the noise considered. For instance, if one uses the 
Von Mises distribution, $P(\theta)\propto \exp\left(\cos(\theta)/\sigma^2\right)$, this non-monotonicity is reduced although it is not
removed.

%%%%%%%%%%%%%%%%%%%%%%%%%%%
\subsection{Linear stability analysis}
\label{subsec:LinearStability_polar}
%%%%%%%%%%%%%%%%%%%%%%%%%%%

We now compute the linear stability of the homogeneous solutions assuming small
fluctuations of the fields around $\rho=\rho_0+\delta\rho$ and $\vec{w}=\vec{w}_0 + \delta\vec{w}$.
The linearized hydrodynamic equations are
\begin{subequations}
\begin{eqnarray}
	\partial_t \delta\rho &=& -\vec{\nabla}\cdot\delta\vec{w} \;,\\
	\partial_t \delta\vec{w}&=& \partial\mu^1\vec{w}_0\delta\rho - \frac{\xi}{2}\vec{w}_0(\vec{w}_0\cdot\delta\vec{w}) + D_B\vec{\nabla}(\vec{\nabla}\cdot\delta\vec{w})+ D_T\triangle\delta\vec{w} \nonumber \\
	&\,&-\lambda_1(\vec{w}_0\cdot\vec{\nabla})\delta\vec{w} - \lambda_2\vec{w}_0(\vec{\nabla}\cdot\delta\vec{w})  - 2\lambda_3\vec{\nabla}(\vec{w_0}\cdot\delta\vec{w}) \\
	&\,& +(\mu^1[\rho_0] - \frac{\xi}{4}\vert \vec{w}_0 \vert^2)\delta\vec{w} - \frac{1}{3}\vec{\nabla}\delta\rho \;, \nonumber
\end{eqnarray}
\end{subequations}
with $\partial\mu^1 = \partial \mu^1/\partial \rho$. 
Using Fourier transform in space ($\vec{q}$ is the wave-vector) 
these linear equations become
\begin{subequations}
\begin{eqnarray}
	\label{eq:LinFourierContinuity}
	\partial_t \delta\rho &=& -\imath \vec{q}\cdot\delta\vec{w} \;,\\
	\label{eq:LinFourierPolar}
	\partial_t \delta\vec{w}&=& \partial\mu^1\vec{w}_0\delta\rho - \frac{\xi}{2}\vec{w}_0(\vec{w}_0\cdot\delta\vec{w}) - D_B\vec{q}(\vec{q}\cdot\delta\vec{w})- D_T q^2 \delta\vec{w} \nonumber \\
	&\,&-\lambda_1\imath (\vec{w}_0\cdot\vec{q})\delta\vec{w} - \lambda_2\imath \vec{w}_0(\vec{q}\cdot\delta\vec{w})  - 2\lambda_3\imath \vec{q}(\vec{w_0}\cdot\delta\vec{w}) \\
	&\,& +(\mu^1[\rho_0] - \frac{\xi}{4}\vert \vec{w}_0 \vert^2)\delta\vec{w} - \frac{\imath }{3}\vec{q}\delta\rho \;. \nonumber
\end{eqnarray}
\end{subequations}
Solving this eigenvalue problem yields four solutions $s_i=s_i(\vec{q},\vec{w}_0)$, with $i=1,\dots,4$.
The system is linearly unstable whenever any real part of these solutions is positive.

%The continuity equation~\eqref{eq:LinFourierContinuity} immediately gives $\delta\rho=-\imath\frac{\vec{q}\cdot\delta\vec{w}}{s}$, and 
In the disordered case ($\vec{w}_0=\vec{0}$) the global rotational invariance implies that only wave-vectors parallel to the perturbation
$\delta\vec{w}\parallel\vec{q}$ evolve and can destabilize the system. 
Inserting the solution of~\eqref{eq:LinFourierContinuity} in~\eqref{eq:LinFourierPolar} we obtain the following relations 
between the wavevectors and the eigenvalues
\begin{eqnarray}
	&\left(s +  D_Tq^2-\mu^1 \right)^2 = 0 \;,\\
	&s^2 + s\left( (D_T+D_B)q^2-\mu^1 \right) + \frac{1}{3}q^2=0,\qquad q=\vert \vec{q} \vert \;.
\end{eqnarray}
Defining $D=D_T+D_B$ to lighten notations, the independent solutions are 
\begin{eqnarray}
	s_r&=&\mu^1-D_Tq^2 \;,\\
	s_\pm &=& \frac{\mu^1 - Dq^2}{2}\pm\frac{1}{2}\sqrt{ \mu^2 - 2\mu^1 Dq^2 - \frac{1}{3}q^2 + D^2q^4 } \;.
\end{eqnarray}
They are negative for all $D_T>0$,  $D_T+D_B>0$ and $\mu^1<0$.
Consequently, the disordered state is stable in all the region where $\mu^1<0$, as in the $2$ dimensional case.

%%%%%%%%%%%%%%%%%%%%%%%%%%%
\begin{figure}[t!]
	\centering
	\includegraphics[scale=0.5]{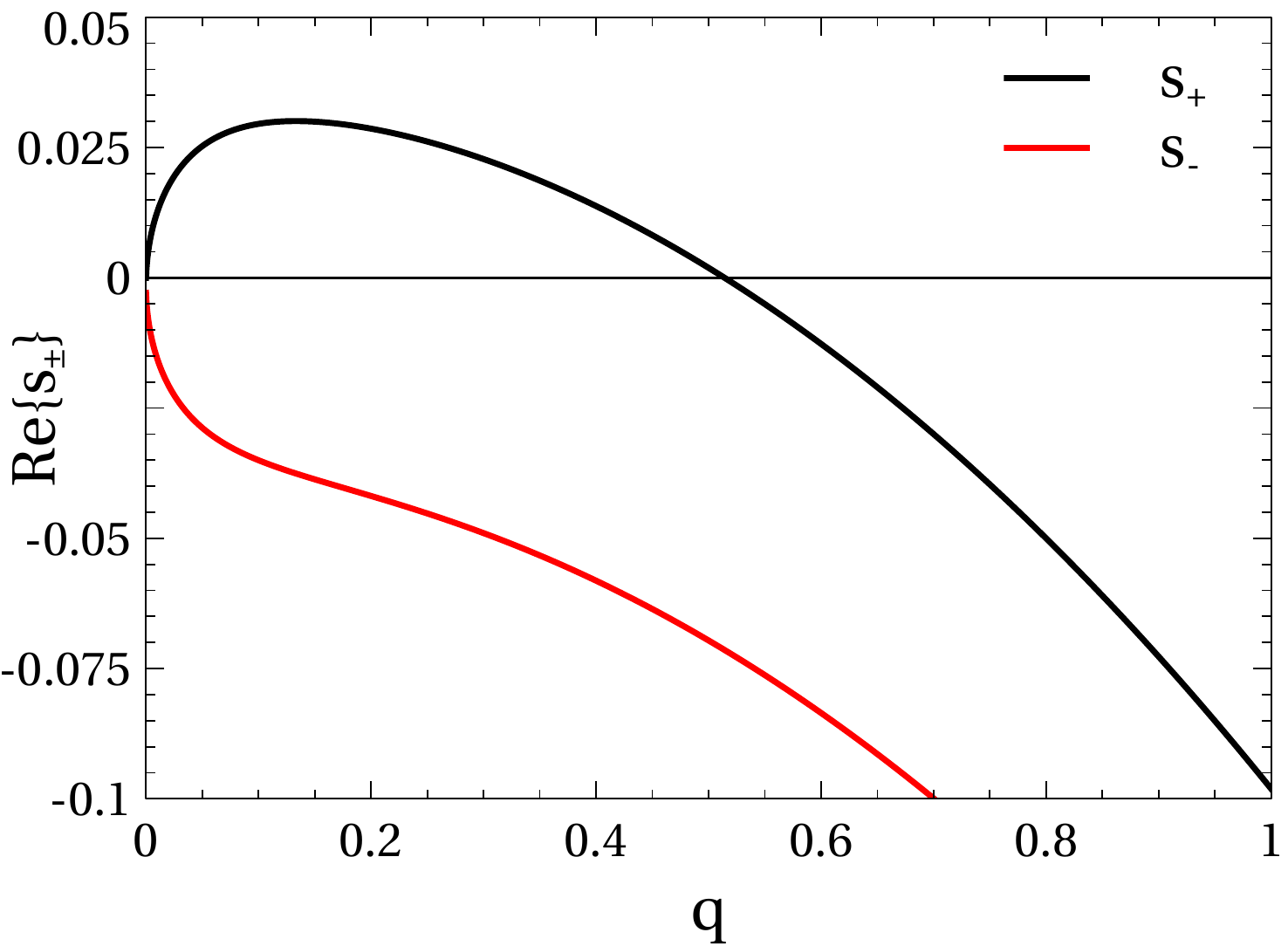}
	\caption{Behavior of the real part of the eigenvalue governing the long wavelength linear stability of the homogeneous ordered solution of the polar hydrodynamic equations, given by Eq.~\eqref{eq:wave_stability}. Parameters: $\rho_0=1$ and $\sigma=0.55$.
			Solid black line: real part of $s_+$. 
			Solid red line: real part of $s_-$.}
	\label{fig:Lin_wave}
\end{figure}
%%%%%%%%%%%%%%%%%%%%%%%%%%%

Considering only longitudinal perturbations of the ordered state $\vec{w}_0=w_0\vec{e}\neq\vec{0}$, we have $\vec{q}=q\vec{e}$ and
$\delta\vec{w}=\delta w\vec{e}$. The equations for this family of perturbations are
\begin{subequations}
\begin{eqnarray}
	\partial_t \delta \rho &=& -\imath q \delta w \;,\\ 
	\partial_t \delta w &=& w_0\partial\mu^1\delta\rho - \frac{\xi}{2}w_0^2\delta w - \frac{\imath}{3}q\delta\rho - Dq^2\delta w - \lambda w_0 \imath q\delta w \;,
\end{eqnarray}
\end{subequations}
where $\lambda = \lambda_1+\lambda_2 + 2\lambda_3$.
After some algebra, the corresponding equation for the eigenvalues \footnote{The other two relations come from the transverse perturbations.} reads
\begin{equation}
	s^2 + s\left(\frac{\xi}{2}w_0^2 + Dq^2 + \imath\lambda w_0q\right) + \left( \frac{1}{3}q^2 + \imath \partial\mu^1 w_0 q \right)=0 \;.
\end{equation}
Defining $B(q)=\frac{\xi}{2}w_0^2 + Dq^2 + \imath\lambda w_0q$ and $C(q)=\frac{1}{3}q^2 + \imath \partial\mu^1 w_0 q$,
the two solutions of this equation are
\begin{equation}
	s_\pm^l = -\frac{B(q)}{2} \pm \frac{1}{2}\sqrt{B^2(q) - 4C(q)} \;.
\label{eq:wave_stability}
\end{equation}
The real part of the $(-)$ solution is negative, while the real part of the $(+)$ solution has a range of wave-vectors for which it is positive
as shown in Figure~\ref{fig:Lin_wave}.
Hence, the ordered phase is longitudinally unstable in the region close to the transition line.
At small wave-vectors the solutions behave as
\begin{eqnarray}
	s_-^l &=& -\frac{\xi}{2}w_0^2 + \imath w_0\left( \frac{2\partial\mu^1}{w_0^2\xi} - \lambda \right)q - \left(D + 8\frac{\left(\delta\mu^{1}\right)^2}{w_0^4\xi^3} - 4\frac{\lambda\partial\mu^1}{w_0^2\xi^2} - \frac{2}{3w_0^2\xi}\right)q^2 \;,\nonumber \\
	s_+^l &=& - 2\imath \frac{\partial\mu^1}{w_0\xi}q + \left( 2\frac{\left(\delta\mu^{1}\right)^2}{w_0^2\xi^2} - \frac{\lambda\partial\mu^1}{\xi} - \frac{1}{6} \right)\frac{4}{w_0^2\xi}q^2	\;.
\end{eqnarray}
The real part of the $(+)$ solution is driven by a $q^2$ term that grows at small wave-vectors on a scale $\frac{1}{\mu^1}$ 
which is diverging at the transition.
The resulting instability is thus due to the $\delta\mu$ term, 
which comes from the density dependence of the linear coefficient $\mu^1[\rho]$ in the polar field equation.

%%%%%%%%%%%%%%%%%%%%%%%%%%%
\begin{figure}[t!]
	\centering
	\includegraphics[scale=0.35]{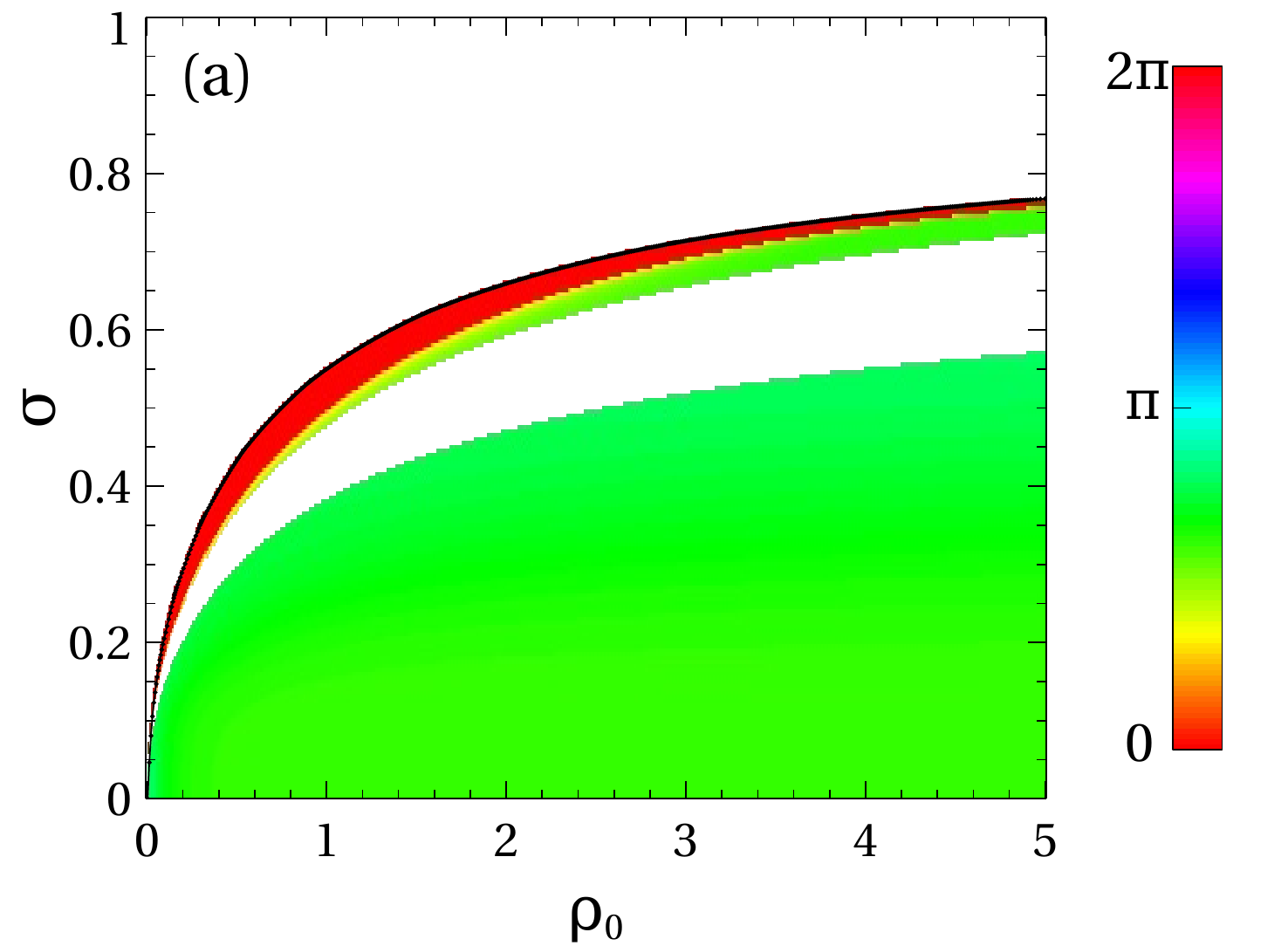}
	\includegraphics[scale=0.35]{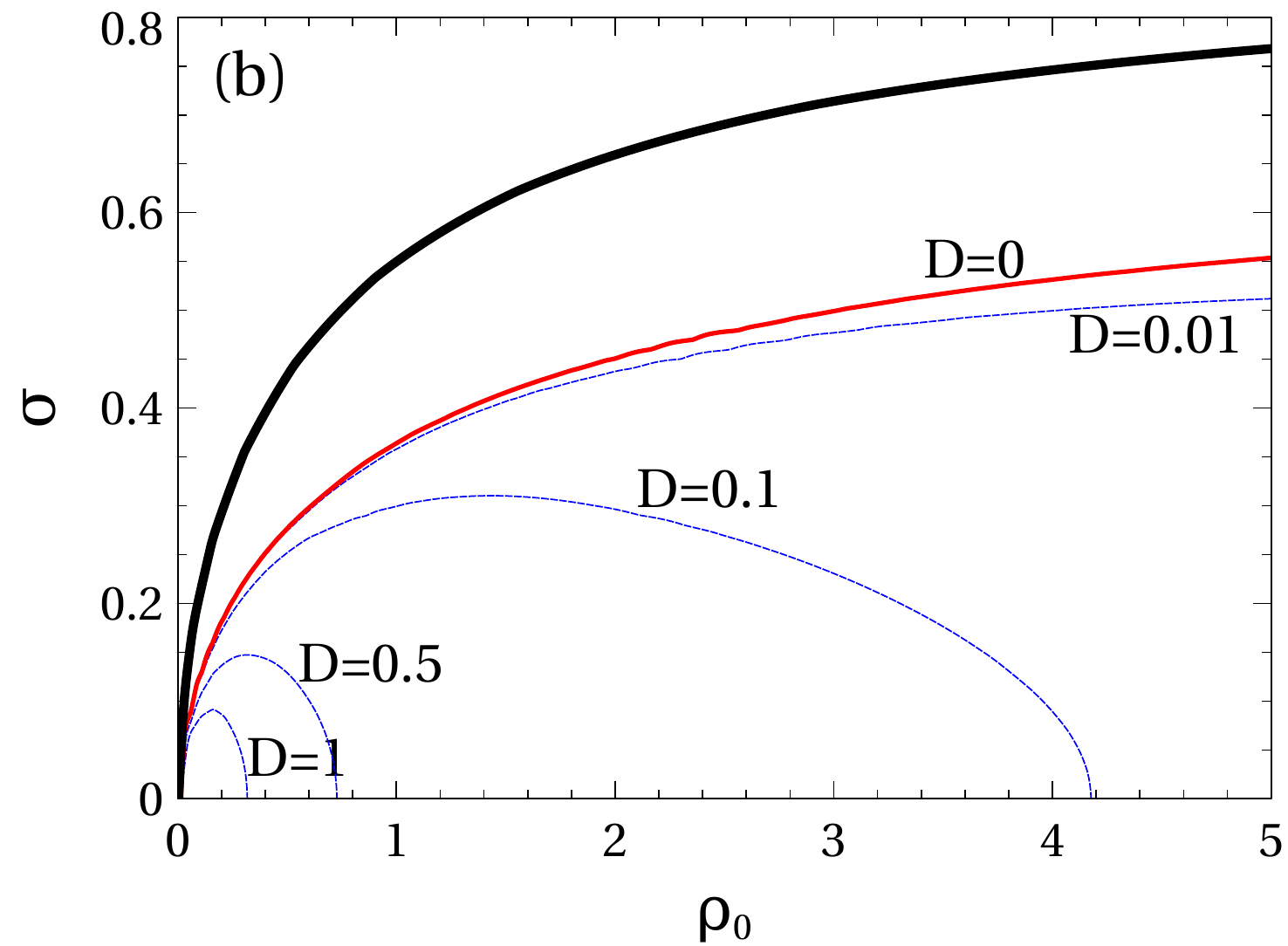}
	\caption{(a): phase diagram of the hydrodynamic equations for the polar class. Solid black line: $\mu^1 [\rho]=0$. Below this
	line the homogeneous ordered solution exists. It is linearly unstable in the colored regions, where 
			the color represents the direction of the most unstable mode.
			In the red region close to the transition the instability is parallel to the global order $\vec{w}_0$. 
			The green region of transversal modes deep in the ordered phase corresponds to the spurious instability described in the text. 
			(b): variation of the upper limit of spurious instability with the spatial diffusion constant $D$.
			Solid red line: $D=0$. 
			Dashed blue lines:  $D=0.01,0.1,0.5,1$.}
	\label{fig:LinearStability_Polar}
\end{figure}
%%%%%%%%%%%%%%%%%%%%%%%%%%%

The full linear stability analysis of the homogeneous solutions was also computed numerically
(Figure~\ref{fig:LinearStability_Polar}(a)).
Close to the order/disorder threshold (black line), the ordered solution is linearly unstable with respect to longitudinal perturbations at a finite wavelength. Deeper in the ordered phase the ordered solution becomes linearly stable. 
At even lower noise, a second instability appears.
This instability is likely to be an artifact of the truncation procedure
\footnote{We actually studied semi-numerically the linear stability of the homogeneous ordered solution in the 2D case {\it at the kinetic level} and found that this instability is not present then \cite{Kinetic_paper}.}.
Moreover, like in 2D \cite{peshkov2014boltzmann}, its impact strongly depends on the presence of spatial diffusion:
 adding spatial diffusion directly at the kinetic level,
its domain in parameter space shrinks rapidly (Fig.~\ref{fig:LinearStability_Polar}(b)).
We hereafter call it the \textit{spurious} instability.

%%%%%%%%%%%%%%%%%%%%%%%%%%%
\section{Hydrodynamic equations for nematic alignment}
\label{sec:HE_active_nematic}
%%%%%%%%%%%%%%%%%%%%%%%%%%%

In this section we derive the hydrodynamic equations for 3D Vicsek-style systems with nematic alignment, represented by the active nematics and the rods classes.

%%%%%%%%%%%%%%%%%%%%%%%%%%%
\subsection{Boltzmann equations for classes with nematic alignment}
\label{subsec:Boltzmann_nematic}
%%%%%%%%%%%%%%%%%%%%%%%%%%%

%%%%%%%%%%%%%%%%%%%%%%%%%%%
\subsubsection{Kinetic equations for active nematics and rods}
\label{subsubsec:}
%%%%%%%%%%%%%%%%%%%%%%%%%%%

In the active nematics class particles reverse their velocity on short time scales, the computation sketched in Section~\ref{subsec:Kinetic} shows that the resulting Boltzmann equation has no drift term but only spatial diffusion with coefficients $D_0=\frac{1}{3}$ and $D_1=3D_0$.
Moreover, as this problem possesses a full nematic symmetry, only even modes (with respect to $l$ index) need to be considered.
Hence, for active nematics the Boltzmann equation becomes
\begin{eqnarray}
	\partial_tg^{2l}_m & = & D_0\Delta g_m^{2l} + D_1{\cal D}_m^{2l}\left[\left\{g_m^{2l}\right\}\right] \;,\nonumber \\
	& & +\left[P_{0}^{2l}-1\right]g_{m}^{2l}+\sum_{l_1,m_1}\sum_{l_2=|l-l_1|,\,m_2}^{l+l_1} {J^{\rm A}}^{\,2l,2l_1,2l_2}_{m,m_1,m_2}\;g_{m_1}^{2l_1}g_{m_2}^{2l_2} \;,
	\label{Kin_eq_AN}
\end{eqnarray}
where the coefficients ${J^{\rm A}}^{\,2l,2l_1,2l_2}_{m,m_1,m_2}$ are the ones computed in Sec.~\ref{subsec:Decomposition} using the apolar kernel \eqref{apolar_kernel} and the nematic alignement rule \eqref{nematic_alignment}.

On the contrary, in the rods class particles reverse their directions of motion on a finite timescale $a$ such that no spatial diffusion is attained at the kinetic level.
Their dynamics is therefore propagative and the resulting Boltzmann equation reads 
\begin{eqnarray}
	\partial_tg^l_m & = & {\cal T}_m^l\left[\left\{g_m^l\right\}\right] -a\left(1 + (-1)^l\right)g_{m}^l +\left[P_{0}^l-1\right]g_{m}^l \nonumber \\
	& & +\sum_{l_1,m_1}\,\sum_{l_2=|l-l_1|,\,m_2}^{l+l_1}{J^{\rm R}}^{\,l,l_1,l_2}_{m,m_1,m_2}g_{m_1}^{l_1}g_{m_2}^{l_2} \;,
	\label{Kin_eq_rods}
\end{eqnarray}
with ${J^{\rm R}}^{\,l,l_1,l_2}_{m,m_1,m_2}$ evaluated considering the polar kernel \eqref{Kernel} and nematic alignment rule \eqref{nematic_alignment}.
Here, as the motion possesses polar symmetry, we can not set the odd modes to zero.

The large reversal rate diffusive limit \eqref{Kin_eq_AN} can be retrieved from the propagative hierarchy \eqref{Kin_eq_rods} with finite reversal.
To do this we need to temporarily reintroduce $v_0$, the microscopic velocity of particles, which was set to $1$ in Section \ref{subsec:Kinetic} 
when the Boltzmann equation was de-dimentionalized.
Indeed, in the infinite reversal rate limit the odd field equations (with respect to $l$ index) possess a diverging damping term $\sim -2a$ and can thus be enslaved to the even fields.
The latter then acquire an effective diffusion coefficient that scales like $\frac{v_0^2}{a}$.
In order to keep it finite, we assume that $v_0^2\underset{a\to\infty}{\sim} a$.
After enslaving the odd modes, the hierarchy \eqref{Kin_eq_rods} at ${\cal O}\left(\frac{v_0^2}{a}\right)$ is formally the same as \eqref{Kin_eq_AN}, with $D_0 = \frac{v_0^2}{6a}$ and $D_1 = 3D_0$.
Therefore, in the following, when we refer to the large reversal rate limit for the rods we will implicitly consider
a nondimensionalization where $v_0^2\sim a$ instead of $v_0=1$ as previously set for the polar class.
For numerical evaluations, we have considered the scaling form $v_0 = \sqrt{1 + a}$ such that $v_0 = 1$ when $a = 0$ and $v_0^2\sim a$ when $a$ becomes large.

%%%%%%%%%%%%%%%%%%%%%%%%%%%
\subsubsection{Homogeneous solutions of the Boltzmann equation}
\label{subsubsec:Hom_solution_B_eq}
%%%%%%%%%%%%%%%%%%%%%%%%%%%

In this section we compute the homogeneous solutions of the Boltzmann equation considering nematic alignment (both active nematics and rods classes).
Since the rods hierarchy \eqref{Kin_eq_rods} does not show any homogeneous solution with non-zero odd modes, the two classes are formally the same at this stage.
Therefore we will focus on the active nematics class, the results of this section being easily generalized to the rods class.

%%%%%%%%%%%%%%%%%%%%%%%%%%%%%%%%
\begin{figure}[t!]
	\centering
	\includegraphics[scale=0.5]{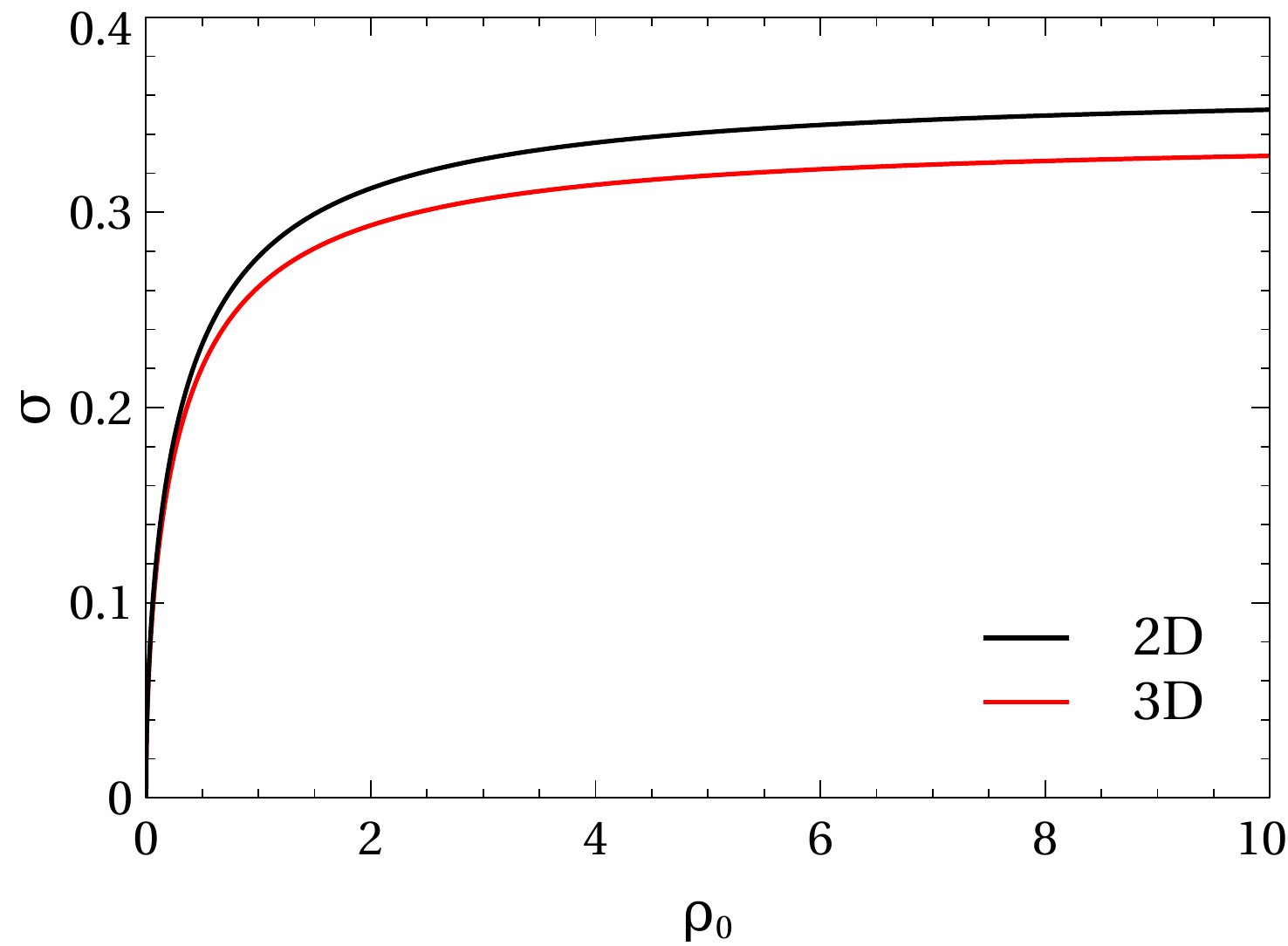}
	\caption{The transition line ($\mu^2=0$) delimiting the stability region of the homogeneous disordered solution in the density-noise plane for the active nematics class. The solution is unstable to homogeneous perturbations below the black line in 2D, and below the red line in 3D.}
	\label{fig:AN_transitions}
\end{figure}
%%%%%%%%%%%%%%%%%%%%%%%%%%%%%%%%

The Boltzmann hierarchy \eqref{Kin_eq_AN} has the trivial homogeneous disordered solution: $g_0^0 = \rho_0$, $g^{2l}_m = 0$ $\forall l > 0$ $\forall m$.
Homogeneous perturbations of this solution are
\begin{equation}
	g^0_0=\rho=\rho_0+\delta\rho,\quad g^{2l}_{m}=\delta g^{2l}_{m}\quad \forall l>0 \;,\; \forall m
\end{equation}
and linearizing the Boltzmann hierarchy around this particular solution gives the linear coefficients 
\begin{equation}
	\begin{split}
	\partial_t\delta g^{2l}_m(\vec{r},t)&=\left[\left(P_{0}^{2l}-1\right)+\left(J^{\,2l,0,2l}_{m,0,m}+J^{\,2l,2l,0}_{m,m,0}\right)\rho_0\right]\delta g^{2l}_{m}\\
	&\equiv\mu^{2l}_{m}[\rho_0]\delta g^{2l}_{m} \;.
	\end{split}
\label{lin_Kin_eq}
\end{equation}
As in the polar case, the coefficients $\mu^{2l}_m$ do not depend on the index $m$, and we thus omit it thereafter.
The first three coefficients are
\begin{subequations}
\begin{eqnarray}
	\mu^2&=&(P^2_{0}-1)+\frac{2}{15}\left((8+3\sqrt{2})P^2_{0}-\frac{68}{7}\right)\rho_0 \;,\\
	\mu^4&=&(P^4_{0}-1)+\frac{1}{3}\left(\frac{1}{\sqrt{2}}P^4_{0}-\frac{920}{131}\right)\rho_0\; (\le0) \;,\\
	\mu^6&=&(P^6_{0}-1)-\frac{1}{5}\left(\frac{59}{24\sqrt{2}}P^6_{0}+\frac{952}{143}\right)\rho_0\; (\le0) \;.
\end{eqnarray}
\end{subequations}
Only the $\mu^2$ coefficient, corresponding to the nematic field, can change sign while the others are negative.
The disordered solution is stable at large noise and small densities and it becomes unstable to homogeneous 
perturbations below the line   $\sigma_t(\rho_0)$ defined by $\mu^2=0$, as shown in Fig~\ref{fig:AN_transitions}.
Like for the polar case studied in the previous section, 
the transition line in 3D is at lower noises than in the $2$ dimensional case.

\begin{figure}[t!]
	\centering
	\includegraphics[scale=0.4]{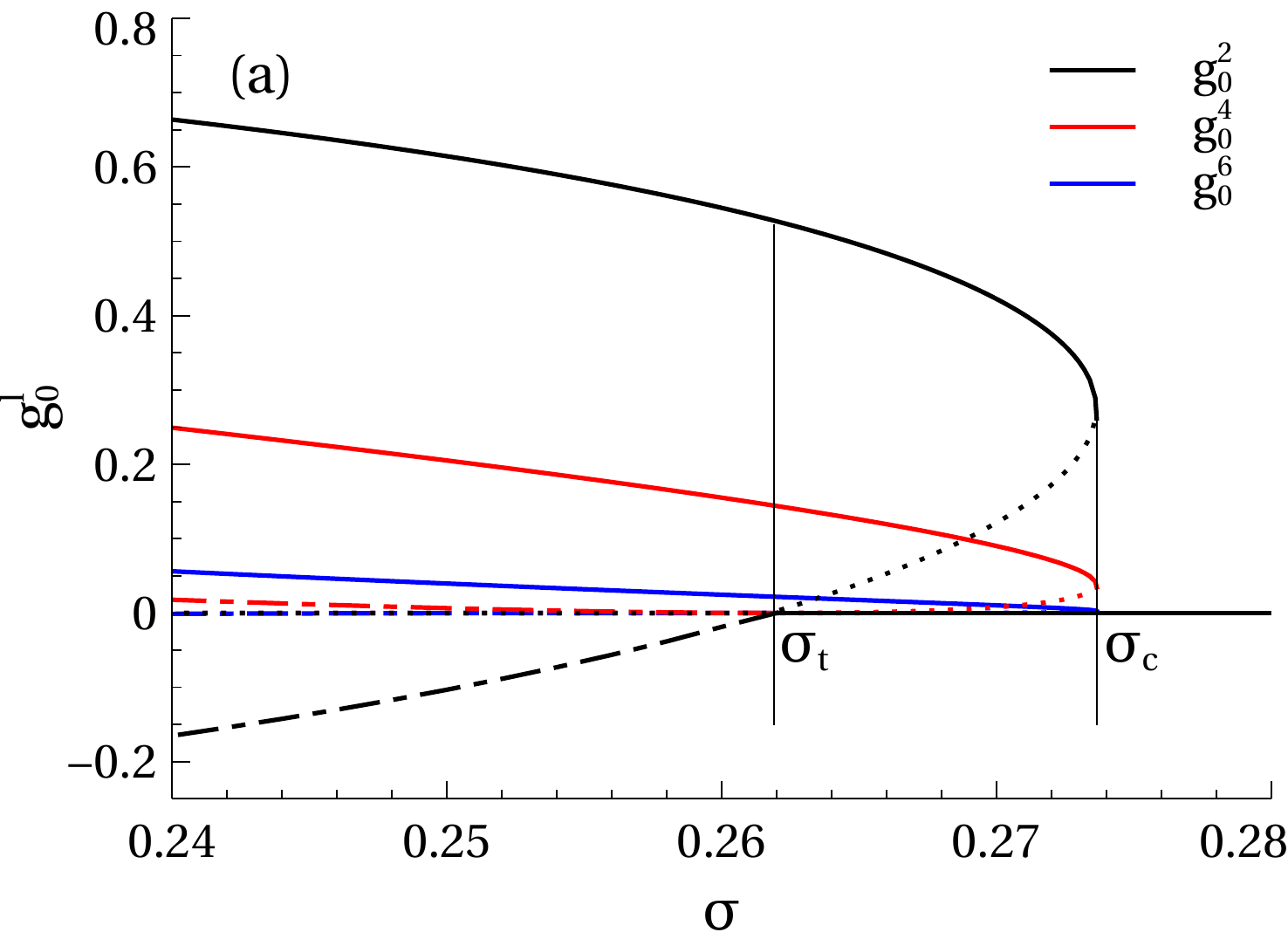}
	\includegraphics[scale=0.4]{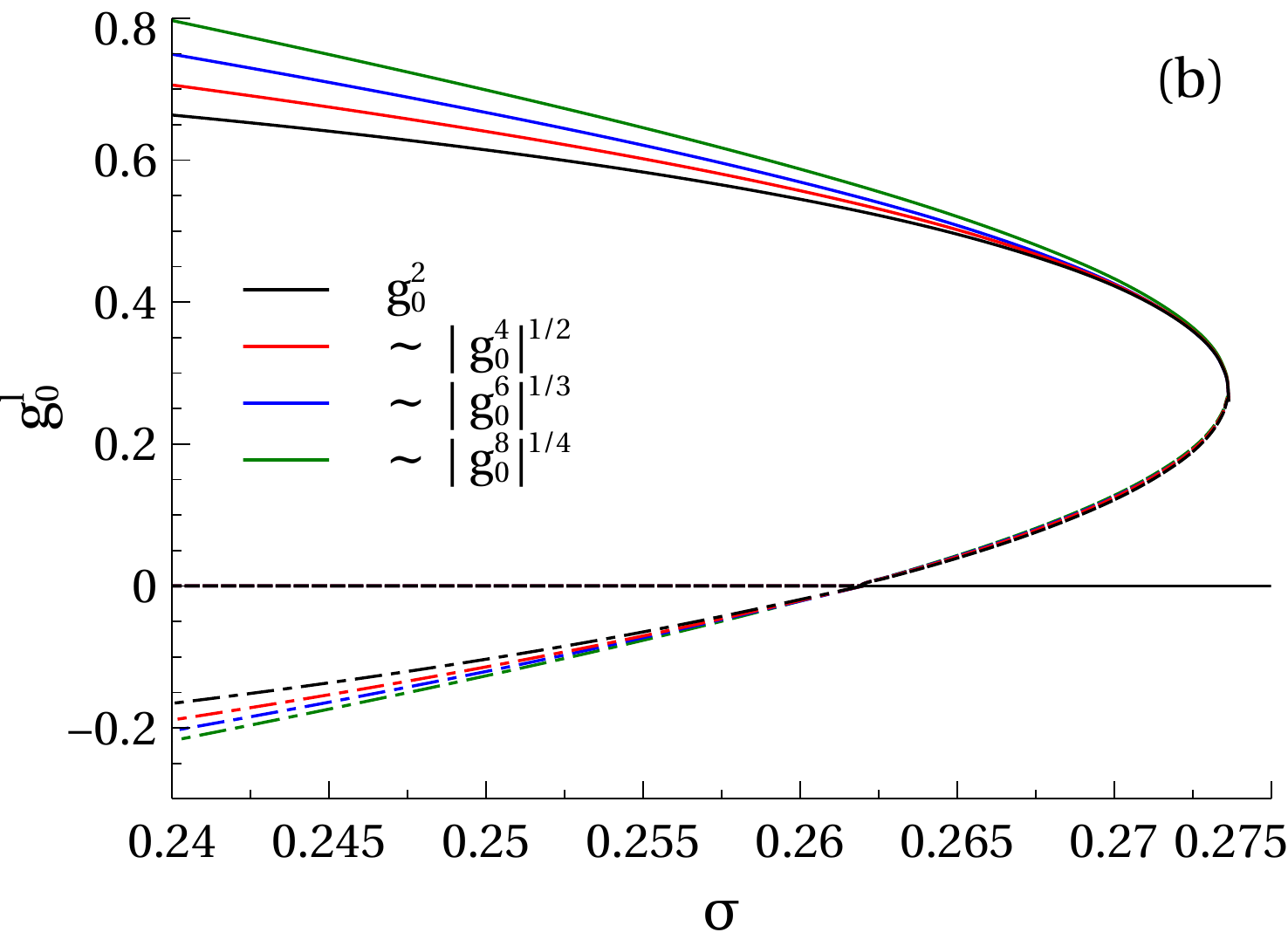}
	\caption{Numerical evaluation of the homogeneous ordered solution of the Boltzmann equation for the active nematics class, truncating the hierarchy at the $10^{\rm th}$ mode.
		(a) shows the discontinuous character of the transition with the bistability region between $\sigma_t$ and $\sigma_c$.
		The full lines are the values of the modes following the uniaxial solution, while the dotted lines are the values of the modes along the biaxial solution at the transition. The biaxial solution below $\sigma_t$ are shown with a dot-dashed line.
		(b) shows the rescaled modes following the scaling ansatz \eqref{eq:scaling_AN3D}, giving a good collapse close to the transition, although the first non zero mode $g^2_0$ is not small. For convenience the positive values of the modes $g^4_0,g^6_0,\ldots$, following the biaxial solution (negative $g^2_0$) are shown
		with a minus sign in order to highlight the validity of the scaling.}
	\label{fig:AN_10_mode}
\end{figure}

As in the polar case, assuming that the nematic order is along the $z$ direction we can set all the $m\ne0$ modes to zero for the numerical evaluation of the HO solution of the Boltzmann equation (see Section~\ref{subsec:relation_fields_SH} for details).
Figure~\ref{fig:AN_10_mode}(a) shows the result for the first $10$ even modes of the Boltzmann hierarchy \eqref{Kin_eq_AN} for $\rho_0 = 1$ as function of the noise.
This computation was done using the Newton method in order to capture both stable and unstable solutions.
Decreasing the noise the disordered solution becomes unstable for $\sigma \le \sigma_t$ and the $l > 0$ modes jump discontinuously to a positive value.
Then a hysteresis loop can be built increasing the noise up to $\sigma_c > \sigma_t$ defining an upper bound for the existence of the ordered solution.
Therefore, active nematics in three dimensions exhibit a discontinuous transition with coexistence of disordered and homogeneous ordered solutions at the mean field level.
As for equilibrium liquid crystals, we will show that this can be understood at the hydrodynamic level from symmetry reasons~\cite{degennes1995physics}.
We also see that equations~\eqref{Kin_eq_AN} admit solutions with a negative order parameter which will be discussed in the following and are always unstable to homogeneous perturbations\footnote{These solutions are stable considering only the $m=0$ modes as we do for the computation of the HO solution. 
This is because below $\sigma_t$ the instability is located on the other components of the nematic field (those for which $m\ne0$), as shown in 
Section~\ref{subsec:Homogeneous_active_nematic}.}.

%%%%%%%%%%%%%%%%%%%%%%%%%%%%%%%%
%%%%%%%%%%%%%%%%%%%%%%%%%%%%%%%%

%%%%%%%%%%%%%%%%%%%%%%%%%%%
\subsection{Derivation of the hydrodynamic equations}
\label{subsec:Hydro_derivation_nematic}
%%%%%%%%%%%%%%%%%%%%%%%%%%%

The following sections are dedicated to the derivation of the hydrodynamic equations for classes with nematic alignment.
As in Section~\ref{subsec:HE_der_polar}, they are obtained using a scaling parameter $\varepsilon$ that allows to truncate and close the Boltzmann hierarchy with an ansatz compatible with the symmetries of the problem.

%%%%%%%%%%%%%%%%%%%%%%%%%%%
\subsubsection{Hydrodynamic equations for the active nematics class}
\label{subsubsec:Hydro_active_nematics}
%%%%%%%%%%%%%%%%%%%%%%%%%%%

The scaling ansatz necessary to truncate and close the infinite Boltzmann hierarchy \eqref{Kin_eq_AN} for active nematics relates space and time \textit{diffusively}:
\begin{equation}
	\nabla^2 \approx \nabla^{*^2} \approx \partial^2_{ij}\approx \partial_t \approx \varepsilon^2 \;,
\end{equation}
with $i,j=x,y,z$.
Close to and below the transition line $\sigma_t(\rho_0)$, we assume that the nematic field saturates to a small value $\varepsilon$, therefore, balancing the hierarchy terms we can obtain the self-consistent scaling
\begin{equation}
	g^{2l}_m \approx \varepsilon^{ l }, \qquad \forall l>0, \quad \forall m \;.
	\label{eq:scaling_AN3D}
\end{equation}
As usual, the first non trivial order is then $\varepsilon^3$, leaving the density field $\rho$, the nematic field $g^2_m$ and $g^4_m$.
The dynamics of the $g^4_m$'s is completely slaved to the lower order modes.

After lengthy algebraic calculations, similar to those shown in the previous section for ferromagnetic alignment, 
the continuity equation in real space takes the simple form
\begin{equation}
	\partial_t\rho=D_0\Delta\rho+D_1\left(\mathbf{\Gamma}\,\textbf{:}\,\textbf{Q}\right) \;,
	\label{eq:Continuity_AN}
\end{equation}
where the traceless symmetric matrix $\mathbf{\Gamma}$ is 
\begin{equation}
	\mathbf{\Gamma}=\frac{1}{3}\begin{pmatrix} 2\partial_{xx}^2-\partial_{yy}^2-\partial_{zz}^2 & 3\partial_{xy}^2 & 3\partial_{xz}^2\\
	3\partial_{xy}^2 & 2\partial_{yy}^2-\partial_{xx}^2-\partial_{zz}^2 & 3\partial_{yz}^2\\
	3\partial_{xz}^2 & 3\partial_{yz}^2 & 2\partial_{zz}^2-\partial_{xx}^2-\partial_{yy}^2
	\end{pmatrix}\;,
	\label{Gamma}
\end{equation}
and the Frobenius inner product between rank $2$ tensors is defined by $\textbf{A}\,\textbf{:}\,\textbf{B}= \sum_{\alpha,\beta} A_{\alpha\beta}B_{\alpha\beta}$.

Knowing the relation between the second angular mode of the single-particle distribution $f$ and the nematic order parameter, explicited in Eq.~\eqref{eq:Nem_order_parameter}, we can write the hydrodynamic equations in real space in the following compact form:
\begin{equation}
\partial_t \textbf{Q} = \frac{2D_1}{15}\mathbf{\Gamma}\rho+\left[\mu^2[\rho]-\xi(\textbf{Q}\,\textbf{:}\,\textbf{Q})\right]\textbf{Q} + \alpha\left[\textbf{Q}\textbf{Q}\right]_\text{ST} + D_0\Delta \textbf{Q} + \frac{4D_1}{7} \left[\mathbf{\Gamma}\textbf{Q}\right]_\text{ST} \;,
\label{Hydro_AN_Real}
\end{equation}
where $\left[\textbf{A}\right]_\text{ST}=\frac{1}{2}(\textbf{A}+\textbf{A}^t)-\frac{1}{3}\textbf{I}(\mathrm{Tr} \textbf{A})$ is the symmetric traceless part of the tensor $\textbf{A}$ and $\textbf{I}$ is the identity matrix.
The dependencies of the coefficients in terms of the microscopic parameters $\rho_0$ and $\sigma$ are listed in 
Table~\ref{tab:AN_parameters}.
%%%%%%%%%%%%%%%%
\begin{table}[t!]
	\centering
\begin{tabular}{ l | c | c }
   & $3$D & $2$D \\
  \hline	
   & & \\
   $\mu^2[\rho]$ & $P_0^2-1+\frac{2}{15}\left((8+3\sqrt{2})P_0^2-\frac{68}{7}\right)\rho$  & $P_2-1 + \frac{8}{3\pi}\left( (2\sqrt{2}-1)P_2-\frac{7}{5}\right)\rho$  \\
   & & \\
   $\xi {\,(>0)}$ & $\frac{393\sqrt{2}(304-11(-32+203\sqrt{2})P_0^2)(16+(384+35\sqrt{2})P_0^4)}{528220\left( (393\sqrt{2} + 131\rho_0 ) P_0^4 - 393\sqrt{2}-920\sqrt{2}\rho_0 \right)}$ & ${ \frac{4}{45\pi}}\frac{(1+15P_2)(9(1+6\sqrt{2})P_4 - 13)}{315\pi(1-P_4) + 8(21P_4 + 155)\rho_0}$ \\
   & & \\
   $\alpha$ & $\frac{2}{49}\left(4+(-16+21\sqrt{2})P_0^2\right) \; {(>0)}$ & $0$ \\
   & & \\
   $D_0$ & $\frac{1}{3}$ & $\frac{1}{2}$ \\
   & & \\
   $D_1$ & $3D_0$ & $2D_0$ \\
\end{tabular}
\caption{
Comparison of the hydrodynamic coefficients of active nematics equations between the $3$ and $2$ dimensional cases \cite{bertin2013mesoscopic}. The functional form of the linear ($\mu^2$) and the cubic ($\xi$) parameters is comparable between the two cases. The quadratic term ($\alpha$) is absent in 2D because of rotational invariance while it gets a non zero (positive for rod shaped particles) value in 3D.
In the right column the $P_i$ parameters are the moments of the angular noise distribution in 2D, analogous to the $P^i_0$ in 3D.
}
	\label{tab:AN_parameters}
\end{table}
%%%%%%%%%%%%%%%%
Once more Eq.~\eqref{Hydro_AN_Real} has the familiar Ginzburg-Landau structure and a linear coupling to the density field.
Note, however, the presence of the quadratic term in the field tensor $\textbf{Q}$ with coefficient $\alpha$ in addition to anisotropic spatial diffusion, allowed by the symmetries of the system
\footnote{In $2$ dimensions both these terms are not allowed by rotational symmetry. 
Although using the tensorial notation of Eq.\eqref{Hydro_AN_Real} the reason why these terms are not allowed in 2D is not evident, it is easy to show that $[\textbf{Q}\textbf{Q}]_{\rm{ST}}=0=[\mathbf{\Gamma}\textbf{Q}]_{\rm{ST}}$ in 2D.}.
Unlike in 2D and because of the presence of anisotropic spatial diffusion, the structure of Eq.~\eqref{Hydro_AN_Real} cannot be derived from a free energy in the single Frank constant approximation \cite{bertin2013mesoscopic}
\footnote{The structure of Eq.~\eqref{Hydro_AN_Real} can be obtained from a free energy density $ {\cal F} = aQ_{ij}\partial_{ij}^2\rho + \frac{b}{2}\left(\partial_iQ_{kl}\right)^2 + \frac{c}{2}\left(\partial_kQ_{ik}\right)^2 + ...$, then denoting by $\Gamma_{\rm Q}$ the coupling constant with the nematic order parameter we have $D_0 = \Gamma_{\rm Q} \left(b - \frac{c}{3}\right)$ and $D_1 = \frac{7}{4}\Gamma_{\rm Q} c  = -\frac{15}{2}\Gamma_{\rm Q}a$. In the single Frank constant approximation $c=0$~\cite{frank1958liquid}.}.
Moreover, as in 2D, the active current in Eq.~\eqref{eq:Continuity_AN} cannot be derived from a free energy, therefore Eqs.~\eqref{eq:Continuity_AN} and \eqref{Hydro_AN_Real} cannot be obtained together at equilibrium.

We note that the discontinuous nature of the transition, in principle, prevents us from using a perturbative analysis to truncate the hierarchy around $\sigma_t$.
Indeed when the disordered solution starts to be unstable, the global nematic order, and thus $\varepsilon$, does not go continuously to zero.
However, Fig.~\ref{fig:AN_10_mode}(b) shows that the Ginzburg-Landau ansatz is a good approximation around $\sigma_c$, supporting the enslaving of the higher order modes to the nematic field in this region.
Moreover, as shown in the following, the hydrodynamic equations obtained at the usual third order are well-behaved, with bounded solutions. 
We expect them to continue providing the right qualitative picture. 

%%%%%%%%%%%%%%%%%%%%%%%%%%%
\subsubsection{Hydrodynamic equations for the rods class}
\label{subsubsec:Hydro_rods}
%%%%%%%%%%%%%%%%%%%%%%%%%%%

%%%%%%%%%%%%%%%%
\begin{table}[b!]
	\centering
\begin{tabular}{ l | c | c }
   & $3$D & $2$D \\
  \hline	
   & & \\
   $\mu^1[\rho]\,(<0)$ & $P^1_0 - 1 + \left( \frac{\pi}{8}P^1_0 - \frac{8}{15} \right)\rho -2a$  & $P_1 - 1 + \frac{4}{\pi}\left(P_1-\frac{4}{3}\right)\rho - 2a$  \\
   & & \\
   $\mu^3\,(<0)$ & $P^3_0 - 1 + \left( \frac{20-3\pi}{96}P^3_0 - \frac{208}{315} \right)\rho_0 -2a$  & $P_3 - 1  - \frac{272}{35\pi}\rho_0 - 2a$  \\
   & & \\
   $\beta\,(>0)$& $\frac{\left(\left(-3360 + 315\pi\right)P^1_0 + 512\right)\left(35\left(4+9\pi\right)P^3_0 + 512\right)}{21073920\mu^3}$ & $-\frac{32\left(5P_3+4\right)\left(7P_1-2\right)}{105\pi^2\mu^3}$ \\
   & & \\
   $\zeta\,(>0)$ & $\frac{16}{35} + \frac{3(4-\pi)}{16}P^1_0$ & $\frac{16}{5\pi}$ \\
   & & \\
   $\gamma\,(>0)$ & $\frac{1}{\mu^3} \left[\frac{4}{147} - \frac{5(32-3\pi)}{896}P^1_0\right]$ & $-\frac{4\left(7P_1-2\right)}{21\pi\mu^3}$ \\
\end{tabular}
	\caption{Comparison of the hydrodynamic coefficients of the equation for the polar field in the rods class in $3$ and $2$ dimensions~\cite{peshkov2012nonlinear}. The form of the equation does not change with respect to the $2$ dimensional case and thus no new parameter appears.
	In the right column the $P_i$ parameters are the moments of the angular noise distribution in 2D, analogous to the $P^i_0$ in 3D.}
	\label{tab:rods_polar_parameters}
\end{table}
%%%%%%%%%%%%%%%%

Even though only nematic order arises in the rods class, we retain the polar field in our description because its dynamics depends non-linearly on the density and the nematic fields.
In this case the reversal rate of velocities is sufficiently small so that no diffusive dynamics is attained on the kinetic timescales.
Therefore, we consider the propagative ansatz from Eq.~\eqref{eq:PropagativeAnsatz} for space and time.
Since only nematic order grows in such system, we consider its saturated value as a small parameter $\varepsilon\approx\vert g^2\vert$.
Balancing the modes in the Boltzmann hierarchy, we obtain the following ansatz for the relative strength of the fields
\begin{equation}
	\delta\rho\approx\varepsilon, \quad g^{2l}_m\approx g^{2l-1}_m \approx \varepsilon^{ l }, \qquad \forall l>1, \quad \forall m \;.
\end{equation}
The computation of the terms in the hydrodynamic equations follows the procedure described in the previous sections, 
truncating the Boltzmann hierarchy at order $\varepsilon^3$.
We thus retain equations for the fields up to $l=4$, but the $l=3,4$ fields can be enslaved to the $l=0,1,2$ fields.
After tedious calculations, we obtain the lengthy equations:
\begin{subequations}
\begin{eqnarray}
\partial_t \rho &=& - \partial_i w_i \;,\label{density_rods}\\
\partial_t w_i &=& -\partial_k Q_{ik} - \frac{1}{3}\partial_i\rho + \gamma\left(2Q_{kl}\partial_kQ_{il} + Q_{kl}\partial_i Q_{kl} - \frac{4}{5} Q_{ik}\partial_l Q_{kl}\right) \nonumber \\ 
&\,&+ \left(\mu^1[\rho] - \beta Q_{kl}Q_{lk}\right)w_i + \zeta Q_{ik}w_k - \frac{6}{5}\beta Q_{ik}Q_{kl}w_l \;, \label{w_rods}\\
\partial_t Q_{ij} &=& -\frac{2}{5}\left[\partial_iw_j\right]_\text{ST} + D_I \Delta Q_{ij} + D_A \left[\Gamma_{ik}Q_{kj}\right]_\text{ST} \nonumber \\
&\,& -\kappa\left(w_k\partial_kQ_{ij} + 2\left[w_k\partial_iQ_{kj} -\frac{2}{5}w_i\partial_kQ_{kj}\right]_\text{ST}\right) \nonumber \\
&\,& -\chi\left(\partial_k\left(w_kQ_{ij}\right) + 2\left[\partial_k\left(w_iQ_{kj}\right)-\frac{2}{5}\partial_i\left(w_kQ_{kj}\right)\right]_\text{ST}\right) \nonumber \\
&\,& + \left(\mu^2[\rho] - \xi Q_{kl}Q_{lk}\right)Q_{ij} + \alpha\left[Q_{ik}Q_{kj}\right]_\text{ST} \nonumber \\
&\,&+\omega\left[w_iw_j\right]_\text{ST}+\tau\left(|\vec{w}|^2Q_{ij} + \frac{6}{5}\left[Q_{ik}w_kw_j\right]_\text{ST}\right) \;, \label{Q_rods}
\end{eqnarray}
\label{eq:HYDRO_rods}
\end{subequations}
where implicit summation over repeated indices is assumed.
Although this representation differs from the compact form~\eqref{Hydro_AN_Real} for active nematics, 
we kept explicit notations for tensorial and inner products in order to avoid possible confusions between similar terms.
As before, square brackets take the symmetric traceless part of a tensor: $\left[A\right]_{ST}=\frac{1}{2}(A+A^t)-\frac{1}{3}\textbf{I}(\mathrm{Tr} A)$, and the operator $\mathbf{\Gamma}$, defined in \eqref{Gamma}, is $\Gamma_{ij} = \left[\partial^2_{ij}\right]_\text{ST}$.

The coefficients in the equation for the polar field are listed in Table~\ref{tab:rods_polar_parameters}.
The number of parameters does not change between the $2$ and the $3$ dimensional cases, although terms coupling the polar field with its gradients are still allowed by rotational symmetry.

The coefficients in the equation for the nematic field are listed in Table~\ref{tab:rods_nematic_parameters}.
Like in the case of active nematics, the nematic field equation has 
a Ginzburg-Landau term with a quadratic contribution with coefficient $\alpha$ and shows anisotropic spatial diffusion.
Both are not allowed for symmetry reasons in $2$ dimensions.
The coefficient in front of the anisotropic diffusion term is smaller in modulus than the isotropic diffusion one $\vert D_A\vert<\vert D_I\vert$, this prevents a trivial short wavelength instability to occur and we thus expect these equations to be well behaved.

%%%%%%%%%%%%%%%%
\begin{table}[b!]
	\centering
\begin{tabular}{ l | c | c }
   & $3$D & $2$D \\
  \hline	
   & & \\
   $\mu^2[\rho]$ & $P^2_0 - 1 + \left( \frac{8+3\sqrt{2}}{15}P^2_0 - \frac{68}{105} \right)\rho$  & $P_2 - 1 + \frac{16}{3\pi}\left(P_2(2\sqrt{2}-1)-\frac{7}{5}\right)\rho$  \\
   & & \\
   $\mu^4\,(<0)$ & $P^4_0 - 1 + \left( \frac{1}{6\sqrt{2}}P^4_0 - \frac{460}{693} \right)\rho_0$  & $P_4 - 1 + \frac{16}{15\pi}\left(P_4+\frac{155}{21}\right)\rho_0$  \\
   & & \\
   $D_I\,(>0)$& $-\frac{1}{5\mu^3}$ & $-\frac{1}{4\mu^3}$ \\
   & & \\
   $D_A\,(>0)$& $-\frac{6}{35\mu^3}$ & $0$ \\
   & & \\
   $\kappa$& $-\frac{44 + (24-45\sqrt{2})P^2_0}{735\mu^3}$ & $\frac{8\left(19-P_27(1+\sqrt{2})\right)}{105\pi\mu^3}$ \\
   & & \\
   $\chi$& $-\frac{512 + 35(4+9\pi)P^3_0}{7840\mu^3}$ & $\frac{2\left(5P_3+4\right)}{5\pi\mu^3}$ \\
   & & \\
   $\alpha$ & $\frac{4 + (21\sqrt{2}-16)P^2_0}{49} \, (>0)$ & $0$ \\
   & & \\
   $\xi\,(>0)$& $\frac{\left(304+(352-2233\sqrt{2})P^2_0\right)\left(16+(384+35\sqrt{2})P^4_0\right)}{528220\mu^4}$ & $-\frac{128\left(15P_4+1\right)\left(13-P_29(1+6\sqrt{2})\right)}{4725\pi^2\mu^4}$ \\
   & & \\
   $\omega$ & $\frac{8 + (15\sqrt{2}-32)P^2_0}{50}$ & $\frac{8\left(1 - P_2 3(\sqrt{2}-1)\right)}{3\pi}$ \\
   & & \\
   $\tau$ & $-\frac{\left(44 + (24-45\sqrt{2})P^2_0\right)\left(512+35(4+9\pi)P^3_0\right)}{823200\mu^3}$ & $-\frac{64\left(5P_3+4\right)\left(19-P_27(1+\sqrt{2})\right)}{525\pi^2\mu^3}$ \\
\end{tabular}
	\caption{
	Comparison of the hydrodynamic coefficients of the equation for the nematic field in the rods class in $3$ and $2$ dimensions~\cite{peshkov2012nonlinear}. The functional form of the linear ($\mu^2$) and the cubic ($\xi$) parameters is comparable between the two dimensions, whereas the quadratic term ($\alpha$) is zero in $2$ dimensions because of rotational invariance while it takes a positive value in $3$ dimensions.
	Moreover, anisotropic diffusion appears in 3D, like for the polar class.
	In the right column the $P_i$ parameters are the moments of the angular noise distribution in 2D, analogous to the $P^i_0$ in 3D.
	}
	\label{tab:rods_nematic_parameters}
\end{table}
%%%%%%%%%%%%%%%%

%%%%%%%%%%%%%%%%%%%%%%%%%%%%%%%
\subsection{Homogeneous and periodic solutions}
\label{subsec:Homogeneous_active_nematic}
%%%%%%%%%%%%%%%%%%%%%%%%%%%%%%%

\subsubsection{Homogeneous solutions}

\begin{figure}[t!]
	\centering
	\includegraphics[scale=0.5]{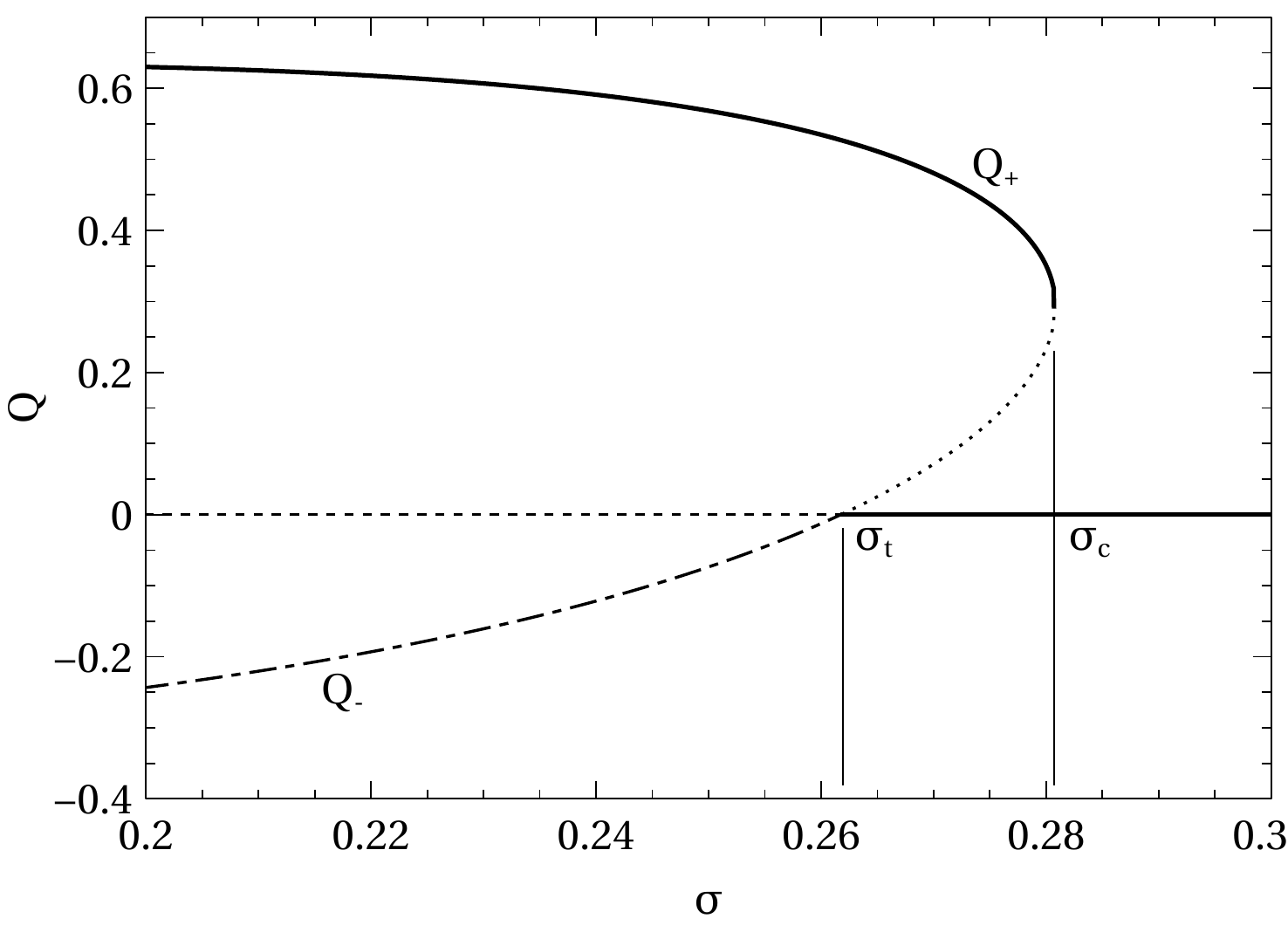}
	\caption{The homogeneous solutions $\bar{Q}$ of the hydrodynamic equations for particles with nematic interactions using the active nematics parameters of Table \ref{tab:AN_parameters} for $\rho_0=1$.
		Like at the kinetic level, there is a region in between $\sigma_t$ and $\sigma_c$ where the disordered and uniaxial solutions are both stable, showing  a ``true'' discontinuous transition.}
	\label{fig:N_order}
\end{figure}

Since no polar order can grow homogeneously in this case,
both the hydrodynamic equations for active nematics \eqref{Hydro_AN_Real} and for rods \eqref{eq:HYDRO_rods} share the same Ginzburg-Landau functional form when setting spatial derivatives to zero 
\begin{equation}
\partial_t \textbf{Q} = \left[\mu[\rho]-\xi(\textbf{Q}\,\textbf{:}\,\textbf{Q})\right]\textbf{Q} + \alpha\left[\textbf{Q}\textbf{Q}\right]_\text{ST} \;,
\end{equation}
with $\mu[\rho] = \mu^2[\rho]$.
This equation admits both the homogeneous disordered ($\rho=\rho_0$, $Q_{ij}=0$ $\forall i,j$) and homogeneous ordered solutions.
The case where all particles' orientations are aligned along a given direction is referred as \textit{uniaxial}.
Assuming without loss of generality that the direction of global order is along the $z$ axis, the nematic tensor reads
\begin{equation}
	\textbf{Q}=\begin{pmatrix} -\bar{Q}/2 & 0 & 0\\ 0 & -\bar{Q}/2 & 0\\ 0 & 0 & \bar{Q} \end{pmatrix} \;,
\end{equation}
where the parameter $\bar{Q}$ solves
\begin{equation}
	\mu+\frac{\alpha}{2}\bar{Q}-\frac{3}{2}\xi\bar{Q}^2=0 \;.
\end{equation}
This quadratic equation has two solutions
\begin{equation}
	\bar{Q}_\pm=\frac{\alpha}{6\xi}\pm\frac{1}{3\xi}\sqrt{\frac{\alpha^2}{4}+6\mu\xi} \;,
	\label{homogeneous_nematic}
\end{equation}
that are both real only for $\alpha^2/4+6\mu\xi\ge0$.
Since $\alpha$ and $\xi$ are strictly positive in the range of density and noise we consider (see Tables~\ref{tab:AN_parameters} and \ref{tab:rods_nematic_parameters}), we can define a critical linear coefficient
\begin{equation}
	\mu_c=-\frac{\alpha^2}{24\xi}\quad(<0) \;,
\end{equation}
such that these solutions exist for $\mu \geq \mu_c$. 
Thus $\sigma_c(\rho_0)$ is defined at the hydrodynamic level by the line where $\mu=\mu_c$, 
two homogeneous solutions stable with respect to homogeneous perturbations coexist in the region $\sigma_t<\sigma<\sigma_c$: 
the disordered solution and $\bar{Q}_+$, as shown  in Fig.~\ref{fig:N_order}.
Therefore, the transition from disorder to order is discontinuous, something already shown at kinetic level in 
Section~\ref{subsubsec:Hom_solution_B_eq}.
Remarkably, like in the polar case, the order is not a monotonous function of the noise, as shown in Fig.~\ref{fig:Cholesteric_AN},
whereas it is in the 2D case (not shown).

%%%%%%%%%%%%%%%%%%%%%%%%%%%%%%%
%%%%%%%%%%%%%%%%%%%%%%%%%%%%%%%

Below the transition line $\sigma_t$ the disordered solution becomes unstable and the solution $\bar{Q}_-$ is negative.
To get a physical insight of this solution we remark that a unit vector $\vec{n}$ can represent both a direction or the Hodge dual of this direction, 
corresponding to any plane orthogonal to $\vec{n}$ in $3$ dimensions.
The dual space can be represented in tensorial notations by
\begin{equation*}
	M_{ij}=\varepsilon_{ijk}n_k \;,
\end{equation*}
where $\{i,j,k\}\in\{1,2,3\}$ and $\varepsilon$ is the Levi-Civita totally antisymmetric tensor.
From the definition of the nematic order~\eqref{eq:Nematic_definition_real}, multiplying $\textbf{M}$ by its transpose, removing the trace, we obtain
\begin{equation}
	\textbf{M}\prescript{\mathrm t}{} {\textbf{M}}-\frac{2}{3}\textbf{I}=-\textbf{Q} \;.
\end{equation}
Thus, the negative tensor $-\textbf{Q}$ states that the order is orthogonal to the direction given by $\vec{n}$, while the positive tensor $\textbf{Q}$ states that the order is parallel to the direction $\vec{n}$.
\footnote{In $2$ dimensions $-\textbf{Q}$ corresponds to a rotation of the space of $\pi/2$ of the order because the geometric object perpendicular to a unit vector is a vector itself, instead of a plane.}

\begin{figure}[t!]
	\centering
	\includegraphics[scale=0.022]{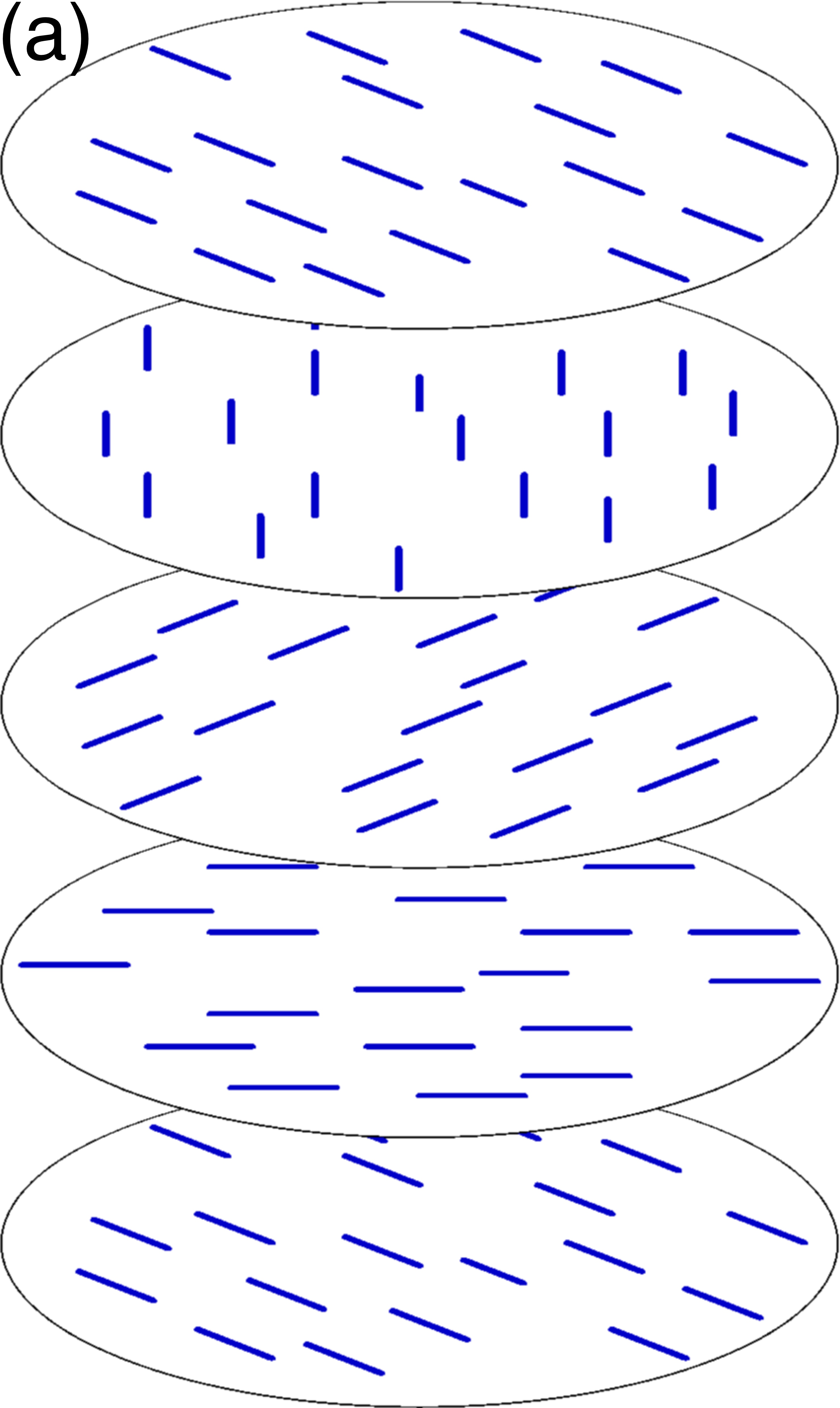}
	$\quad\quad\quad$
	\includegraphics[scale=0.022]{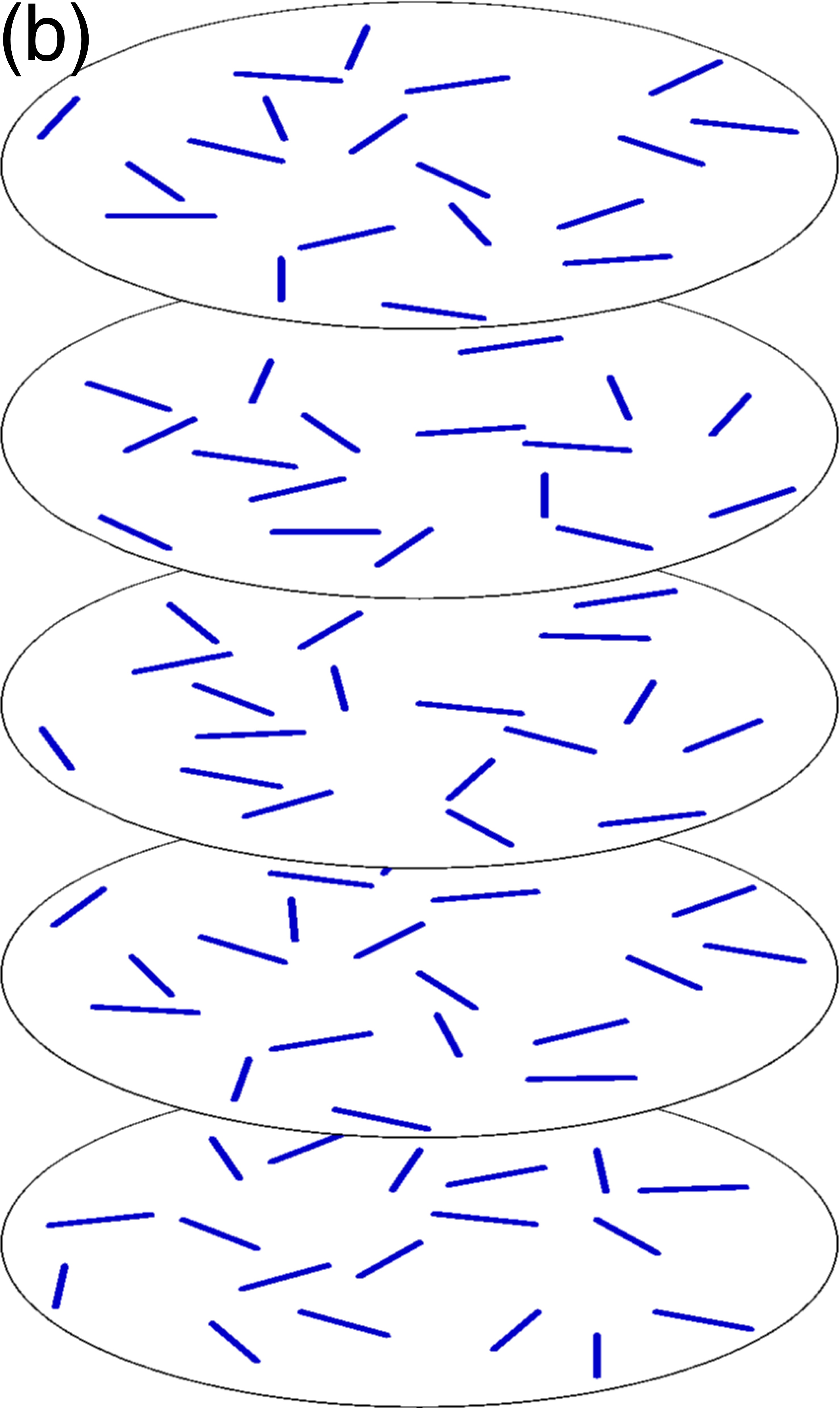}
	$\quad\quad\quad$
	\includegraphics[scale=0.022]{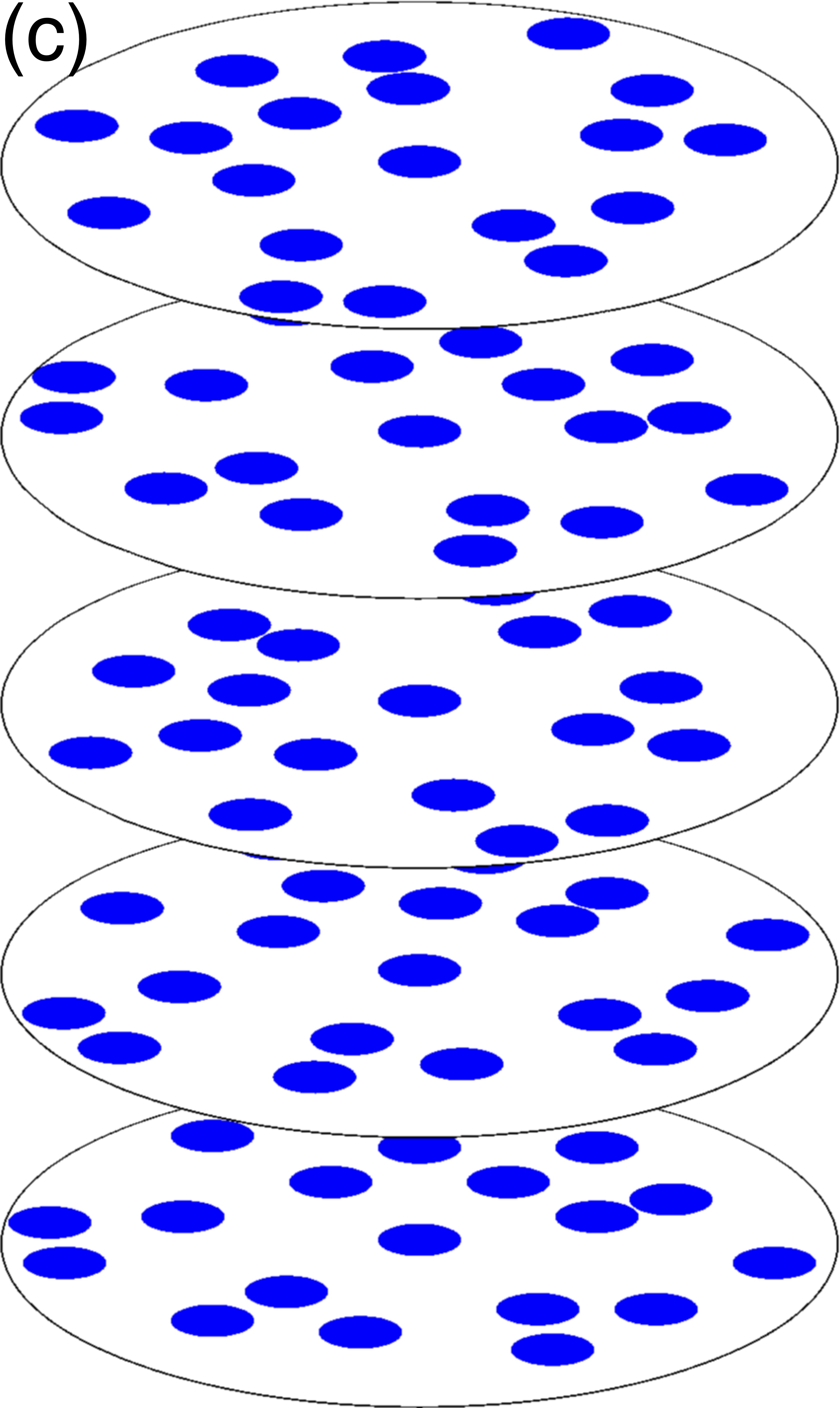}
	\caption{Sketch of configurations of particles in parallel layers leading to a nematic order parameter $\bar{Q}<0$.
	(a): cholesteric configuration where rod-like particles are aligned on parallel planes and order rotates along the third direction.
	In (b) the same particles are still arranged on parallel layers but with random orientations.
	(c): ordering of oblate particles on parallel layers.}
	\label{fig:negative_order}
\end{figure}

Solutions with negative nematic order are physically possible and represent so-called {\it biaxial} phases~\cite{degennes1995physics},  pictured in Fig.~\ref{fig:negative_order}.
There are two typical homogeneous configurations: 
planes of uniaxial particles with random orientations (in the planes), and
oblate particles moving along their axis, with their axes ordered.
An elementary linear stability analysis with respect to homogeneous perturbations shows that $\bar{Q}_-$ is always unstable whenever the quadratic coefficient $\alpha$ is positive.
For instance, in the case of active nematics, the homogeneous linear perturbations around this state evolve as
\begin{eqnarray*}
	\partial_t\delta Q_{zz} &=& \mp\bar{Q}_\pm\sqrt{6\xi(\mu-\mu_c)}\delta Q_{zz} \;,\\
	\partial_t\delta Q_{xx} &=& -\frac{3\alpha}{2} \bar{Q}_{\pm} \delta Q_{xx} + \bar{Q}_\pm\left(\frac{3}{2}\xi\bar{Q}_\pm -\alpha\right)\delta Q_{zz} \;,\\
	\partial_t\delta Q_{xy} &=& -\frac{\alpha}{2} \bar{Q}_{\pm} \delta Q_{xy} \;,\\
	\partial_t\delta Q_{xz} &=& -\frac{\alpha}{2} \bar{Q}_{\pm} \delta Q_{xz} \;,\\
	\partial_t\delta Q_{yz} &=& -\frac{\alpha}{2} \bar{Q}_{\pm} \delta Q_{xy} \;.
\end{eqnarray*}
In the region $\sigma_t<\sigma<\sigma_c$ the solution $\bar{Q}_-$ is positive and $\delta Q_{zz}$ is unstable, while below the transition point $\sigma_t$ it is negative and the instability is transferred to the other directions.
%%%%%%%%%%%%%%%%%%%%%%%%%%%%%%%%%%%%%%%%%%%
%%%%%%%%%%%%%%%%%%%%%%%%%%%%%%%%%%%%%%%%%%%

Physically, this instability relies on particle's shape.
Having a homogeneous negative order along $z$ for rod shaped agents, a small perturbation in the region of $\mu\le0$ brings the system to the disordered phase. 
On the contrary, if $\mu\ge0$ a small perturbation of the order in the ($x$,$y$) plane is not damped, aligning the rods in the same direction favoring the $\bar{Q}>0$ solution.
From the previous linear stability analysis, a stable biaxial nematic state requires the coefficient $\alpha$ to be negative.
This would be the case, for example, considering oblate agents with an alignment rule privileging biaxial phases, as shown in Fig.~\ref{fig:negative_order}(c).
However, this is beyond the scope of this work that deals only with rod shaped particles.

\subsubsection{Periodic solutions: Cholesterics}

The hydrodynamic equations for active nematics and rods possess spatially-periodic solutions (with zero polar field), called cholesteric, that depend on only one direction, say $x$. 
Borrowing from liquid crystals theory, a cholesteric phase is the assembly of identical 2 dimensional homogeneously ordered ``layers" into a helical structure (see Fig.~\ref{fig:negative_order}(a)).
In such solutions, the density and the local norm of the nematic tensor are homogeneous $\rho(x)=\rho_0$, $\|\textbf{Q}\|^2(x)=(\textbf{Q}\,\textbf{:}\,\textbf{Q})(x)=Q_0^2$.
Moreover, as the nematic field is constrained in the plane formed by the $y$ and $z$ axis we impose $Q_{xy}=Q_{xz}=0$ and $Q_{xx}=-\bar{Q}/2$ constant. 
With these constraints, and after some lengthy algebra, the hydrodynamic equations for the density and the nematic fields simplify to
\begin{subequations}
\begin{eqnarray}
	\left(\mu-\alpha\frac{\bar{Q}}{2}-\xi Q_0^2\right)\frac{\bar{Q}}{2}+\alpha\frac{Q_0^2}{3}&=&0 \;, \label{cholesteric_1}\\
	\left(\mu+\alpha Q_{zz}-\xi Q_0^2\right)Q_{zz}+\alpha\left(Q_{yz}^2-\frac{Q_0^2}{3}\right)+D\partial^2_{xx}Q_{zz}&=&0 \;, \label{cholesteric_2}\\
	\left(\mu+\alpha \frac{\bar{Q}}{2}-\xi Q_0^2\right)Q_{yz}+D\partial^2_{xx}Q_{yz}&=&0 \;,
	\label{cholesteric_3}
\end{eqnarray}
\end{subequations}
with $D=D_0-4D_1/21$ (or $D=D_I - D_A/3$ for rods) positive. 
Equation \eqref{cholesteric_1} gives a relation between $\bar{Q}$ and $Q_0$ 
and Eq.~\eqref{cholesteric_3} describes a harmonic oscillator
\begin{equation}
	\partial^2_{xx}Q_{yz}+\omega^2Q_{yz}=0 \;,
\label{eq_Qyz}
\end{equation}
with a twist frequency related to the norm of the nematic field
\begin{equation}
	D\omega^2=\mu+\alpha \frac{\bar{Q}}{2}-\xi Q_0^2 \;,
	\label{omega}
\end{equation}
As we are looking for solutions periodic along $x$, we are interested in the case where $\omega^2$ is positive.
Assuming then that the order is along the $z$-axis at $x=0$ a general solution of \eqref{eq_Qyz} can be written as
\begin{equation}
	Q_{yz}(x)=B\sin(\omega x) \;.
\label{Qyz}
\end{equation}

%%%%%%%%%%%%%%%%%%%%%%%%%
\begin{figure}[t!]
	\centering
	\includegraphics[scale=0.4]{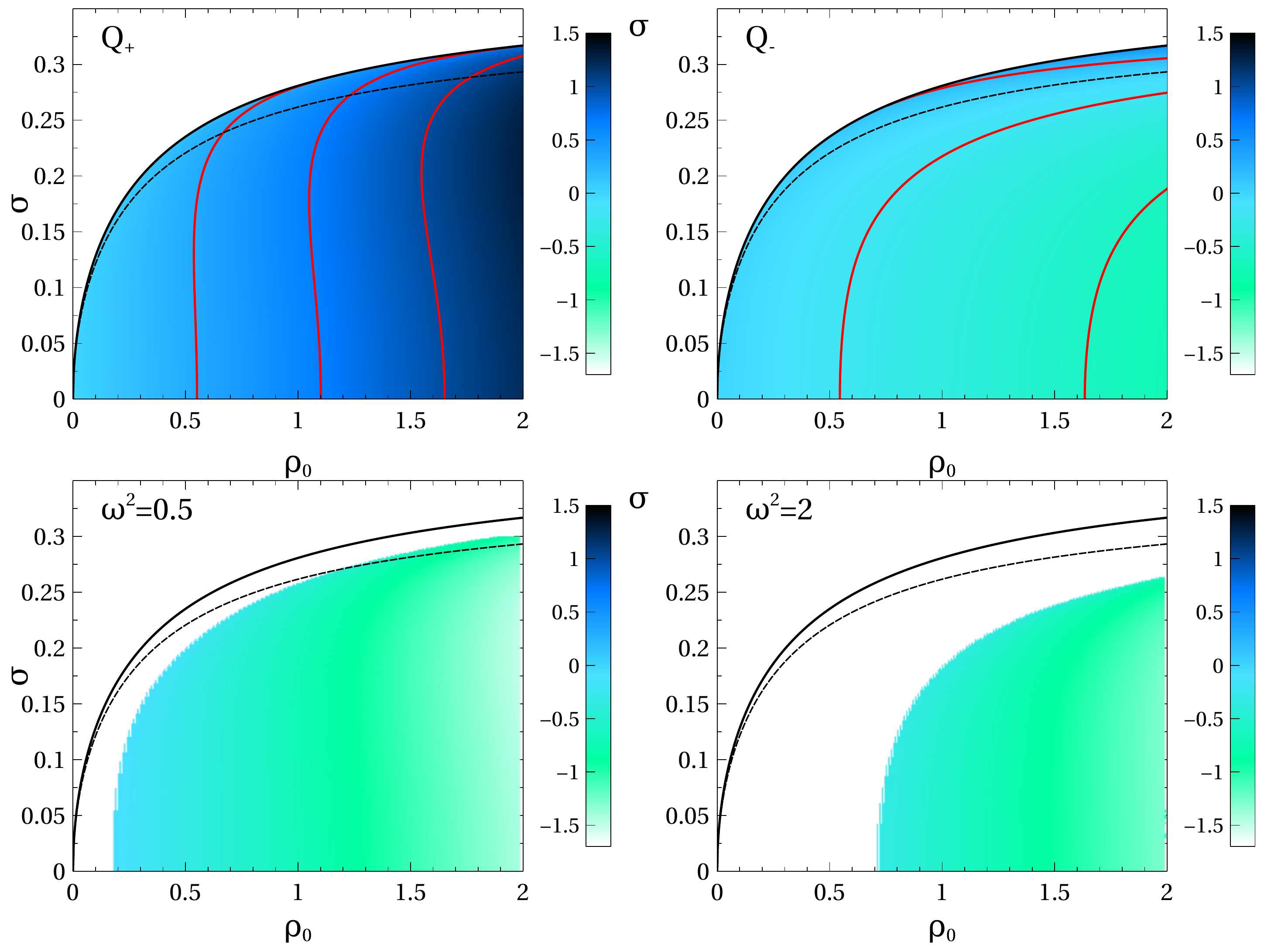}
	\caption{Phase diagrams in the density-noise plane indicating the degree of nematic order (scalar order parameter) for various 
	homogeneous ordered solutions of the hydrodynamic equations \eqref{Hydro_AN_Real}  for the active nematics class. In each panel,
	the solid and dashed black lines correspond to $\sigma_c$ and $\sigma_t$, respectively.
	 Top panels: uniaxial and biaxial solutions $\bar{Q}_+$ and $\bar{Q}_-$ where the red lines mark the contours $\bar{Q}_+ = 0.5,\,0.75,\,1$ and $\bar{Q}_- = 0,\,-0.25,\,-0.5$.
	 Bottom panels: order $Q_0$ of the cholesteric solution for twists $\omega^2=0.5$ and $2$.}
	\label{fig:Cholesteric_AN}
\end{figure}
%%%%%%%%%%%%%%%%%%%%%%%%%

Equations~\eqref{cholesteric_1} and~\eqref{omega} give the expressions of $\bar{Q}$ and $Q_0^2$ as a function of $\omega^2$ and of the hydrodynamic parameters
\begin{eqnarray}
\bar{Q} & = & \frac{\alpha}{6\xi}+\frac{D\omega^2}{2\alpha}+\sqrt{\frac{2}{3\xi}(\mu-\mu_c(\omega^2))} \;,\\
Q_0^2 & = & \frac{1}{\xi}\left(\mu-\frac{3D\omega^2}{4}+\frac{\alpha^2}{12\xi}+\frac{\alpha}{2}\sqrt{\frac{2}{3\xi}(\mu-\mu_c(\omega^2))}\right) \;,\\
\mu_c(\omega^2) & = & D\omega^2-\frac{3\xi}{8\alpha^2}\left(D\omega^2+\frac{\alpha^2}{3\xi}\right)^2 \;.
\end{eqnarray}
Note that in the limit $\omega^2\to0$ the period of the oscillations goes to infinity, and one recovers the homogeneous uniaxial phase.

In order to find a solution to Eq.~\eqref{cholesteric_2} we assume the functional form
\begin{equation}
	Q_{zz}(x)=\lambda\cos(\omega x)+\kappa \;,
\end{equation}
and we obtain
\begin{eqnarray}
	0&=&\left[(\mu-\xi Q_0^2)\kappa+\alpha\left(\kappa^2+\lambda^2-\frac{Q_0^2}{3}\right)\right] \nonumber\\
	&\,&+\lambda\left[\mu-\xi Q_0^2+2\alpha\kappa-D\omega^2\right]\cos(\omega x)+\alpha\left[B^2-\lambda^2\right]\sin^2(\omega x) \;.
\end{eqnarray}
For this equation to be satisfied, all three terms inside square brackets must vanish.
The first two correspond to the same constraint, resulting in $\kappa=\bar{Q}/4$, while the third term gives $\lambda=B>0$.
Finally, knowing the form of the nematic tensor we obtain 
\begin{equation}
	Q_0^2=\frac{3\bar{Q}^2}{8}+2B^2 \;,
\end{equation}
where $B^2$ has to be positive. 
Figure~\ref{fig:Cholesteric_AN} shows the regions of existence of the cholesteric states in the phase diagram ($\rho_0,\sigma$) plane for various twist frequencies $\omega$.
Larger the frequency of the solution is, the more it is confined at higher densities.

\subsection{Linear stability analysis}
\label{subsec:LinearStability_active_nematic}

\subsubsection{Homogeneous uniaxial ordered solution}

\begin{figure}[t!]
	\centering
	\includegraphics[scale=0.4]{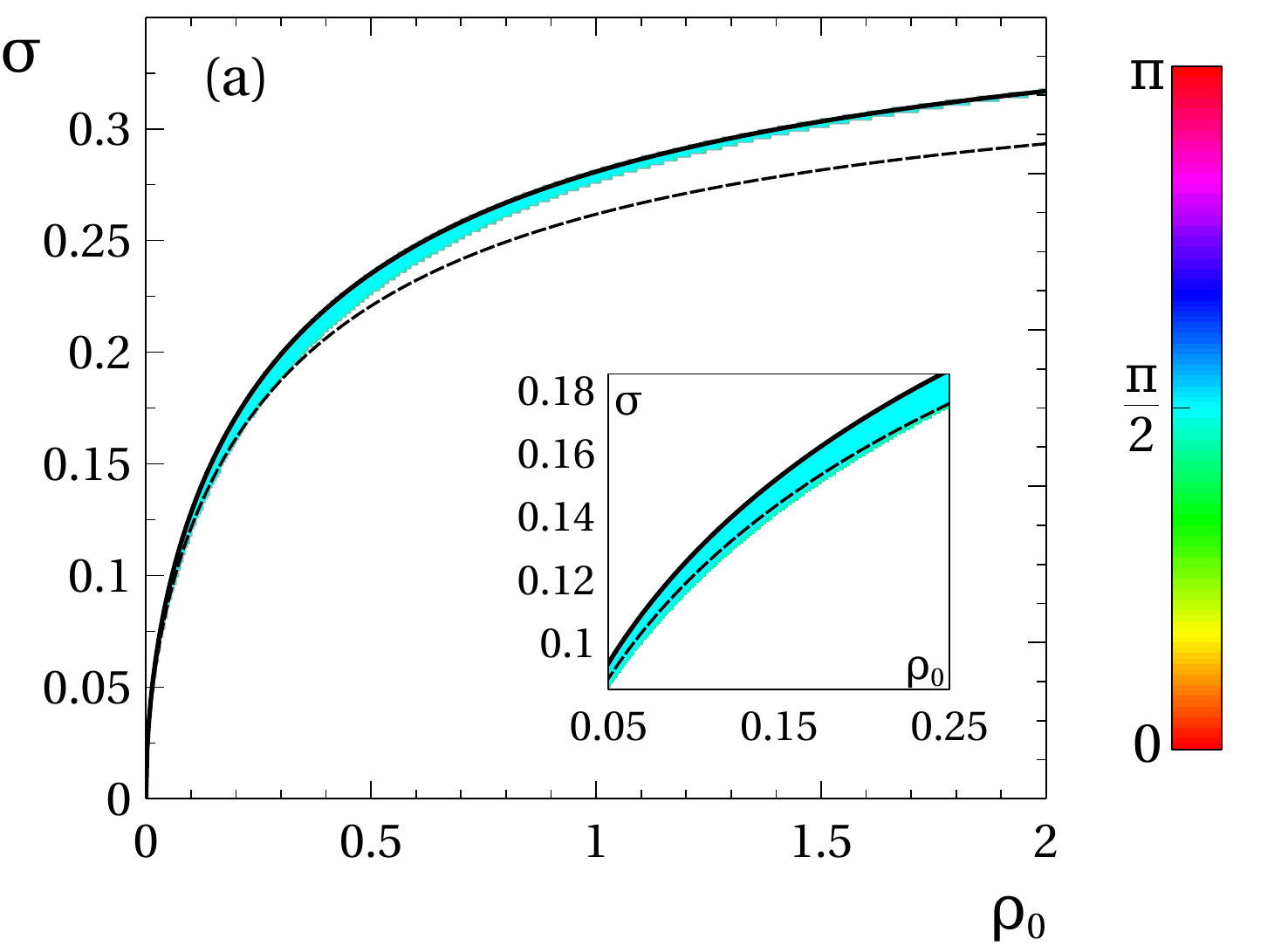}
	\includegraphics[scale=0.4]{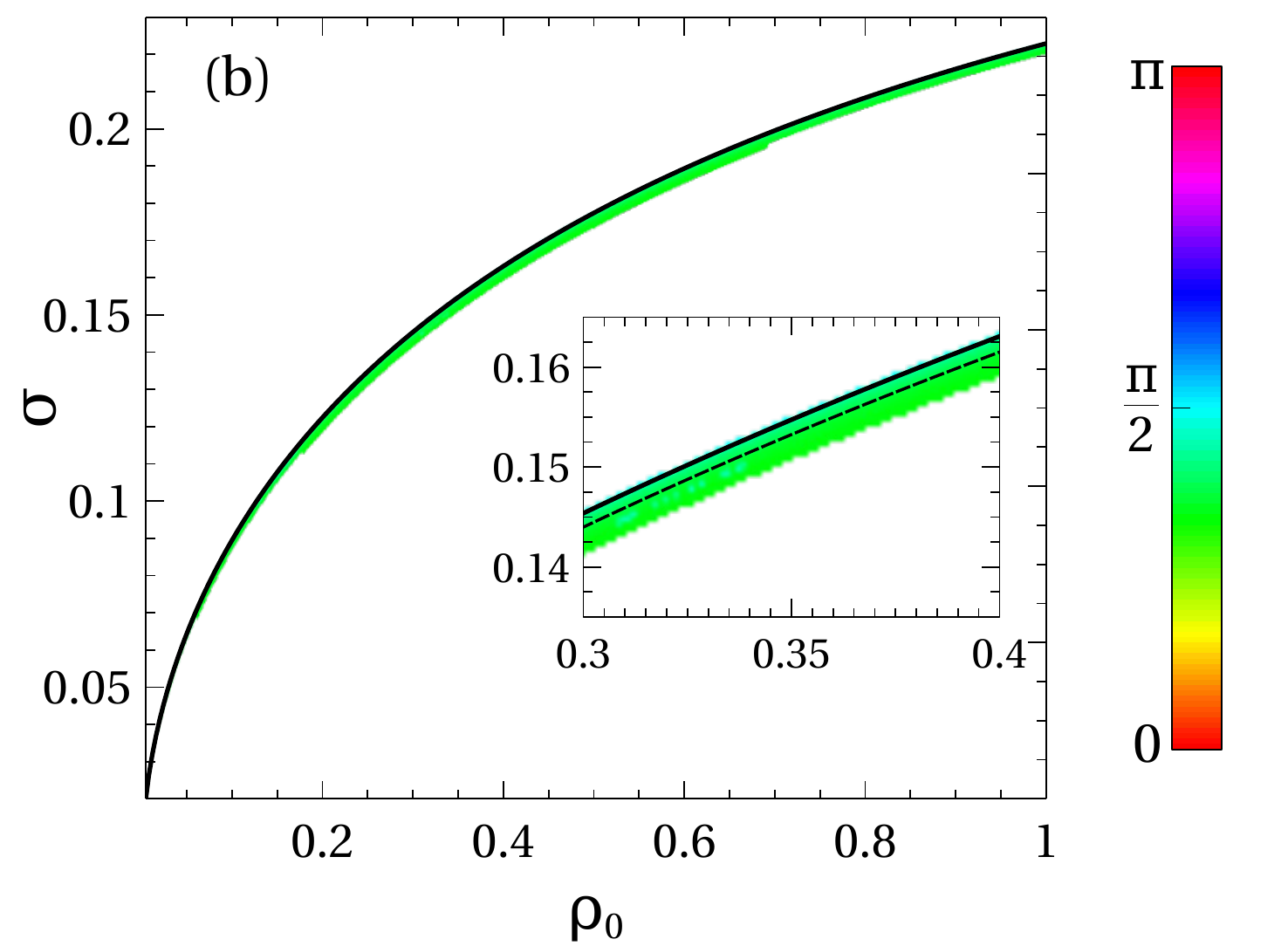}
	\caption{Linear stability of the homogeneous uniaxial ordered solution of the hydrodynamic equations for the nematic classes in the
	density-noise plane. 
		The solid and dashed black lines correspond to $\sigma_c$ and $\sigma_t$, respectively.
			The solution exists below the black solid line. It is linearly unstable in the colored region, and stable below the colored region.
	The color represents the angle between the most unstable wave vector and the direction of order.
	(a): active nematics case (Eqs.\eqref{eq:Continuity_AN} and \eqref{Hydro_AN_Real}). The instability, confined close to the upper transition line, is transversal to the direction of the order of the unperturbed solution. Inset: zoom at low densities where the linear instability covers the mean field bistability region although it does not go deeper in the ordered phase.
	(b): rods case (Eqs.\eqref{eq:HYDRO_REAL} at zero reversal rate ($a=0$)). 
	The instability, confined close to the upper transition line, is nearly transversal to the direction of the order of the unperturbed solution. Inset: zoom close to the transition line in order to show that the linear instability 
	covers all the region between the transition line (dashed black line) and critical line (full black line).
	}
	\label{fig:LinearStability_AN}
\end{figure}

We computed the linear stability analysis of the active nematics hydrodynamic equations \eqref{eq:Continuity_AN}
and \eqref{Hydro_AN_Real} around the uniaxial homogeneous ordered state $\bar{Q}$ semi-numerically (panel a of Fig.~\ref{fig:LinearStability_AN}).
Close to and below the line $\sigma_c$ limiting the existence of the ordered solution a transversal instability appears at finite wavelength, much like in the $2$ dimensional case.
Deeper in the ordered phase, the homogeneous order is stable and, again like in the 2D case, no ``spurious instability'' is found.
The transversal instability region becomes thinner increasing the density 
while the region where both the ordered and the disordered phase exist becomes larger.
Therefore, it is possible to find both the homogeneous phases linearly stable, resulting in a bistability of the system.
Note, however, that non-linear phenomena and strong fluctuations may invalidate this statement.

%%%%%%%%%%%%%%%%%%%%%%%%%%%%%%%%%%%
%%%%%%%%%%%%%%%%%%%%%%%%%%%%%%%%%%%
In the case of the rods equations the linear stability at finite wavelength of the nematic phase resembles that of the active nematics case, as shown in Fig.~\ref{fig:LinearStability_AN}(b). 
The ordered solution is unstable close to $\sigma_c$, but this instability is not purely transversal, 
and involves a component along the order, although not a dominant one.
Increasing the reversal rate of velocities $a$, this instability becomes transversal to the global order.
Moreover, for large values of $a$ the homogeneous ordered solution is stable in the bistability region at large densities, and one retrieves the active nematics structure.
There is, however, a strong difference between the 3D and 2D cases:
in $3$ dimensions one does not find an instability deep in the ordered phase where nematic order triggers polar order and no purely nematic solution is stable.

\subsubsection{Transverse linear stability of cholesteric solutions}

%%%%%%%%%%%%%%%%%%%%%%%%%%%%%
\begin{figure}[t!]
	\centering
	\includegraphics[scale=0.35]{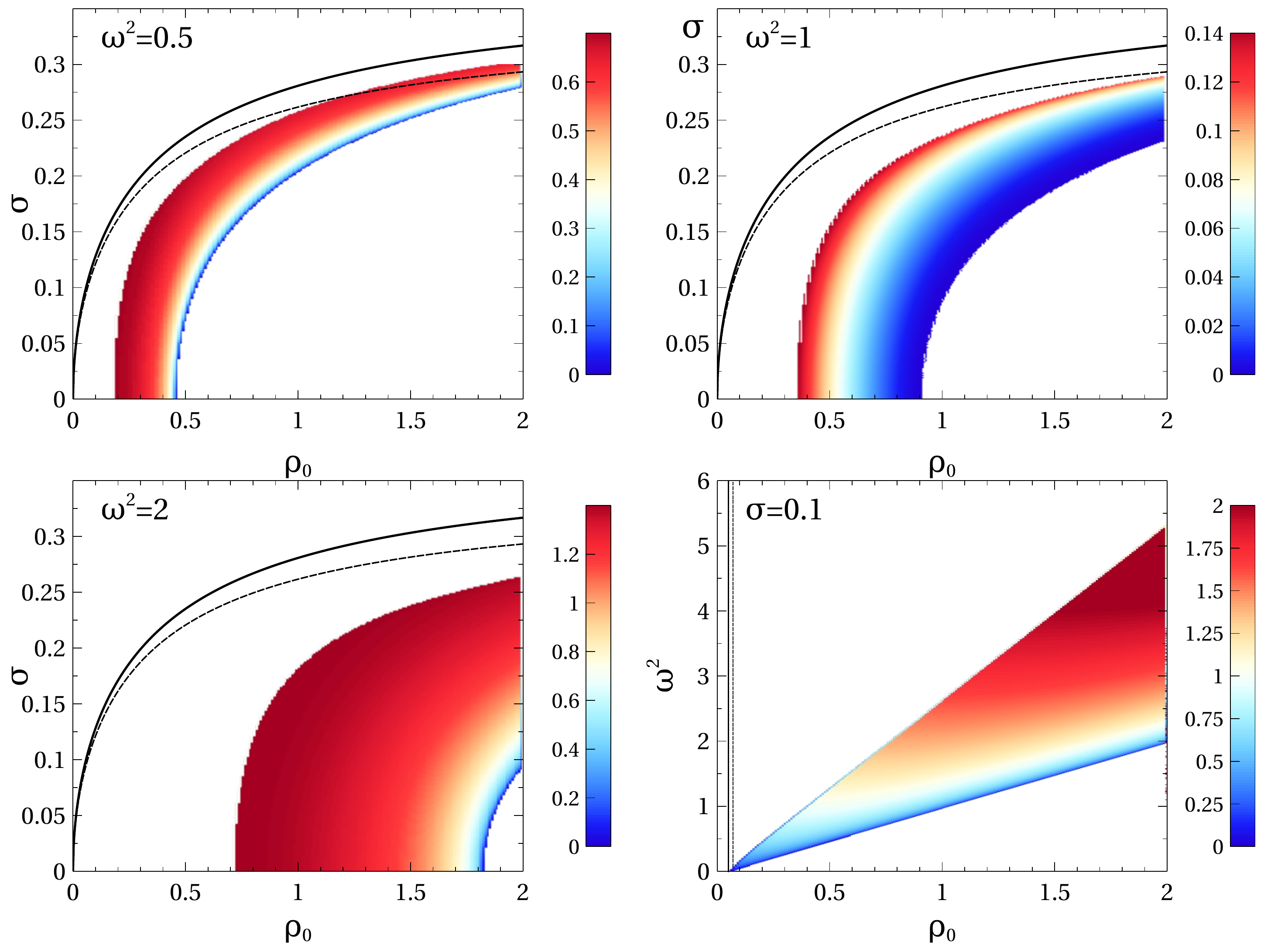}
	\caption{Growth rate of the instability of the cholesteric solution of the active nematics equation for various twists $\omega^2$. The dashed line represents the instability of the disordered solution and the full line above the existence of the homogeneous ordered solution. The first three panels show the linear stability of various fixed twist solutions in the density-noise plane. 
	The fourth, bottom right panel shows the linear stability at fixed noise $\sigma=0.1$ varying twist and density. 
	The lower limit of the colored area corresponds to $\omega^2 = \omega^2_u$, the upper limit of linear stability of the solution.
	}
	\label{fig:Lin_stab_cholesteric_GR}
\end{figure}
%%%%%%%%%%%%%%%%%%%%%%%%%%%%%

The linear stability analysis of the cholesteric solutions is not straightforward because of their spatial dependence, 
leading to non diagonal matrices in Fourier space. It can however be performed
for a family of particular perturbations given by the symmetry of the solutions.
The cholesteric steady solution can be written compactly in the form
\begin{equation}
\mathbf{Q_c}=\frac{\bar{Q}}{4}\mathbf{Q_0}+\frac{B}{2} \left(e^{\imath\omega x}\mathbf{Q_1}+e^{-\imath\omega x}\mathbf{Q_1}^*\right) \;,
\end{equation}
with $\bar{Q}$ and $B$ calculated in the previous section and
\begin{equation}
\mathbf{Q_0}=\begin{pmatrix} -2 & 0 & 0\\ 0 & 1 & 0\\ 0 & 0 & 1\end{pmatrix}\quad \mathbf{Q_1}=\begin{pmatrix} 0 & 0 & 0\\ 0 & -1 & -\imath\\ 0 & -\imath & 1\end{pmatrix} \;. \label{Q_0&Q_1}
\end{equation}
We consider perturbations of the form of spatially-varying amplitudes.
The nematic field then becomes
\begin{equation}
	\mathbf{Q}=A_0(x,t)\mathbf{Q_0}+A_1(x,t)e^{\imath\omega x}\mathbf{Q_1}+A_1^*(x,t)e^{-\imath\omega x}\mathbf{Q_1}^* \;,
	\label{Q_amplitudes}
\end{equation}
where $A_0$ is real and $A_1$ is complex.
For this family of perturbations the twist $\omega$ is kept constant such that they are transversal to the cholesteric axis.
In the active nematics setting, the coupled equations for the density $\rho$ and the amplitudes $A_0$ and $A_1$ are
\begin{subequations}
\begin{eqnarray}
	\partial_t\rho & = & D_0\partial^2_{xx}\rho-2D_1\partial^2_{xx}A_0 \;,\\
	\partial_t A_0 & = & \left(\mu[\rho]-\alpha A_0-2\xi\left(3A_0^2+4|A_1|^2\right)\right)A_0 +\frac{4}{3}\alpha|A_1|^2\nonumber \\
	&\,&+\left(D_0+\frac{4D_1}{21}\right)\partial^2_{xx}A_0-\frac{2D_1}{45}\partial^2_{xx}\rho \;,\\
\partial_t A_1 & = & \left(\mu[\rho]+2\alpha A_0-2\xi\left(3A_0^2+4|A_1|^2\right)\right)A_1 \nonumber \\
	&\,&+\left(D_0-\frac{4D_1}{21}\right)\left(\partial_x+\imath\omega\right)^2A_1 \;.
\end{eqnarray}
\end{subequations}
Figure~\ref{fig:Lin_stab_cholesteric_GR} shows the numerical evaluation of the linear stability of this set of equations around the fixed point
\begin{equation}
	\rho = \rho_0 \quad ; \quad \bar{A}_0 = \frac{\bar{Q}}{4} \quad ; \quad \bar{A}_1 = \frac{B}{2} \;.
	\label{sol_cholesteric_A}
\end{equation}
For $\omega=0$, we of course recover the phase diagram shown in Fig.~\ref{fig:LinearStability_AN}.
For finite $\omega$ we observe that the cholesteric solutions are linearly unstable close to their existence line, 
while they are stable deeper in the ordered phase. 
Note, however, that the instability region grows quickly with twist.

%%%%%%%%%%%%%%%%%%%%%%%%%%%%%
\begin{figure}[t!]
	\centering
	\includegraphics[scale=0.027]{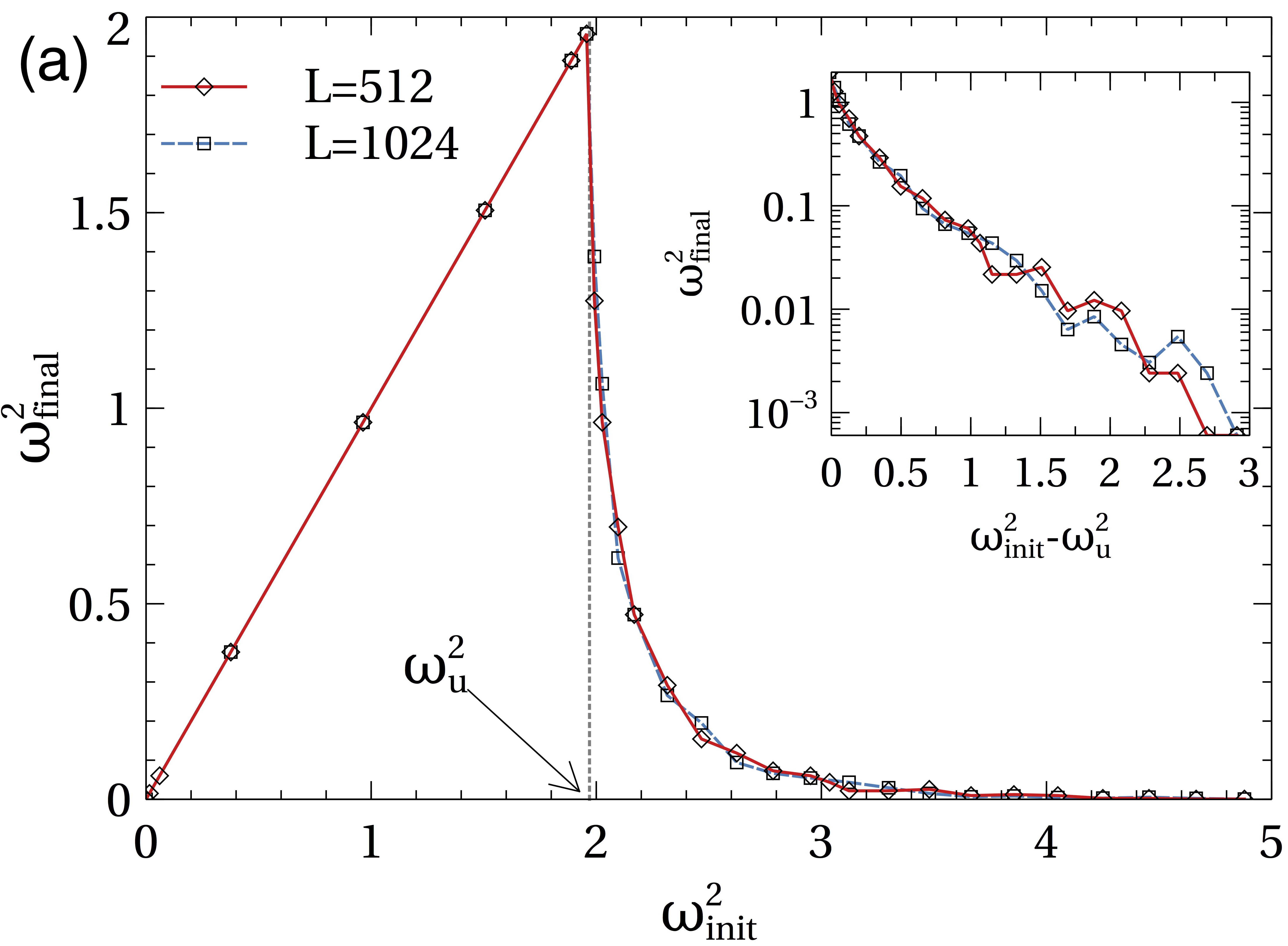} \;
	\includegraphics[scale=0.27]{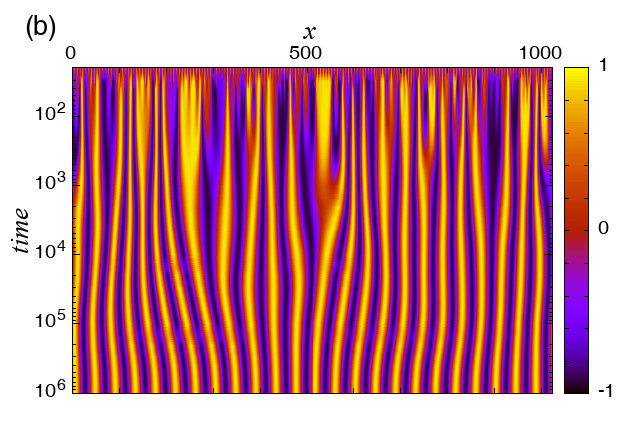}
	\caption{
		(a): The final twist of the cholesteric state $\omega_{\rm final}^2$ found in numerical simulations of the one dimensional version of the active nematics equations~\eqref{eq:Continuity_AN} and \eqref{Hydro_AN_Real} (see text), as function of the twist of the initial cholesteric configuration $\omega_{\rm init}^2$ for system sizes $L=512$ and $1024$.
		The cholesteric solutions are stable as long as $\omega_{\rm init}^2$ is smaller than the linear stability threshold $\omega_{\rm u}^2$, while above it they are unstable and converge to a new cholesteric solution whose twist depends on $\omega_{\rm init}^2 - \omega_{\rm u}^2$.
		The inset suggest a possible exponential decay of $\omega_{\rm final}^2$ as function of $\omega_{\rm init}^2 - \omega_{\rm u}^2$.
		(b): Space time variations of the $Q_{yz}$ component of the nematic tensor ($\sim B\sin(\omega x)$ for a cholesteric solution) starting from a cholesteric configuration with $\omega_{\rm init}^2 - \omega_{\rm u}^2 \simeq 1.69$ and ending with $\omega_{\rm final}^2 \simeq 0.02 $ at system size $L=1024$.
		All the simulations are performed at $\rho_0=2$ and $\sigma=0.1$ ($\omega_u^2\simeq 1.97$), with an Euler scheme with resolutions $dt=0.01$ and $dx=\frac{1}{8}$.
}
	\label{fig:Cholesteric_1D_simulation}
\end{figure}
%%%%%%%%%%%%%%%%%%%%%%%%%%%%%

To get an idea of the respective stability of cholesteric solutions, 
we performed numerical simulations of the hydrodynamic equations \eqref{eq:Continuity_AN} 
and \eqref{Hydro_AN_Real} but in their reduced one-dimensional form, {\it i.e.} setting $\partial_{y}$ and $\partial_z$ to $0$, 
and using a periodic domain in $x$.
As expected, the cholesteric solutions with $\omega^2 < \omega_u^2$, 
where $\omega_u^2$ is the threshold given by the linear stability analysis performed above, 
are stable (Figure~\ref{fig:Cholesteric_1D_simulation}(a)).
On the other hand in the linear instability region, we find that
a cholesteric initial configuration of twist $\omega^2_{\rm init}$
typically settles to a stable, lower-twist, cholesteric solution (see Fig.~\ref{fig:Cholesteric_1D_simulation}(b)).
We find that the final cholesteric solution possesses a well defined twist $\omega^2_{\rm final}$ 
independent of system size (for large-enough domains).
This final twist decreases quickly when $\omega^2_{\rm init}$ increases, {\it i.e.} when the solution is more and more unstable, 
as shown in the last panel of Fig.~\ref{fig:Lin_stab_cholesteric_GR}. 
Finally, as we can only simulate finite systems, the homogeneous ordered uniaxial solution is always reached for large-enough unstable initial twist.

For rods one also needs to consider perturbations of the polar field $\vec{w}$. Those compatible with \eqref{Q_amplitudes} read
\begin{equation}
	\mathbf{w} = F_0(x,t) \mathbf{w_0} + F_1(x,t)e^{\frac{\imath\omega x}{2}}\mathbf{w_0} + F_1^*(x,t)e^{-\frac{\imath\omega x}{2}}\mathbf{w_1}^*
\end{equation}
where the complex vectors are
\begin{equation}
	 \mathbf{w_0} = \begin{pmatrix} 1 \\ 0 \\ 0 \end{pmatrix} \quad \quad \mathbf{w_1} = \begin{pmatrix} 0 \\ -\imath \\ 1 \end{pmatrix}
\end{equation}
and the amplitudes $F_0$ and $F_1$ are equal to zero in the unperturbed state.

Inserting the perturbed cholesteric solution into the hydrodynamic equations for rods \eqref{eq:HYDRO_rods} we obtain coupled equations for the density and the amplitudes
\begin{subequations}
\begin{eqnarray}
\partial_t \rho & = & - \partial_x F_0 \;,\\
\partial_t F_0 & = & 2\partial_x A_0 - \frac{1}{3}\partial_x \rho + \gamma\partial_x\left(\frac{27}{5}A_0^2 + 4|A_1|^2\right) \nonumber \\
& &  + \left[\mu^1[\rho] - 2\zeta A_0 - 2\beta\left(\frac{27}{5}A_0^2 + 4|A_1|^2\right)\right] F_0 \;, \\
\partial_t F_1 & = & \left[\mu^1[\rho]  - \frac{4\beta}{5}\left(9A_0^2 + 16|A_1|^2\right)\right] F_1 - \frac{24}{5}\beta A_0A_1F_1^* + \zeta\left(A_0F_1 + 2A_1F_1^* \right)\,,\;\;\;\;\;\;\;\;\;\; \\
\partial_t A_0 & = & \frac{2}{15}\partial_x F_0 + \left(D_\text{I} + \frac{2}{3} D_\text{A}\right)\partial_{xx}^2 A_0 - \frac{9}{5}\left(\kappa F_0\partial_x A_0 + \chi \partial_x(F_0A_0)\right) \nonumber \\
& & + \left[\mu^2[\rho] - \alpha A_0 - 2\xi \left(3A_0^2 + 4|A_1|^2 \right)\right] A_0 + \frac{4}{3}\alpha|A_1|^2 \nonumber\\
& &  - \frac{\omega}{3}\left(F_0^2 - 2|F_1|^2\right)+ \tau\left[ \frac{9}{5}\left( F_0^2 + 4|F_1|^2\right)A_0 - 8 \Re\left( A_1^* F_1^2 \right)\right] \;, \\
\partial_t A_1 & = & \left(D_\text{I} - \frac{2}{3} D_\text{A}\right)(\partial_{x} + \imath\omega)^2 A_1 - \kappa A_0 (\partial_{x} + \imath\omega)A_1 - \chi (\partial_{x} + \imath\omega) (A_0A_1)  \nonumber \\
& & + \left[\mu^2[\rho] + 2\alpha A_0 - 2\xi \left(3A_0^2 + 4|A_1|^2 \right)\right] A_1 \nonumber \\
& & + \left( \omega + \frac{12}{5}\tau A_0\right)F_1^2 + \tau\left(F_0^2 + \frac{44}{5}|F_1|^2\right)A_1 \;.
\end{eqnarray}
\end{subequations}
Linear stability analysis of these equations gives results similar to the active nematics case, shown in Fig.~\ref{fig:Lin_stab_cholesteric_GR}, {\it i.e.} a region of instability close to the limit of existence of the solution whose extension grows with $\omega^2$.
This instability region also increases with the reversal rate, but its size saturates in the limit of large $a$ such that the active nematics picture is recovered.

%%%%%%%%%%%%%%%%%%%%%%%%%%%%%%%%%%%%%
\section{Conclusion}
\label{sec:conclusion}
%%%%%%%%%%%%%%%%%%%%%%%%%%%%%%%%%%%%%

%%%XXX discuss effect of reversal rate in general?

%%% mention that results on cholesteric are partial and will have to be confirmed in full 3D simulations

We have derived the hydrodynamic equations of the three main classes of dry, aligning, dilute active matter in three spatial dimensions, and compared them to their two-dimensional counterparts. We have used the Boltzmann-Ginzburg-Landau approach that, by construction, yields well-behaved partial differential equations governing the main physical fields. 
For convenience, we first treated the polar class with ferromagnetic alignment, then the cases with nematic alignment, \textit{i.e.} the fast-velocity-reversal limit of active nematics, and the slow-reversal case, including the zero-reversal ``rods" limit. 

For the polar ferromagnetic class, we find the classic Toner-Tu equations, although here, starting from a Vicsek-style model, we obtain an anisotropic diffusion term not present in 2D using the same approach. 
We also find other differences between 2D and 3D, notably the non-monotonicity of order as a function of noise strength for the 3D homogeneous order solution. 
The linear stability of the spatially-homogeneous solutions of the 3D Toner-Tu equations is similar to that of the 2D case: the ordered solution shows a finite wavelength longitudinal instability near the continuous onset of order, and a residual, ``spurious'' instability deep in the ordered phase.

In the case of nematic alignment, the differences with the 2D case are much more pronounced: first of all, the general scenario departs from the ubiquitous liquid-gas phase separation found in 2D, as the transition to order is found discontinuous even at the mean-field level studied here.
This situation is related to the presence of a quadratic term in the nematic field equation, a term ruled out by rotational symmetry in 2D. 
There is thus a genuine bistability region defining hysteresis loops at the level of homogeneous solutions. 
Nevertheless, we find that the homogeneous ordered solution still retains the generic transversal long-wavelength instability present in 2D, complicating further the mean-field phase diagram.
In addition, we showed that the 3D hydrodynamic equations with nematic alignment support biaxial periodic solutions corresponding to cholesteric configurations. 
We show that these solutions too are generically unstable near their existence limit, and we discuss their relative stability. 
We note finally that, as in 2D, we find no qualitative difference between the active nematics and rods cases at the level considered here.

Naturally, this work now calls for further studies at the nonlinear and fluctuating levels. 

At the nonlinear but still deterministic level, the inhomogeneous solutions of the hydrodynamic equations derived here must be found. 
As in the 2D case, we expect them to exist beyond the narrow band of linear instability of the homogeneous order solution 
where no homogeneous solutions exist. 
In the phase-separation framework described in the introduction, the lines delimiting this region are the spinodal lines. 
The binodal lines, which ultimately delimit the domain of existence of the coexistence phase sketched in Fig.~\ref{fig:Vicsek_phase_diag}, are determined by the existence and stability limits of inhomogeneous solutions.

For the polar case, we expect these inhomogeneous solutions to take the form of travelling dense sheets as observed in microscopic models \cite{chate2008collective}. 
For the nematic alignment cases, as of now, not much is known: the only published account of the structures observed in 3D can be found in \cite{chate2008modeling}, where a dense ordered cylinder with its axis along the global nematic order is shown for the Vicsek-style active nematics model. 

The fluctuating level of either microscopic models or our hydrodynamic equations complemented by stochastic terms remains, as of now, essentially virgin territory. 
Studies of the 3D Vicsek model and others in the same class have revealed the emergence of the dense traveling sheets
mentioned above, but no work has studied in depth the microphase vs macrophase separation scenario found, in 2D,
to distinguish the active Ising and the Vicsek models \cite{solon2015phase}. 
Similarly, the standing of the Toner-Tu theoretical predictions in 3D is unknown.
For nematic alignment, it is fair to say that almost everything remains to be done.
An interesting study of a microscopic model of self propelled rods in 3D has shown the existence and stability of cholesteric solutions that coexist with the homogeneous ordered nematic phase, but it remains rather partial \cite{breier2016spontaneous}.
 %%%XXX other papers in 3D active nematics? 
 
Our ongoing work is devoted to the above endeavours: careful study of 3D Vicsek-style models, search for inhomogeneous
solutions to the hydrodynamic equations derived here, and the eventual complete understanding of 3D dry aligning active matter 
at the fluctuating hydrodynamic level.

\bibliographystyle{ieeetr}
\bibliography{Biblio}

\end{document}